\documentclass[twocolumn]{aastex62}

\usepackage{natbib}

\usepackage{lineno}

\usepackage[T1]{fontenc}

\graphicspath{{./}{figures/}}

\received{August 23, 2022}
\revised{December 2, 2022}
\accepted{December 21, 2022}

\shorttitle{ONC Chemical Modeling}
\shortauthors{Boyden \& Eisner}

\begin{document}

\title{{\bf {\large Chemical modeling of Orion Nebula Cluster disks: evidence for massive, compact gas disks with ISM-like gas-to-dust ratios}}}

\correspondingauthor{Ryan Boyden}
\email{rboyden@arizona.edu}

\author[0000-0001-9857-1853]{Ryan D. Boyden}
\affil{Steward Observatory, University of Arizona, 933 N. Cherry Ave, Tucson, AZ, 85719, USA}

\author{Josh Eisner}
\affil{Steward Observatory, University of Arizona, 933 N. Cherry Ave, Tucson, AZ, 85719, USA}



\begin{abstract}

{The stellar cluster environment is expected to play a central role in the evolution of circumstellar disks. We use thermochemical modeling to constrain the dust and gas masses, disk sizes, UV and X-ray radiation fields, viewing geometries, and central stellar masses of 20 Class II disks in the Orion Nebula Cluster (ONC). We fit a large grid of disk models to $350$ GHz continuum, CO $J=3-2$, and HCO$^+$ $J=4-3$  ALMA observations of each target, and we introduce a procedure for modeling interferometric observations of gas disks detected in absorption against a bright molecular cloud background. We find that the ONC disks are massive and compact, with typical radii $<100$ AU, gas masses $\geq10^{-3}$ $M_{\odot}$, and gas-to-dust ratios $\geq100$. The ISM-like gas-to-dust ratios derived from our modeling suggest that compact, externally-irradiated disks in the ONC are less prone to gas-phase CO depletion than the massive and extended gas disks that are commonly found in nearby low-mass star-forming regions. The presence of massive gas disks indicates that external photoevaporation may have only recently begun operating in the ONC, though it remains unclear whether other cluster members are older and more evaporated than the ones in our sample. Finally, we compare our dynamically-derived stellar masses with the stellar masses predicted from evolutionary models and find excellent agreement. Our study has significantly increased the number of dynamical mass measurements in the mass range $\leq 0.5$ $M_{\odot}$, demonstrating that the ONC is an ideal region for obtaining large samples of dynamical mass measurements towards low-mass M-dwarfs. }

\end{abstract}


\section{\bf{Introduction}} \label{sec:intro}

Protoplanetary disks form in a variety of environments, and the natal environment has the potential to shape the evolution of disks and planetary systems \citep[for a recent review, see][]{Parker20}. The most common environments of star formation are young stellar clusters \citep[e.g., ][]{Lada93, Lada03, Krumholz19}. Understanding how disks evolve in clustered star-formation environments is therefore essential for interpreting the demographics of disks and exoplanets \citep[e.g.,][]{Ndugu18, Winter20b}, and examining how our own solar system fits into the broader exoplanet population \citep[][and references therein]{Adams10}.

A key property of young stellar clusters is the presence of massive O- and B-type stars. Strong UV fields from massive stars launch photoevaporative winds from the surfaces of disks surrounding lower-mass cluster members through a process known as external photevaporation \citep[e.g.,][]{Johnstone98, Storzer99}. In regions of high stellar density and radiation field strength, external photoevaporation acts as the dominant mechanism of disk dispersal \citep{Scally01, Winter18, ConchaRamirez19}, and can effectively deplete the disk of planet-forming material within $<1$ Myr of cluster evolution \citep[][]{Adams04, Clarke07, Haworth18, Winter20a, ConchaRamirez21}. Recent numerical work has shown that even in milder-UV environments, external photoevaporation is expected to play a dominant role in circumstellar disk evolution \citep[][]{Facchini16, Haworth18, Haworth18b, Haworth19}. The rapid reduction of the disk mass, size, and lifetime due to external photoevaporation indicates that if giant planets form in clustered environments, then they must form very early \citep[$<$ 1 Myr; e.g., ][]{Nicholson19, Parker21b}, which appears to be occurring in a number of low-density star-forming regions lacking massive stars \citep[e.g.,][]{Sheehan17, Sheehan18, Alves20, SeguraCox20}.

Spatially-resolved {\it Hubble Space Telescope (HST)} and {\it Karl G. Jansky Very Large Array (VLA)} observations of the Orion ``proplyds'' have provided the most compelling evidence of external photoevaporation in action. These observations reveal ionized cocoons of gas surrounding hundreds of low-mass stars in the Orion Nebula Cluster \citep[ONC; e.g.,][]{Churchwell87, Odell94, Bally98, Bally00, Ricci08}, where the morphologies of the proplyds can be explained by strong far-ultraviolet (FUV) and extreme-ultraviolet (EUV) radiation originating from the O-star $\theta^1$ Ori C \citep[e.g.,][]{Storzer99}. While the proplyds in the ONC have long been regarded as the prototypical evaporating disks, candidate proplyds are now being found in other clusters spanning a range of stellar density, UV irradiation, and age \citep[e.g.,][]{Smith10, Fang12, Kim16, Haworth21b}. The discovery of proplyds outside of the ONC, together with the high photoevaporation rates inferred for many of these systems \citep[$10^{-8} - 10^{-6}$ M$_{\odot}$ yr$^{-1}$;][]{Henney99, Kim16, Haworth21b}, demonstrates that external photoevaporation is operating in a range of Galactic star formation environments.

Submillimeter-wavelength observations provide a means to probe the dust and molecular gas content of externally-irradiated  disks, including those not detected as proplyds. The high sensitivity and resolution achieved by the Atacama Large Millimeter Array (ALMA) has enabled a detailed statistical characterization of disk properties in the nearby clusters of Orion \citep[e.g., ][]{Mann14, Ansdell17, Ansdell20, Eisner16, Eisner18, vanTerwisga20, Otter21}. The mm-wavelength surveys find that dust disks in clustered environments are more compact and substantially less massive than the disks typically found in lower-density regions, including nearby low-mass star-forming regions \citep[e.g., ][]{Andrews13, Ansdell16, Pascucci16} as well as regions in Orion beyond the photoionization field of massive stars \citep[e.g.,][]{vanTerwisga19a, Grant21}. Many of the differences in the inferred disk mass and size distributions are supported by current models of external photoevaporation \citep[e.g.,][]{Winter19, Sellek20, ConchaRamirez21}, although alternative explanations are also emerging \citep[e.g.,][]{Kuffmeier20}. A few of the surveys also find correlations between disk properties and the (projected) distance from the ionizing source \cite[e.g., ][]{Mann14, Ansdell17, Eisner18}, though recent simulations indicate that these trends may not necessarily be attributed to external photoevaporation \citep[][]{Winter19, Parker21a}.

Although recent ALMA surveys have provided important constraints on disk properties in clustered star formation environments, the majority of them focused entirely on measuring dust emission. Molecular gas dominates the mass budget and dynamics of protoplanetary disks, yet it has not been as closely examined in the surveys as the dust content has. Detecting molecular line emission from disks is challenging in clustered environments due to the long integration times required for even the most nearby clusters, and many young clusters also contain extended molecular cloud structures that can obscure line emission from compact sources \citep[e.g.,][]{Eisner16}. Nevertheless, a handful of gas disks have been detected and spatially/spectrally distinguished from the parent cloud  \citep[e.g.,][]{Williams14, Bally15, Ansdell17, Ansdell20, Boyden20}. Recently, \cite{Boyden20} identified 23 gaseous disks in the ONC using the CO $J=3-2$ and HCO$^+$ $J=4-3$ lines. The sample is currently the largest sample of ALMA-detected gas disks in a young cluster, and therefore provides a unique opportunity to probe the thermal, kinematic, and chemical structure of gaseous circumstellar disks influenced by environmental effects such as external photoevaporation. 

ALMA has revolutionized our understanding of gaseous circumstellar disks in nearby low-mass star-forming regions. Large surveys conducted in Taurus, Lupus, and Chameleon I indicate that a majority of nearby disks emit faint CO and CO isotopologue emission \citep[e.g.,][]{WB14, Ansdell16, Long17}, hinting that nearby disks are either gas-poor or universally depleted in gas-phase CO \citep{Miotello17}. Meanwhile, other abundant molecules, such as C$_2$H, CN, and HCO$^+$, continue to be routinely detected in new ALMA surveys \citep[e.g.,][and many others]{Bergner19, Miotello19, vanTerwisga19b, Pegues20, Teague20, Aikawa21, Bergner21, Garufi21, Guzman21, Oberg21, Anderson22}, opening up additional pathways to study the surface density, kinematics, thermal structure, and ionization structure of nearby disks. Thermochemical modeling has proven to be an effective tool for constraining physical/chemical properties from the molecular line observations, such as the gas mass \citep[e.g.,][]{Anderson19, Calahan21a, Calahan21b, Schwarz21, Trapman22}, the gas radius \citep[e.g.,][]{Facchini17, Trapman19}, the CO depletion profile \citep[e.g.,][]{Zhang19, Zhang21}, and the UV, X-ray and cosmic ray radiation fields incident on the disk surface \citep[e.g.,][]{Cleeves15, Cleeves16, Bergin16, Cazzoletti18, Woitke19, Seifert21}. Performing similar analyses on gas disks in clustered star-formation environments, such as the disks found in the ONC, is therefore key to comparing disk properties in low-mass vs. high-mass star-forming regions and examining whether the composition of gaseous circumstellar disks varies as a function of the external environment. 

In this paper, we model the $350$ GHz ($0.87$ mm) continuum, CO $J=3-2$, and HCO$^+$ $J=4-3$ observations of a subset of the disks detected in the ONC by \cite{Boyden20} via a thermochemical disk framework similar to those being employed in studies of nearby disks (see above). We introduce the sample targets and present the ALMA observations used in this study in Section \ref{sec:data}. In Section \ref{sec:modeling}, we describe our thermochemical disk model and outline a systematic parameter study that demonstrates how the model can be used to constrain the disk masses, disk morphologies, and UV and X-ray radiation fields from the ALMA observations. In Section \ref{sec:fitting}, we fit a large grid of thermochemical models to the ALMA observations and constrain the dust and gas masses, disk sizes, UV and X-ray radiation fields, viewing geometries, and central stellar masses of all targets in our sample. The results of our fitting are presented in Section \ref{sec:fitting:results}, and discussed in Section \ref{sec:discussion}. Finally, we summarize our main takeaways in Section \ref{sec:conclusions}.

\section{\bf{Sample}} \label{sec:data}

\begin{figure*}[ht!]
	\epsscale{1.18}
	\plotone{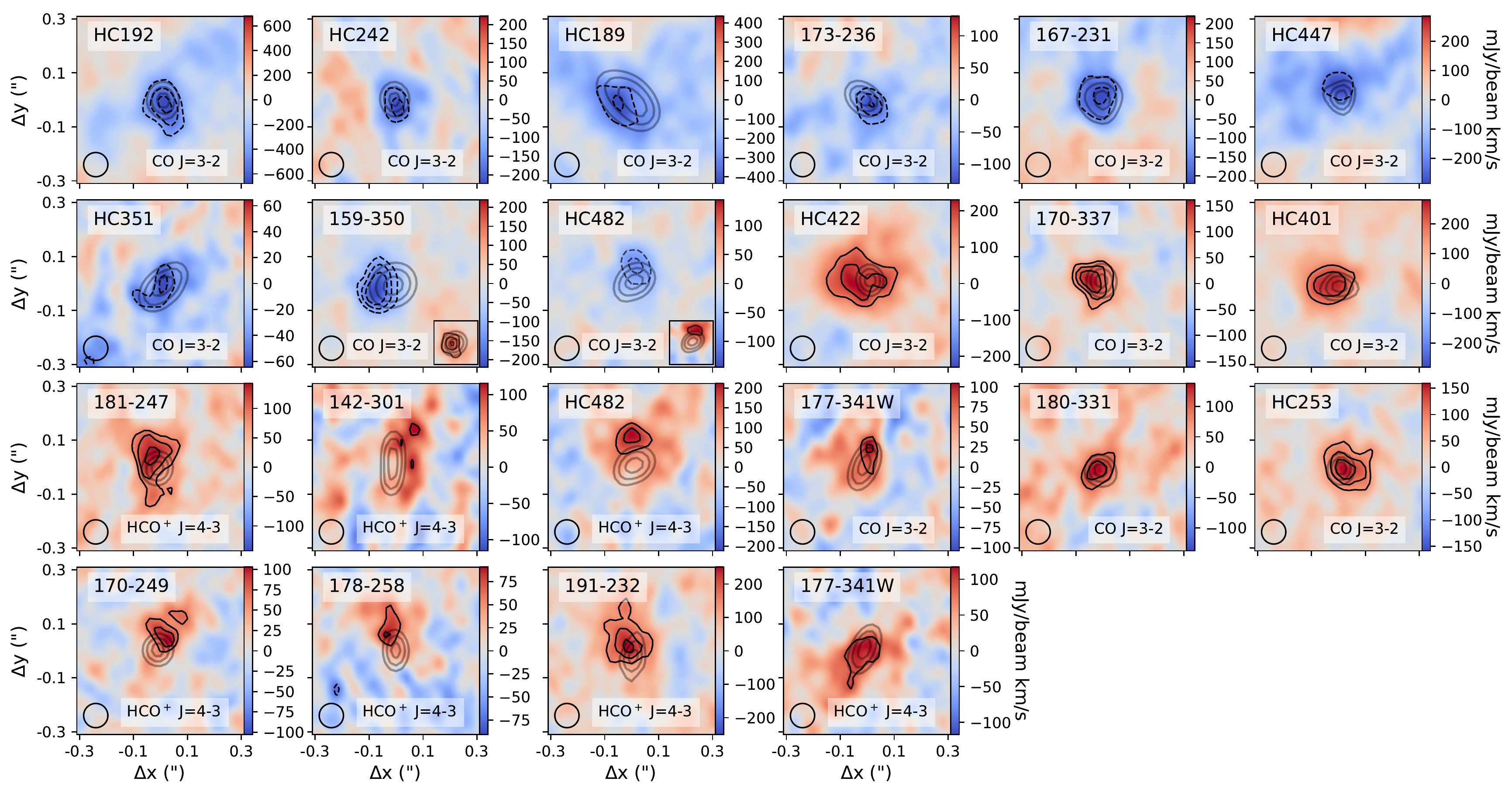}
	\vspace{-0.3cm}
	\caption{Moment 0 (integrated intensity) maps of our selected ONC disks. The name of each target and the gas tracer being plotted are specified in each panel. Black contours start at $3\sigma$ and increase in increments of $\sigma$. Grey contours show the $350$ GHz continuum emission towards each target, with contours drawn at $50\%$, $70\%$, and $90\%$ of the maximum continuum flux. The full width at half maximum of the ALMA synthesized beam is shown in the bottom-left corner of each panel. Each moment 0 map is generated from the velocity ranges specified in Table  \ref{tab:detection_info}. Because 159-350 and HC482 are detected in CO emission in some velocity channels and CO absorption in other velocity channels, we generate moment 0 maps over two different velocity ranges (see Table  \ref{tab:detection_info}) in order to highlight both the emission and absorption features. In the main panels, we plot the moment 0 maps generated from velocities $6.0-11.5$ km s$^{-1}$ for 159-350 and $6.5-10.5$ km s$^{-1}$ for HC482. In the smaller sub-panels, we plot the moment 0 maps generated from velocities $0.0-6.0$ km s$^{-1}$ for 159-350 and $2.5-6.5$ km s$^{-1}$ for HC482.}
	\label{fig:data_mom0}
\end{figure*}

\begin{figure*}[ht!]
	\epsscale{1.18}
	\plotone{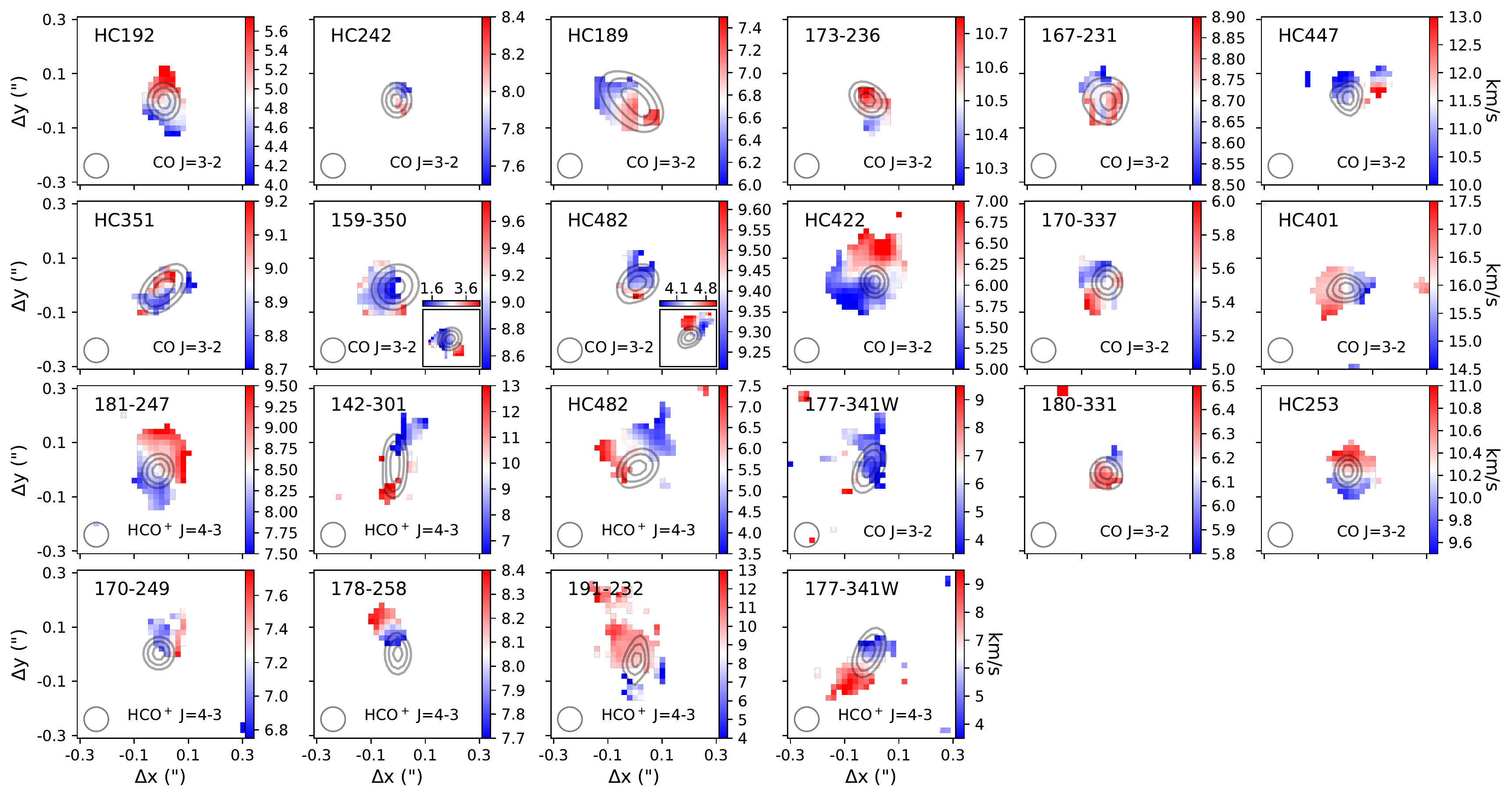}
	\vspace{-0.3cm}
	\caption{Moment 1 (intensity-weighted velocity) maps of our selected targets, generated using $\geq$3$\sigma$ emission/absorption from the channel maps. The layout of this plot is similar to that of Figure \ref{fig:data_mom0}.
	\label{fig:data_mom1}}
\end{figure*}

\begin{deluxetable*}{lccCRCCCCC}
\tablenum{1}
\tabletypesize{\footnotesize}
\tablewidth{0pt}
\tablecaption{Observed properties of ALMA-detected sources\tablenotemark{ }\label{tab:detection_info}}
\tablehead{ 
    \colhead{Source} & 
    \colhead{R. A.}  &
    \colhead{Decl.}  &  
    \colhead{d ($\theta^1$ C)} &  
    \colhead{$F_{dust}$}  & 
    \colhead{Channels} & 
    \colhead{$I_{CO \ J = 3-2}$}  & 
    \colhead{$I_{HCO^+ \ J = 4-3}$}  & 
    \colhead{$F_{CO \ J = 3-2}$}  & 
    \colhead{$F_{HCO^+ \ J = 4-3}$} \\ 
    \colhead{}        &  
    \colhead{(J2000)} & 
    \colhead{(J2000)}    & 
    \colhead{(pc)}     & 
    \colhead{(mJy)}   &
    \colhead{(km s$^{-1}$)}  &        
    \colhead{(mJy beam$^{-1}$)}  & 
    \colhead{(mJy beam$^{-1}$)}  & 
    \colhead{(Jy km s$^{-1}$)}  & 
    \colhead{(Jy km s$^{-1}$)}  
}
\colnumbers
\startdata 
170-337        &  
5 35 16.97     &  
-5 23 37.15    &   
0.031             &
13.1 \pm 3.0  & 
3.0 - 9.5         &  
60 \pm 8         &   
|F| < 27  & 
0.44 \pm 0.09    &   
0.08 \pm 0.06  \\ 
%
180-331        &  
5 35 18.03     &  
-5 23 30.80    & 
0.049          &   
1.5  \pm 1.0   &  
5.0 - 15.0     &   
36 \pm 8    &   
|F| < 28   & 
0.46 \pm 0.09  &    
|F| < 0.004   \\  
%
177-341W       &  
5 35 17.66     &  
-5 23 41.00    &  
0.050          &   
2.8  \pm 2.7   &  
2.5 - 11.0     &  
35 \pm 8    &   
45 \pm 9  & 
0.28 \pm 0.08  &   
0.56 \pm 0.09  \\ 
%
159-350\tablenotemark{a}        &  
5 35 15.96     & 
-5 23 50.30    &   
0.055          &  
43.1 \pm 8.5   &  
0.0 - 6.0      & 
45 \pm 10    &   
|F| < 30  & 
0.80 \pm 0.11  &  
|F| < 0.003    \\ 
               &                      
               &                                                    &                     
               &                             
               &  
6.0 - 11.5     & 
-50 \pm 12    &   
|F| < 31  & 
-0.55 \pm 0.10 &  
|F| < 0.004    \\ 
%
HC401          &  
5 35 16.08     &  
-5 22 54.10    &   
0.056          &  
1.2 \pm 0.2    & 
13.5 - 20.0    & 
51 \pm 14    &   
|F| < 27  &  
0.58 \pm 0.15  & 
|F| < 0.004    \\ 
%
HC253          &  
5 35 18.21     &  
-5 23 35.90    &  
0.057          &  
6.4 \pm 0.2    &  
7.5 - 13.0     & 
40 \pm 8    &   
|F| < 24  &  
0.56 \pm 0.08  & 
0.41 \pm  0.08 \\ 
%
178-258        &  
5 35 17.84     &  
-5 22 58.15    &    
0.062          &   
5.7 \pm 0.2    & 
6.0 - 10.5     & 
|F| < 45    &   
43 \pm 9  & 
|F| <  0.004   & 
0.15 \pm  0.06 \\ 
%
170-249        &  
5 35 16.96     &  
-5 22 48.51    &
0.068          & 
11.3 \pm 1.9   & 
5.0 - 9.5      & 
|F| < 36    & 
44 \pm 9  &    
|F| < 0.005    &  
0.25 \pm 0.06  \\ 
%
%
HC422          &  
5 35 17.38     &  
-5 22 45.80    &    
0.077          &   
6.0 \pm 0.2    &  
3.0 - 9.5      & 
105 \pm 15    &   
|F| < 28  & 
1.85 \pm 0.15  &  
0.12 \pm 0.06  \\ 
%
142-301        &  
5 35 14.15     &  
-5 23 0.91     &    
0.079          &  
2.5 \pm 2.0    &  
4.5 - 13.5     &  
|F| < 42    & %
41 \pm 10  & 
|F| < 0.006    &  
0.51 \pm 0.10  \\ 
%
HC351          &  
5 35 19.07     &  
-5 23 7.50     & 
0.081          &   
4.1 \pm 0.2    &  
8.0 - 10.5     & 
-37 \pm 10     & %
|F| < 26  & 
-0.32 \pm 0.05 &  
0.02 \pm 0.04  \\ 
%
181-247        &  
5 35 18.08     &  
-5 22 47.10    &    
0.084          &   
4.7 \pm 0.7    &  
6.0 - 11.5     & 
|F| < 36    & 
55 \pm 9  &  
|F| < 0.005    & 
0.76 \pm 0.08  \\ 
%
HC242          &  
5 35 13.80     &  
-5 23 40.20    &   
0.084          &  
31.5 \pm 0.9   & 
6.0 - 13.0     & 
-67 \pm 21     & %
|F| < 58       & 
-0.65 \pm 0.15 &  
|F| < 0.006    \\ 
%
HC189          &  
5 35 14.53     &  
-5 23 56.00    &     
0.085          &   
46.3 \pm 1.9   &  
0.0 - 13.0     & 
-76 \pm 19     & %
|F| < 62       &
-4.55 \pm 0.31 & 
-2.84 \pm 0.28 \\ 
%
173-236        &  
5 35 17.34     &  
-5 22 35.81    &    
0.095          &  
18.1 \pm 2.2   &  
9.0 - 13.5     & 
-38 \pm 13     & 
|F| < 26       & 
-0.51 \pm 0.08 & 
0.11 \pm 0.05  \\ 
%
HC447          &  
5 35 15.89     &  
-5 22 33.20    & 
0.098          &    
1.6 \pm 0.3    & 
5.0 - 15.5     & 
-57 \pm 16     & 
|F| < 32 & 
-1.95 \pm 0.23 &   
|F| < 0.006    \\ 
%
167-231        &  
5 35 16.73     &  
-5 22 31.30    &   
0.100          &    
3.4 \pm 0.2    &  
7.0 - 13.5     &
-55 \pm 14     & 
|F| < 29 & 
-0.86 \pm 0.15 &  
|F| < 0.004    \\ 
%
HC192          &  
5 35 13.59     &  
-5 23 55.30    &    
0.105          &  
12.5 \pm 4.8   &  
0.0 - 11.0     & 
-101 \pm 24     & 
|F| < 84 &
-2.99 \pm 0.38 &  
-1.78 \pm 0.36 \\ 
%
191-232        &  
5 35 19.13     &  
-5 22 31.20    &  
0.127          &    
1.1 \pm 0.1    & 
4.0 - 19.0     &    
|F| < 45    & 
43 \pm 10  &
|F| < 0.01     & 
1.19 \pm 0.14  \\ 
%
HC482\tablenotemark{a}          &  
5 35 18.85     &  
-5 22 23.10    &   
0.135          &     
5.5 \pm 0.3    &
2.5 - 8.0      &                          
               &  
83 \pm 18  &
                   &
0.79 \pm 0.10  \\ 
              &                               
              &                              
              &                
              &                             
              & 
2.5 - 6.5     &  
59 \pm 12     &                       
              & 
0.32 \pm 0.09 &                          
              \\ 
              &                               
              &                               
              &               
              &                             
              & 
6.5 - 10.5    & 
-60 \pm 15     &
                     &
-0.35 \pm 0.07&                           
              \\ 
\enddata
\tablenotetext{ }{{\bf Notes.} Column (1): cluster member name, where proplyds are indicated with six-digit IDs, and Near-IR sources not detected as proplyds are labeled with ``HC'' and additional digits. Columns (2) and (3): phase center coordinates. Column (4): projected distance from  $\theta^1$ Ori C. Column (5): free-free corrected $350$ GHz dust continuum fluxes, taken from \citet{Eisner18}. Column (6): velocity channels used to compute the CO $J=3-2$ and/or HCO$^+$ $J=4-3$ moment  maps shown in Figures \ref{fig:data_mom0} and \ref{fig:data_mom1}. Columns (7) and (8): peak CO $J=3-2$ and HCO$^+$ $J=4-3$ channel map fluxes, with the uncertainties corresponding to the local rms in the peak intensity velocity channel. A negative flux corresponds to a detection in absorption rather than in emission. Columns (9) and (10): integrated CO $J=3-2$ and HCO$^+$ $J=4-3$  fluxes, taken from \cite{Boyden20}.} 
\tablenotetext{a}{{159-350 and HC482 are detected in CO emission, CO absorption, and/or HCO$^+$ emission depending on the velocity channel. For these sources, we use separate rows to list the velocity ranges over which the emission and absorption are seen. The separate rows also provide the peak and integrated line fluxes for the respective emission and absorption.}}
\vspace{-0.7cm}
\end{deluxetable*}

Our sample consists of $20$ circumstellar disks in the ONC that were recently detected in CO $J=3-2$ and/or HCO$^+$ $J=4-3$ with ALMA by \cite{Boyden20}: HC192, HC242, HC189, 173-236, 167-231, HC447, HC351, 159-350, HC482, HC422, 170-337, 181-247, 142-301, 177-341W, HC253, HC401, 170-249, 178-258, 191-232, and 180-331. All targets are known cluster members that have been detected previously at near-IR and mm wavelengths \citep[e.g., ][]{Hillenbrand00, Mann14, Eisner18}. Names with six-digit IDs refer to systems with optically identified proplyds, while sources not detected as proplyds are labeled with ``HC'' and three digits.

Our sample likely consists of Class II disks. The extinction levels measured towards our targets are relatively small, with typical values $A_V \lesssim 10$ mag \citep[e.g., ][]{Hillenbrand00, Fang21}. These values are much lower than the extinction levels implied by spherical distributions of dust \citep[as expected for younger protostars; e.g.,][]{Beckwith90} with mm-fluxes comparable to the values measured for each target (see Table \ref{tab:detection_info}). The mm-detected emission likely arises from flattened spatial distributions of dust, though we do not rule out the existence of some remnant non-spherical envelopes, which may persist in the late Class I or early Class II phase.

All targets were observed in a Cycle 4 ALMA program (2015.1.00534.S; PI: Eisner) that mosaicked the central $1\rlap{.}'5$ $\times$ $1\rlap{.}'5$ region of the ONC at Band 7, covering the $350$ GHz continuum as well as the CO $J=3-2$ and HCO$^+$ $J=4-3$ lines. 104 dusty disks were detected in the $350$ GHz continuum by \cite{Eisner18}, and of these 104 detected sources, our sample targets are the ones that were also detected in the CO $J=3-2$ and/or HCO$^+$ $J=4-3$ channel maps in more than one velocity channel with a peak intensity detection threshold of $>3\sigma$. 

We constructed $350$ GHz continuum images and CO $J=3-2$ and HCO$^+$ $J=4-3$ data cubes centered around each target by extracting $1''$ $\times$ $1''$ sub-images from the larger $1\rlap{.}'5$ $\times$ $1\rlap{.}'5$ images of the central ONC generated by \cite{Eisner18}. These larger images were generated with a 100 k$\lambda$ {\it uv} cut to improve the spatial filtering of extended emission in the region, and we performed no additional data reduction or calibration on the extracted sub-images. For a detailed description of the continuum data reduction and imaging, we refer the reader to \cite{Eisner18}. For additional information regarding the reduction and imaging of the CO $J=3-2$ and HCO$^+$ $J=4-3$ observations, we refer the reader to \cite{Boyden20}.

Figures \ref{fig:data_mom0} and \ref{fig:data_mom1} show the CO $J=3-2$ and/or HCO$^+$ $J=4-3$ moment 0 (integrated intensity) and moment 1 (intensity-weighted velocity) maps of our sample targets, plotted over contours of continuum emission. The moment 1 maps are generated using $\geq$$3\sigma$ emission or absorption from the channel maps, and the moment 0 maps are generated using all data cube points (i.e., no flux clip or $\sigma$ clip). The angular resolution is uniform across the sample, with a synthesized beam area of $\sim 0\rlap{.}''09 \times 0\rlap{.}''1$ and beam position angle of about $-5^{\circ}$.  This corresponds to a linear resolution of $\sim$35 AU at the canonical $\sim$400 pc distance to Orion \citep[e.g., ][]{Kounkel17, Kounkel18, Grob18}, which we assume as the distance to all of our targets. The CO $J=3-2$ and HCO$^+$ $J=4-3$ data cubes cover velocities ranging from $0.0-20.0$ km $s^{-1}$ with 0.5 km s$^{-1}$ channels. In Table \ref{tab:detection_info}, we provide additional information pertaining to each individual target, including the coordinates, projected distance from $\theta^1$ Ori C, and observed properties of the continuum and line emission.

Of the 20 ONC disks in our sample, all are detected in the $350$ GHz continuum, 15 are detected in the CO $J=3-2$ channel maps, 7 are detected in the HCO$^+$ $J=4-3$ channels maps, and 2 are detected in both sets of channel maps (177-341W and HC482). A subset of the CO $J=3-2$ detections are also detected at $>3\sigma$ in the HCO$^+$ $J=4-3$ moment 0 maps but not in the HCO$^+$ $J=4-3$ channel maps \citep[see][]{Boyden20}. In those cases, the HCO$^+$ observations fall below our threshold for kinematic thermochemical modeling, so we do not include them in our sample. 

9 of the CO $J=3-2$ detections are seen in {\it absorption} against the warm background, meaning that their observed emission is {\it negative} (see Figure  \ref{fig:data_mom0}). These sources are:  HC192, HC242, HC189, 173-236, 167-231, HC447, HC351, 159-350, and HC482. All 9 absorption detections are located in regions of the ONC where there is significant large-scale emission from the background molecular cloud \citep{Boyden20}, and they are seen in absorption because their CO brightness temperatures are less than the localized CO brightness temperature of the background molecular cloud, as discussed further in Appendix \ref{appendix:a}. Typically, we detect these cluster members exclusively in absorption. However, 159-350 and HC482 are seen in absorption in some velocity channels and in emission in other velocity channels. 
This occurs because the CO background temperature is also velocity dependent, and when the brightness temperature of the disk exceeds the cloud background temperature in a velocity channel, it will be seen in emission in that channel rather than in absorption (see Appendix \ref{appendix:a}).

We exclude cluster members HC756/7, HC361, and HC771 from our sample \citep[see][]{Boyden20}, even though they meet our criteria for kinematic modeling. HC756/7 is a misaligned binary system that exhibits two distinct velocity gradients in CO absorption. Our modeling procedure is not designed to handle complex velocity structures associated with misaligned binaries (see Section \ref{sec:modeling}), and so we defer a detailed analysis of HC756/7 to future work. 

HC361 is detected in CO emission along velocity channels that are higher than the typical velocities where other cluster members are detected, and the observed CO emission is asymmetric such that it peaks away from the continuum emission by more than two beam sizes, unlike what is seen with the other CO emission detections. We suspect that the CO emission from this source may trace non-disk structure, such as an outflow, or other extended structure from the background cloud. Thus, we exclude HC361 from our sample. 

HC771 is detected in HCO$^+$ absorption, and as discussed in Appendix \ref{appendix:a}, we are currently unable to accurately model HCO$^+$ absorption since there are no reliable estimates of the HCO$^+$ background intensity in the ONC. We therefore defer an analysis of HC771 to a future study that involves obtaining direct measurements of the HCO$^+$ background intensity.

\section{\bf Thermochemical Models} \label{sec:modeling}

We use the thermochemical code {\tt RAC2D} \citep{Du14} to construct a model disk and explore the effects of model parameters on simulated observations of CO $J=3-2$ emission, HCO$^+$ $J=4-3$ emission, and thermal dust emission at $350$ GHz ($0.87$ mm). {\tt RAC2D} has been utilized in a variety of studies to model the molecular line emission of nearby disks \citep[e.g., ][]{Bergin16, Du17, Zhang19, Zhang21, Calahan21a, Calahan21b, Schwarz21}, and a detailed description of the code can be found in \cite{Du14}. 

While the properties of the dust, CO, and HCO$^+$ have been investigated in previous studies \citep[e.g.,][]{Cleeves14, Miotello16, Schwarz18, Ballering19}, those studies often adopted models tailored to the massive, extended disks commonly found in nearby, low-density star-forming regions \citep[e.g.,][]{Ansdell16, Ansdell18}. Disks in the ONC are thought to be more compact than disks in lower-density regions \citep[e.g.,][]{Eisner18, Boyden20}. Furthermore, most disk-bearing stars in the ONC are M-dwarfs \citep[e.g.,][]{Hillenbrand97, Fang21}, whereas earlier chemical modeling focused on disks surrounding solar-type T-Tauri stars. These differences have motivated us to generate our own set of thermochemical models, which we can then use to interpret the ALMA observations.

\subsection{Physical Model}\label{sec:modeling:disk}

Our model disk follows the basic setup of a viscously evolving accretion disk \citep{Lynden74}. We assume an azimuthally symmetric distribution of gas surrounding a pre-main-sequence star of mass $M_*$, with global surface density profile described as  
\begin{equation}\label{Eq:1}
    \Sigma_g = \Sigma_0 \Big(\frac{r}{r_c}\Big)^{-\gamma} \exp\bigg[ - \Big(\frac{r}{r_c}\Big)^{2-\gamma} \bigg] \ 
\end{equation}
between an inner radius $R_{in}$ and an outer edge $R_{out}$. $r_c$ is the radius at which the surface density profile transitions from a power law to an exponential taper, which we assume scales with the disk outer edge as $r_c = 0.8 \ R_{out}$. $\gamma$ is the surface density power-law exponent, and the normalization constant $\Sigma_0$ is related to total mass of the gas, $M_g$, by integrating Equation \ref{Eq:1} over all space. The vertical structure of the gas is governed by hydrostatic equilibrium (see Section \ref{sec:modeling:thermochemical_calculations}), while the gas kinematics are described by a Keplerian velocity field.

We model the dust disk using two grain size populations: a small grain population composed primarily of micron-sized grains, and a large grain population composed primarily of mm-sized grains \citep[e.g.,][]{Andrews11, Cleeves13, Miotello16, Zhang19, Calahan21a}. Both populations are composed of a mixture of $80\%$ astronomical silicates \citep{Draine84} and $20\%$ graphite, and follow an MRN size distribution \citep[][]{Mathis77} where $n(a) \propto a ^{-3.5}$. We adopt a minimum grain size of $0.005$ $\mu$m, a maximum grain size of $a_{max, \mu m}$ for the small grain population, and a maximum grain size of $a_{max, mm}$ for the large grain population. The total mass of the dust, $M_d$, is equivalent to the sum of the masses of the large and small grain populations. We parametrize the relative masses of each grain population via the parameter $f_{mm}$. Specifically, $f_{mm}$ is defined as the fraction of the total dust mass allocated to the large grain population, while $(1 - f_{mm})$ is defined as the fraction of the total dust mass allocated to the small grain population.

For our modeling, we assume that the gas, small grains, and large grains are all spatially co-located. This means that we do not account for the potential vertical settling of mm-sized grains or the potential radial variations between small and large grain populations, for which there is observational evidence in a number of nearby disks \citep[e.g.,][]{Trapman20, Rich21, Law22, Villenave22}, and is important for modeling spatially-resolved, multi-wavelength continuum emission \citep[e.g.,][]{Andrews11, Cleeves16, Sheehan17b, Ballering19}. As we demonstrate in Section \ref{sec:modeling:pstudy}, the chemistry and excitation conditions of our molecular lines of interest depend mainly on the photodissociation and photoionization of molecules by UV and X-ray photons, and the presence of large grains in the upper/outer layers of the disk does not pose a strong influence on those reactions. 
While mm-sized grains can influence the continuum opacity to UV and X-ray photons, we find that this influence is small compared with the influences of the gas and the smaller, micron-sized grains.  
We therefore opt to simplify the model setup by coupling the density profiles of the gas, the small grain, and large grain populations. This implies that mass ratio of gas and small $+$ large grains is uniform throughout all regions of our model disk.

\subsection{Ionization sources}\label{sec:modeling:ionization_sources}

\begin{figure}[ht!]
	\epsscale{1.2}
	\plotone{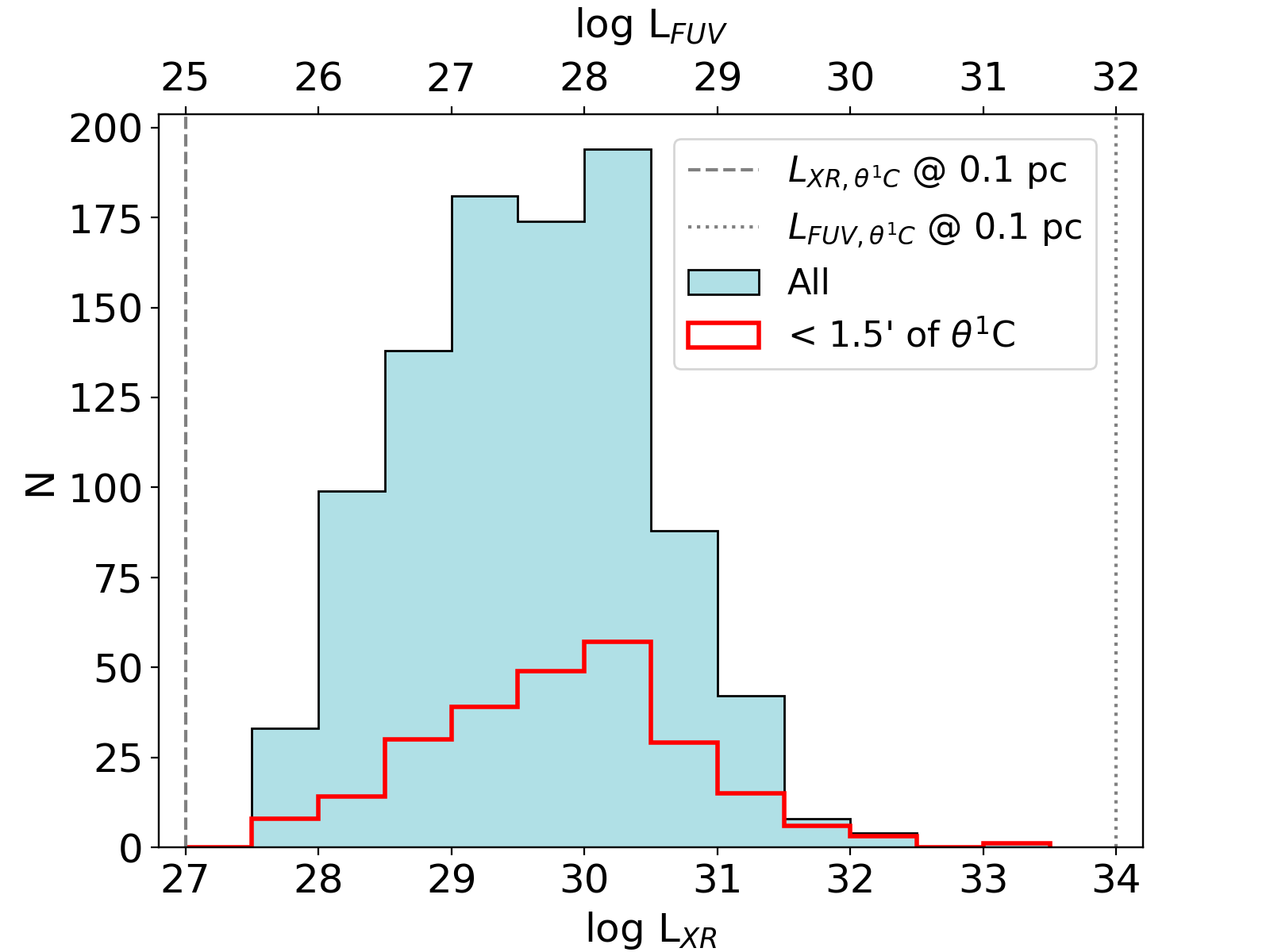}
	\caption{Distribution of measured X-ray luminosities of young stars in the Orion Nebula, taken from \cite{Getman05}. 
	Blue histograms show the distribution obtained from the entire dataset, while the red histograms show the distribution for the central $1\rlap{.}'5$ region of the ONC, where our ALMA detections are located. The bottom x-axis indicates the binned X-ray luminosities in logspace. In the top x-axis, we show the corresponding FUV luminosities assuming that all of the X-ray active sources are pre-MS M-dwarfs, which are typically $\sim 100$ times less luminous at FUV wavelengths than at X-ray wavelengths \citep[e.g.,][]{Shkolnik14}. For comparison, the dotted grey line indicates what the luminosity of the FUV emission from  $\theta^1$ Ori C would be at $d = 0.1$ pc (c.f. Table \ref{tab:detection_info}), whereas the dashed grey line indicates what the luminosity of the X-ray emission from  $\theta^1$ Ori C would be at $0.1$ pc. 
	\label{fig:COUP_ONC}}
\end{figure}

Ionizing UV photons, X-rays, and cosmic rays shape the physical and chemical landscape of circumstellar disks by heating the gas and dust, ionizing the gas, and driving various chemical reactions. In dense, clustered environments like the ONC, disks are exposed to internally-sourced UV and X-ray emission from their central stars due to accretion and stellar magnetospheric activity \citep{Feigelson99, Shkolnik14, Hartmann16, Waggoner22}, 
as well as externally-sourced UV and X-ray emission from other cluster members \citep{Fatuzzo08, Adams12}. In particular, the massive O-star $\theta^1$ Ori C has a total FUV luminosity of $\sim 10^{38}$ ergs s$^{-1}$ between $930$ and $2000$ \AA, and this external radiation can be comparable to or even dominate over the internal radiation produced from lower-mass cluster members (see Figure \ref{fig:COUP_ONC}).  At X-ray wavelengths, however, the emission from massive stars is weaker, and comparable to the emission produced from lower-mass cluster members. Based on the measured distribution of X-ray luminosities in the ONC \citep[][see Figure \ref{fig:COUP_ONC}]{Getman05}, we expect most of the X-rays incident on our sample disks to be generated internally, with only small contributions from nearby cluster members.

The presence of strong internal and external FUV emission in the ONC warrants clarification on whether we need to include multiple FUV ionization sources in our thermochemical modeling. We explored the competing effects of internally- vs. externally-sourced FUV radiation on our modeling of compact CO $J=3-2$ and HCO$^+$ $J=4-3$ emission by generating models with accretion-UV emission from the central star as the source of ionizing FUV photons, and models with UV emission from nearby massive stars as the source of ionizing FUV photons. These models are presented in Appendix \ref{appendix:ionization}, and show that while some differences can arise when we treat the ionizing radiation as internal or external, the differences manifest at levels below the sensitivity and resolution of our ALMA observations. Given the low significance of these differences relative to our data quality, we opt to treat the ionizing UV radiation as a single parameter that encapsulates contributions from both the central star and external environment. By analogy, we use a single parameter to describe the X-ray emission generated from a dominant internal source and weaker external source(s).

Finally, we consider ionization by Galactic cosmic rays (CRs) in our modeling. Galactic CRs act as a form of external ionization, and they are capable of providing additional ionization in the dense, cold midplane layers of the disk \citep{Umebayashi81}.

\subsection{Thermochemical calculations and synthetic observations}\label{sec:modeling:thermochemical_calculations}

{We input our model disk surface density profile into {\tt RAC2D}, which executes all thermochemical calculations used in this study. For each model run, {\tt RAC2D} first solves the dust temperature, radiation field, and vertical structure through iterative Monte Carlo simulations of dust radiative transfer. After a self-consistent solution is obtained from the dust radiative transfer step, } {\tt RAC2D} then simultaneously evolves the gas chemistry and thermal structure. For a detailed description of the dust radiative transfer calculations and gas heating and cooling mechanisms, we refer the reader to \cite{Du14}. Following the discussion from section \ref{sec:modeling:ionization_sources}, the photon sources that we consider in the thermochemical calculations are: 1) blackbody radiation from the central star, assuming an effective temperature $T_{eff}$ and radius $R_{*}$; { 2) X-ray emission in the range $0.1 - 10$ keV, with a total luminosity $L_{XR}$; 3) FUV emission in the range $930-2000$ \AA, with a total luminosity $L_{UV}$; } and 4) ionizing Galactic CRs, with ionization rate $\zeta_{CR}$ and attenuation length of $96$ g cm$^{-2}$ \citep[][]{Umebayashi81}.

Our chemical network includes the full set of gas-phase reactions from the UMIST12 database \citep{McElroy13}. Additional reaction networks include photodissociation, self-shielding of H$_2$ and CO, adsorption, thermal desorption, photodesorption by UV photons and cosmic rays, two-body reactions on the dust grain surface, and recombination of ions with charged grains. We initialize the disk chemical composition using the species abundances listed in Table \ref{tab:init}, and then let the gas temperature and chemical evolution run for $1$ Myr. This duration is enough for the gas temperature and abundances of key molecular species to reach a steady state, and $1$ Myr of chemical evolution mirrors the expected $\sim 1$ Myr age of the ONC \citep{Hillenbrand97, Fang21}. 

Using the computed gas temperature profiles and species abundances, we generate synthetic observations of CO $J=3-2$ and HCO$^+$ $J=4-3$ using the ray-tracing module of {\tt RAC2D}. This module returns the line intensity profiles as well as the line optical depths, both of which we track in our analysis. We assume the gas is in local thermodynamic equilibrium (LTE). All image cubes are simulated at a distance of $400$ pc with a pixel size of $0\rlap{.}''02$ ($8$ AU), channel width of $0.25$ km s$^{-1}$, {image size of $1''$ $\times$ $1''$,} and velocity range of $\pm 100$ km s$^{-1}$. Moreover, the viewing angle parameters, inclination and position angle ($i$ and $\theta$), and systemic velocity ($v_{sys}$) are also specified during the ray-tracing step. 

We note that {\tt RAC2D} simulates line emission and dust continuum emission together, reflective of actual ALMA observations. We perform a continuum subtraction to separate the line emission from the continuum. Our chosen velocity range is broad enough to render plenty of line-free channels to perform an effective continuum subtraction. 

For each set of synthetic CO $J=3-2$ and HCO$^+$ $J=4-3$ observations, we construct a representative model $350$ GHz continuum flux by averaging the line-free continuum emission over all velocity channels, and then calculating the total flux of the spectrally-averaged image. We choose to track the total continuum flux, rather than the full continuum image, because the observed continuum images of our targets are compact, and hence provide little information beyond the total flux.

\subsection{Exploration of Model Parameters}\label{sec:modeling:pstudy}

\begin{deluxetable}{lR}
\tabletypesize{\small}
\tablewidth{5pt}
\tablenum{2}
\tablecaption{Initial Chemical Abundances \label{tab:init}}
\tablehead{ \colhead{Species} &  \colhead{Abundance}}
\startdata 
H$_2$            &  5.0 \times 10^{-1} \\
He               &  9.0 \times 10^{-2} \\
CO               &  1.4 \times 10^{-4} \\
N                &  7.5 \times 10^{-5} \\
H$_2$O (ice)  \ \ \ \ \ \ \ \ \ \  \ \ \ \ \ \  \ \ \ \ \  &  1.8 \times 10^{-4} \\
S                &  8.0 \times 10^{-8} \\
Si$^+$           &  8.0 \times 10^{-9} \\
Mg$^+$           &  7.0 \times 10^{-9} \\
Fe$^+$           &  3.0 \times 10^{-9} \\
P                &  3.0 \times 10^{-9} \\
F                &  2.0 \times 10^{-8} \\
Cl               &  4.0 \times 10^{-9} \\
\enddata
\end{deluxetable}

\begin{deluxetable}{lll}
\tabletypesize{\footnotesize}
\tablenum{3}
\tablecaption{Model Parameters and their Fiducial Values \label{table:fiducial}}
\tablehead{\vspace{-.125in}}
\startdata 
            &  Stellar Parameters   & \\
$M_*$       &   Stellar mass & $0.3$ $M_{\odot}$      \\
T$_{eff}$   &   Effective  temperature &$3500$  K    \\     
$R_*$       &    Stellar radius & $1$  $R_{\odot}$   \\   
\hline
                  &  Disk Parameters  & \\ 
$M_{gas}$         &  Disk gas mass & $10^{-3}$ $M_{\odot}$   \\  
$M_{dust}$        &  Disk dust mass &  $0.01$ $M_{gas}$       \\  
$R_{in}$          &  Disk inner radius & $1$  AU          \\     
$R_{out}$         &  Disk outer radius &$50$  AU        \\   
r$_{c}$      &   Disk characteristic radius &$0.8 R_{out}$   \\  
$\gamma$          &   Surface density power-law index &$1.0$   \\      
$a_{min}$         &   Minimum grain size & $0.005$ $\mu$m        \\
$a_{max, \mu m}$  &   Maximum size of small grains &$1.0$ $\mu$m   \\
$a_{max, mm}$     &   Maximum size of large grains & $1.0$ mm        \\
$f_{mm}$          &   Dust mass fraction in large grains & 0.85     \\
$i$          &   Inclination & $45^{\circ}$     \\
 \hline
             &  Ionization Parameters   & \\
$L_{UV}$      &  FUV luminosity & $10^{30}$ ergs s$^{-1}$  \\
$L_{XR}$     &   X-ray luminosity & $10^{30}$  ergs s$^{-1}$ \\   
$\zeta_{CR}$ &  Cosmic ray ionization rate & $1.6 {\text{\scriptsize  $ \times $} } 10^{-19}$ s$^{-1}$   \\
 \hline
             &  Chemistry Parameters    & \\
$X_{CO}$      &  Initial CO abundance & $1.4  {\text{\scriptsize  $\times$} }  10^{-4}$  \\
H$_{2}^*$      &  Vibrationally-excited H$_2$ chemistry & Excluded  \\
\enddata
\tablenotetext{}{\scriptsize References for fiducial stellar parameter values: \cite{Hillenbrand97, Getman05, Fang21}. References for fiducial disk parameter values: \cite{Andrews11, Cleeves13, Mann14, Miotello16, Eisner18, Ballering19, Zhang19, Boyden20, Calahan21a}. References for fiducial ionization parameter values: \cite{Johnstone98, Storzer99, Getman05, Adams12, Cleeves13, Shkolnik14, Cleeves15, Rab18}. References for fiducial chemistry parameter values: \cite{Bolatto13, Lacy17, Kamp17}.}
\vspace{-0.3cm}
\end{deluxetable}

\begin{figure}[ht!]
	\epsscale{1.15}
	\plotone{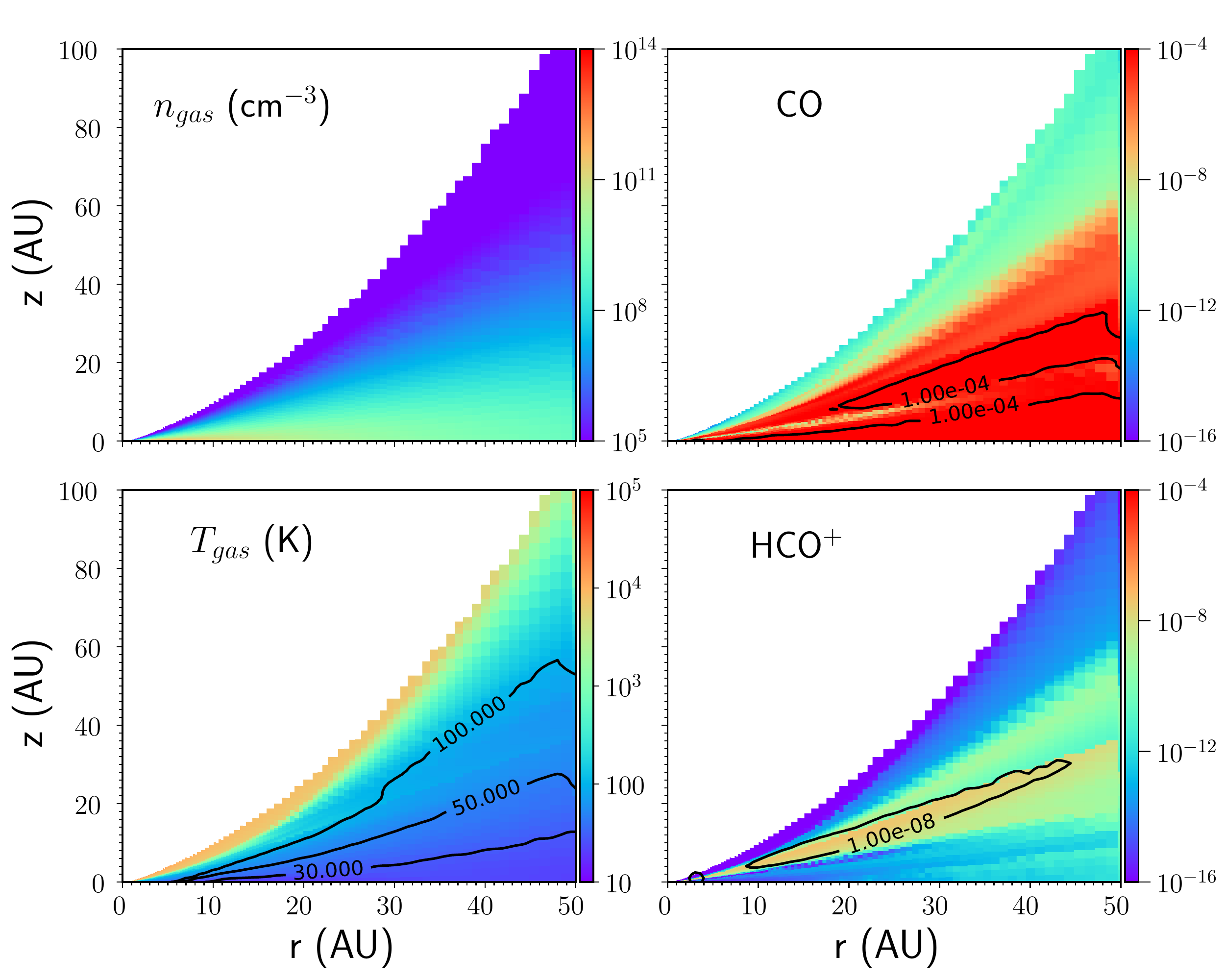}
	\caption{Gas density, gas temperature, CO abundance and HCO$^+$ abundance profiles of our fiducial model, generated with the thermochemical code {\tt RAC2D}. 
	\label{fig:fiducial}}
\end{figure}

\begin{figure*}[ht!]
	\epsscale{1.15}
	\plotone{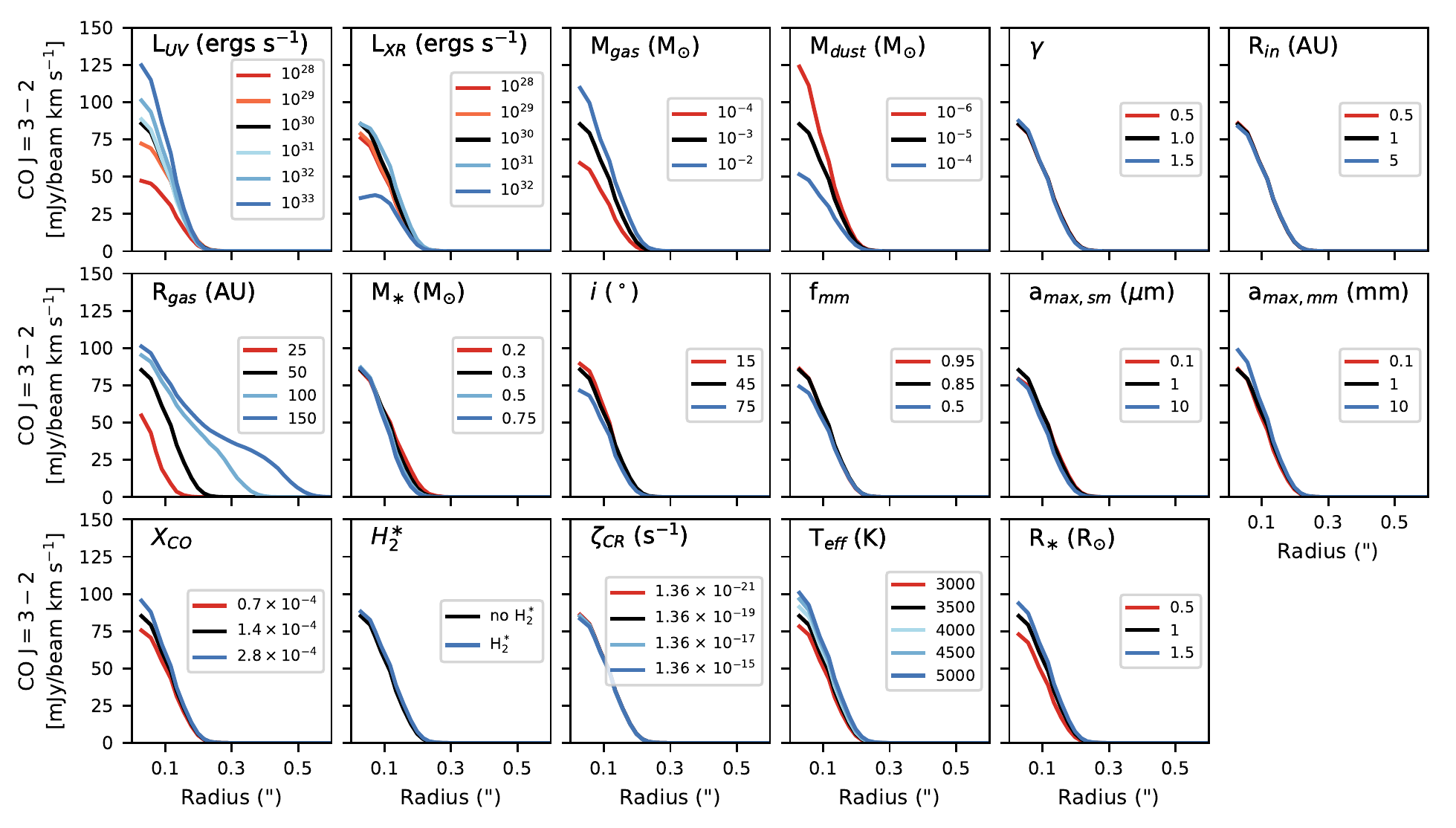}
	\caption{demonstration of the effects of varying model parameters on the simulated integrated intensity profiles of CO $J=3-2$. 
	Each column shows the results for a single parameter. The black-colored profiles in each panel correspond to the intensity profiles of our fiducial model. All model images are generated using the thermochemical code {\tt RAC2D}. We simulate the images assuming a distance of $400$ pc, and convolve the model images with a $0\rlap{.}''09$ Gaussian beam. 
	\label{fig:pstudy}}
\end{figure*}

\begin{figure*}[ht!]
	\epsscale{1.15}
	\plotone{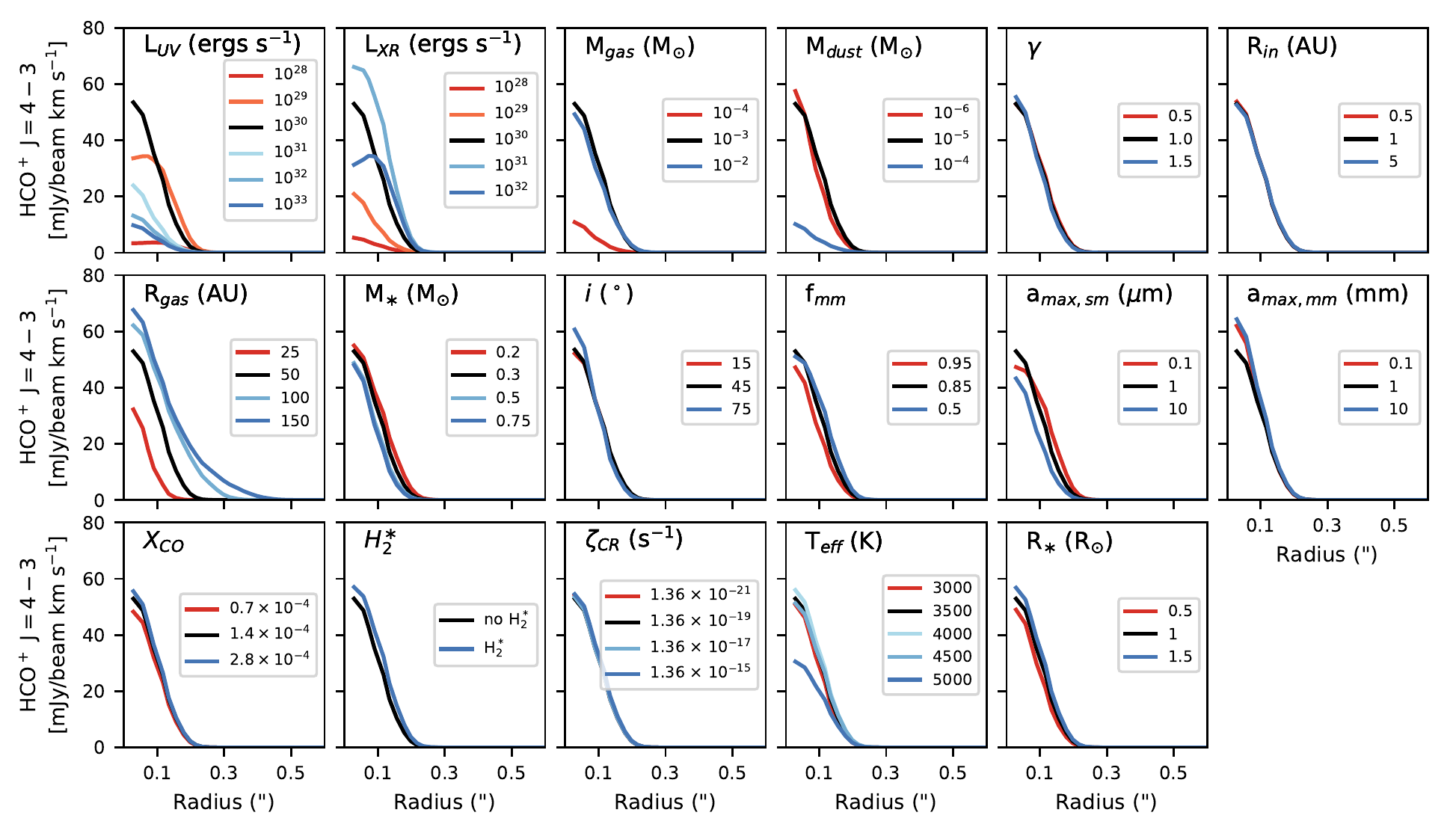}
	\caption{Same as Figure \ref{fig:pstudy}, but for HCO$^+$ $J=4-3$.
	\label{fig:pstudy2}}
\end{figure*}

\begin{figure*}[ht!]
	\epsscale{1.15}
	\plotone{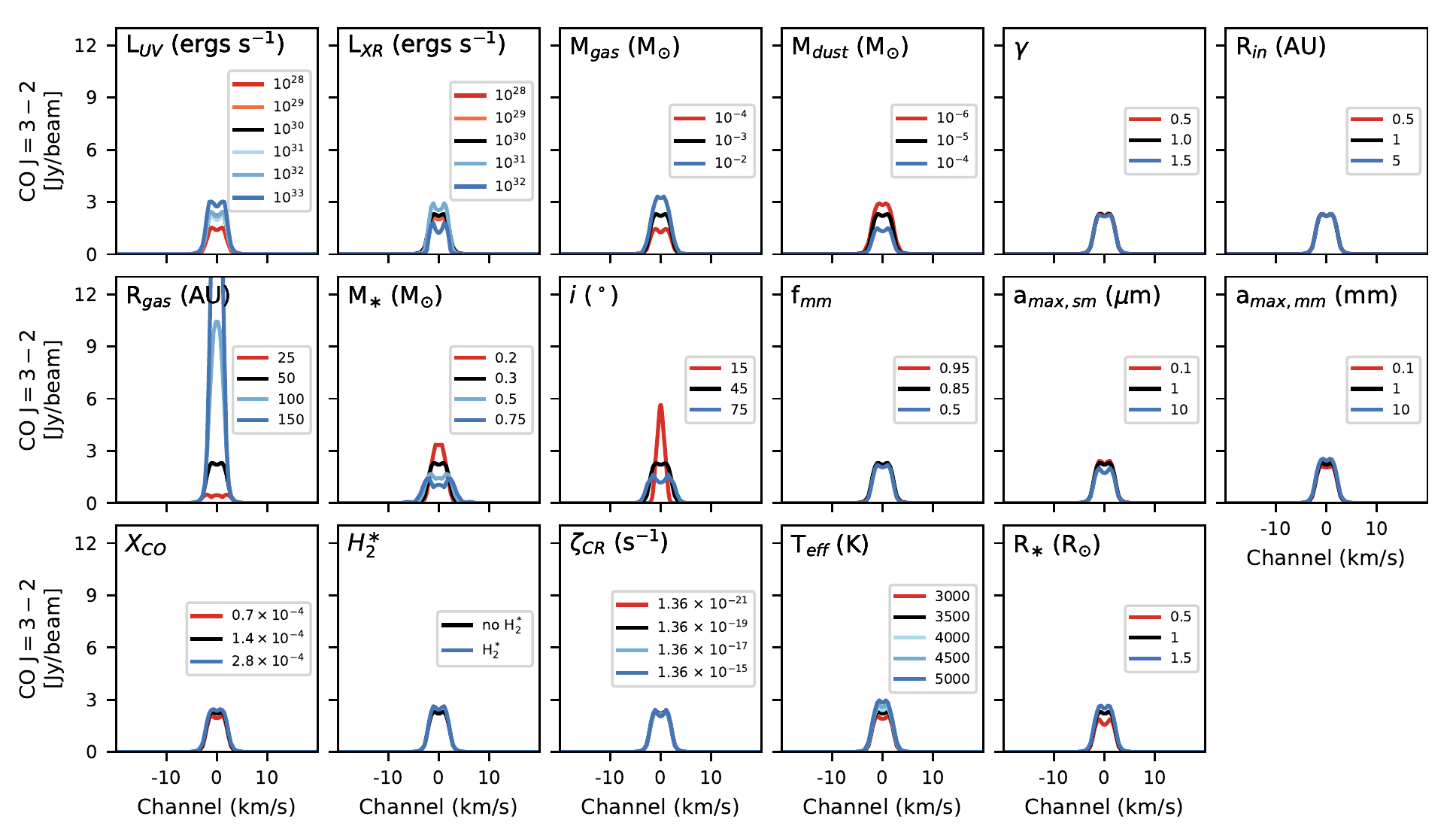}
	\caption{Demonstration of the effects of varying model parameters on the simulated line profiles of CO $J=3-2$. The layout of this plot is similar to that of Figure \ref{fig:pstudy}.
	\label{fig:pstudy_spec1}}
\end{figure*}

\begin{figure*}[ht!]
	\epsscale{1.15}
	\plotone{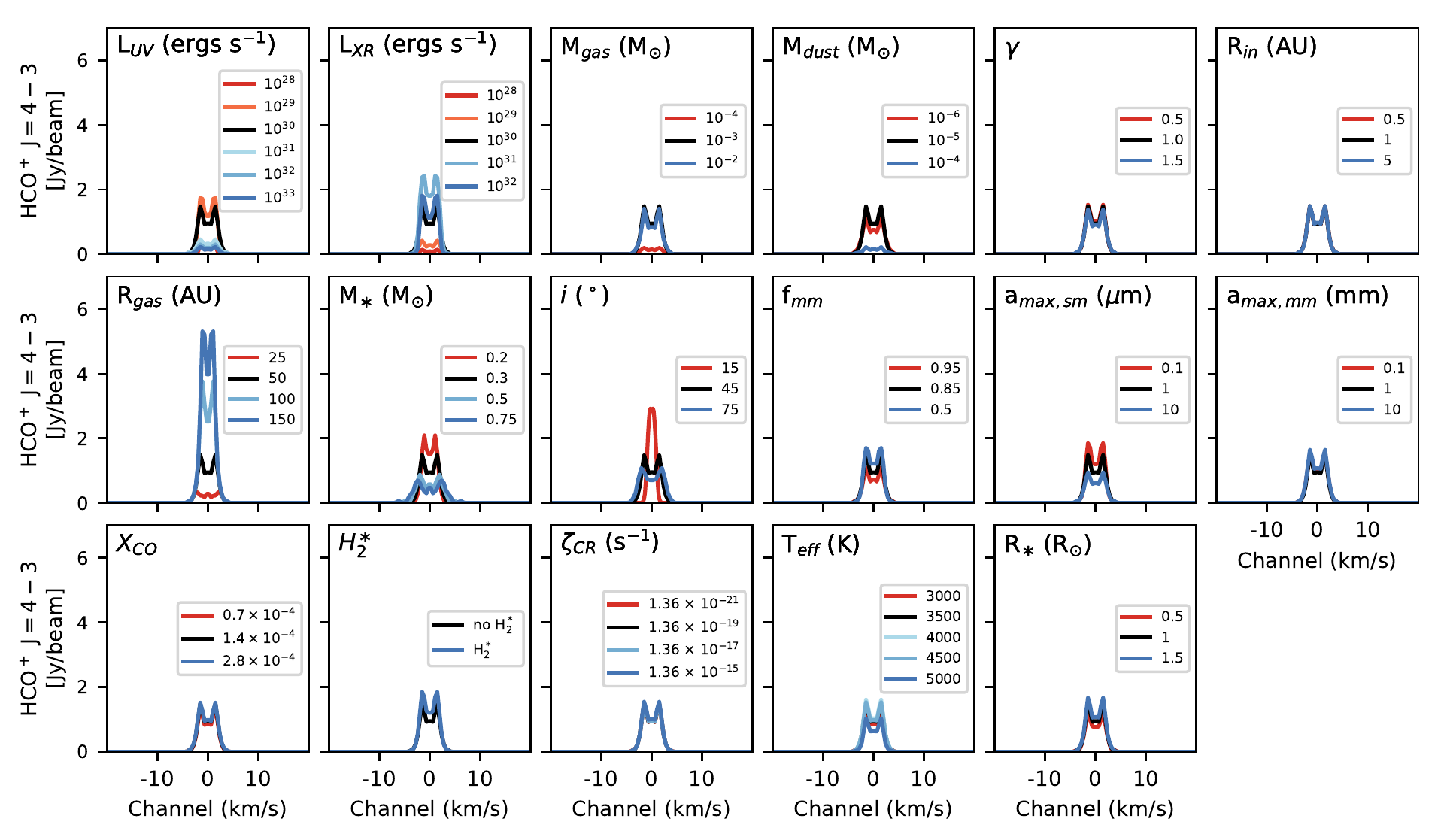}
	\caption{Same as Figure \ref{fig:pstudy_spec1}, but for HCO$^+$ $J=4-3$.
	\label{fig:pstudy_spec2}}
\end{figure*} 

\begin{figure}[ht!]
	\epsscale{1.1}
	\plotone{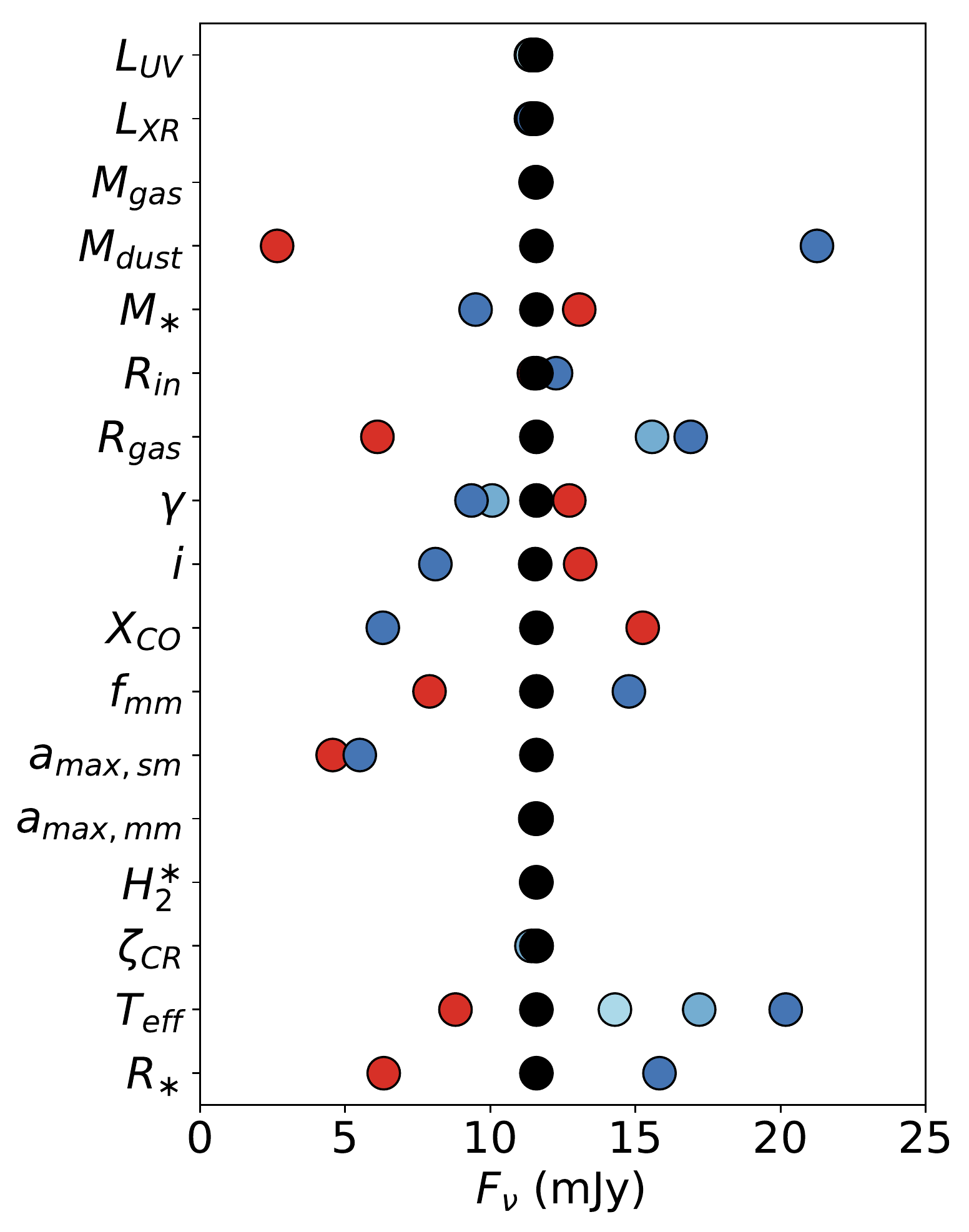}
	\caption{Demonstration of the effects of varying model parameters on the simulated $350$ GHz continuum flux. The color-coding for each model parameter value is the same as in Figures \ref{fig:pstudy} - \ref{fig:pstudy_spec2}.
	\label{fig:cont_fluxes_pstudy}}
\end{figure}

\begin{figure}[ht!]
	\epsscale{1.1}
	\plotone{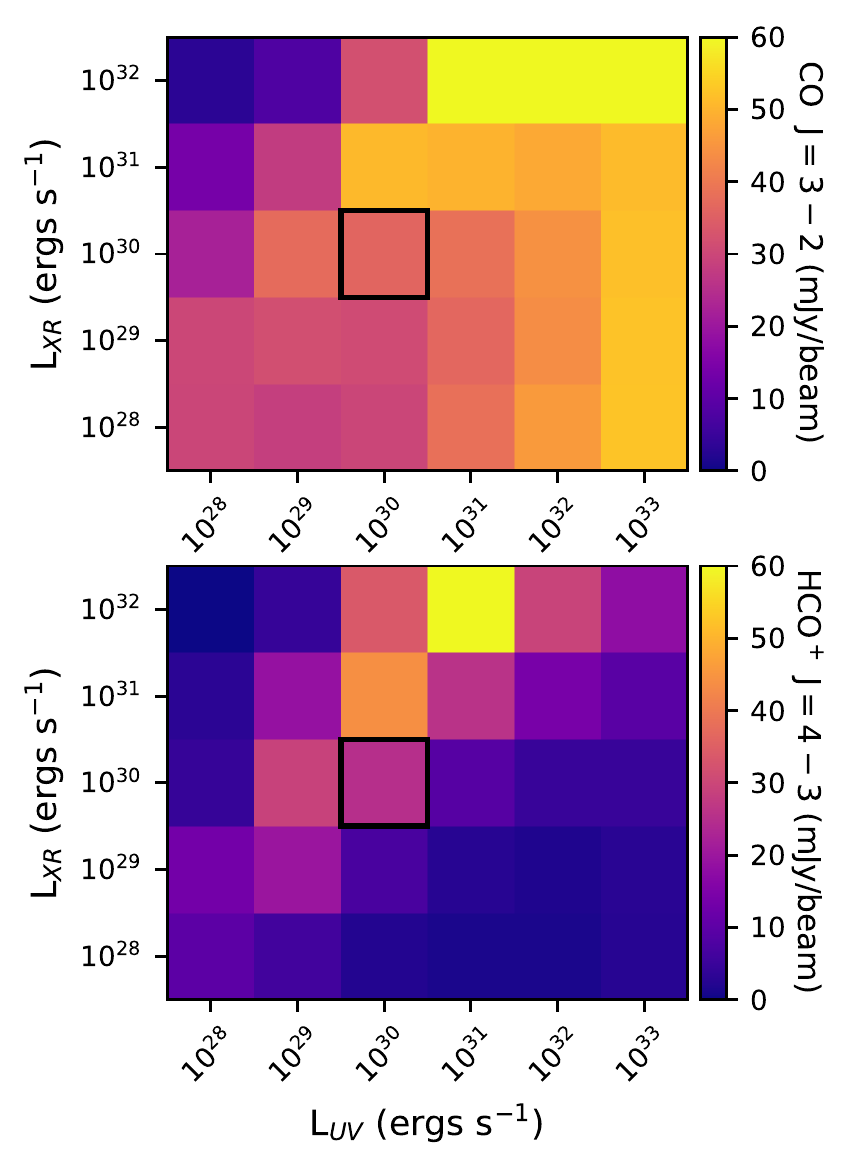}
	\caption{Peak CO $J=3-2$ signal (top), and peak HCO$^+$ $J=4-3$ signal (bottom) of our model disk plotted as a function of the FUV and X-ray luminosities. The black square indicates the fiducial values of $L_{UV}$ and $L_{XR}$.
	\label{fig:checker_GO_LXR}}
\end{figure}

\begin{figure}[ht!]
	\epsscale{1.1}
	\plotone{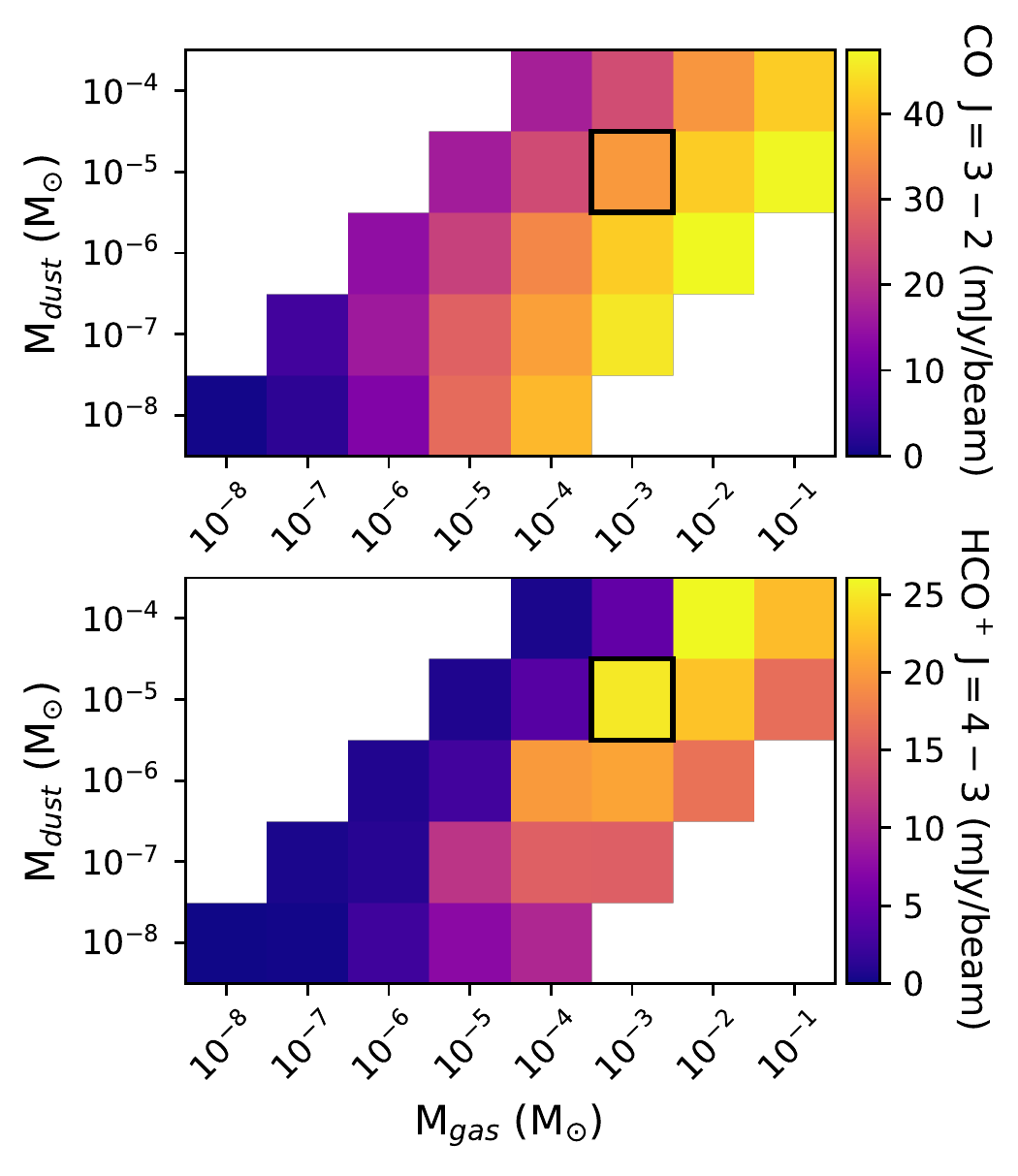}
	\caption{Peak CO $J=3-2$ (top) and HCO$^+$ $J=4-3$ (bottom) signals of our model disk plotted as a function of the disk dust and gas masses. The black square indicates the fiducial dust and gas masses.
	\label{fig:checker_Mgas}}
\end{figure}

Here we vary the input parameters of our thermochemical disk model described above to illustrate the effects of model parameters on the synthetic CO $J=3-2$ emission, HCO$^+$ $J=4-3$ emission, and $350$ GHz continuum flux. While many of the parameters have been explored individually in previous studies \citep[e.g.,][]{Walsh13, Cleeves14, Champion17, Rab18, Schwarz18, Schwarz21, Calahan21b, Haworth21a}, here we examine them collectively to quantify the relative importance of each parameter on the observations.

We vary parameters about a baseline model with fiducial parameter values listed in Table \ref{table:fiducial}. Our baseline model consists of a compact disk surrounding a pre-main-sequence star with stellar properties that are typical for M-dwarfs in the ONC \citep[e.g.,][]{Hillenbrand97, Getman05, Shkolnik14, Fang21}. The fiducial disk radius and total dust mass are comparable to the measured radii and dust masses of our targets \citep[e.g., ][]{Eisner18, Boyden20}, and we assume a fiducial gas-to-dust mass ratio of 100. The baseline model also has $\gamma = 1$, $R_{in}  = 1$ AU, $a_{min} = 0.005$ $\mu$m, $a_{max, \mu m} = 1$ $\mu$m,  $a_{max, mm} = 1$ mm, and $f_{mm} = 0.85$  \citep[e.g.,][]{Andrews11, Cleeves13, Miotello16, Zhang19, Calahan21a}. For the baseline ionization field, we adopt intermediate-level UV and X-ray radiation field strengths, and a cosmic-ray ionization rate that is comparable to the expected value for an accreting pre-main-sequence star  \citep[e.g.,][]{Cleeves13, Cleeves15}. Moreover, we assume a fiducial initial CO abundance that is comparable to expected CO abundances in the ISM, and we do not consider reactions involving vibrationally-excited molecular hydrogen (H$_2^*$) in our baseline chemical network \citep[c.f.,][]{Kamp17}. In Figure \ref{fig:fiducial}, we show the gas density, gas temperature, CO abundance, and HCO$^+$ abundance profiles of our baseline model.

Figures \ref{fig:pstudy} and  \ref{fig:pstudy2} show the results of varying model parameters on the synthetic integrated CO $J=3-2$ and HCO$^+$ $J=4-3$ emission.  In Figures \ref{fig:pstudy_spec1} and  \ref{fig:pstudy_spec2}, we show the effects of model parameters on the synthetic CO $J=3-2$ and HCO$^+$ $J=4-3$ line profiles.  We note that these spatially/spectrally integrated parameter study plots are meant to summarize the effects of model parameters on the simulated CO $J=3-2$ and HCO$^+$ $J=4-3$ channel map emission, and in Section \ref{sec:fitting} we use the full synthetic channel maps to model our ALMA observations. The effects of model parameters on the synthetic $350$ GHz continuum flux are shown in Figure \ref{fig:cont_fluxes_pstudy}.  In Figure \ref{fig:checker_GO_LXR}, we plot the peak CO $J=3-2$ and HCO$^+$ $J=4-3$ intensities as a function of both the UV and X-ray radiation fields, and in Figure \ref{fig:checker_Mgas} we show the peak intensities as a function of both the disk dust and gas masses. These 2-D parameter study plots are generated to explore sensitivities to the UV and X-ray ionization balance and to the gas-to-dust mass ratio, both of which are found to influence the model CO and HCO$^+$ emission.

For a brief summary of our parameter space exploration, we refer the reader to Figures \ref{fig:pstudy} - \ref{fig:checker_Mgas}. In Appendix \ref{appendix:a1}, we provide the full explanations of the effects of the model parameters on the synthetic continuum, CO, and HCO$^+$ observations.

\

\

\

\

\section{ \bf Modeling of Orion Nebula Cluster Disks}\label{sec:fitting}

Our parameter space exploration, as outlined in Figures \ref{fig:pstudy} - \ref{fig:checker_Mgas} and in Appendix \ref{appendix:a1}, demonstrates that the simulated $350$ GHz continuum, CO $J=3-2$, and HCO$^+$ $J=4-3$ emission of our thermochemical disk model exhibit strong sensitivities to select physical/chemical parameters. The $350$ GHz continuum emission is most sensitive to changes in the dust mass, whereas the CO $J=3-2$, and HCO$^+$ $J=4-3$ emission are equally sensitive to changes in the dust and gas masses, as well as changes in the UV and X-ray radiation field strengths. Changes in the disk size can also impact the model continuum and line emission, while changes in the stellar mass are mainly imprinted on the line emission. Many other explored parameters affect the disk chemistry and thermal and/or ionization structure in ways consistent with previous studies (see Appendix \ref{appendix:a1}). However, most of them have a relatively insignificant effect on the synthetic continuum, CO, and HCO$^+$ emission unless we consider unrealistic regions of parameter space (e.g., extremely-high cosmic-ray ionization rates).

Therefore, we expect that with currently-available ALMA data, we can constrain the dust and gas masses, disk sizes, UV and X-ray radiation fields, and central stellar masses of our selected ONC targets. 
In this section, we describe a large grid of thermochemical models and a procedure that fits synthetic observations of the thermochemical models to the $350$ GHz continuum, CO $J=3-2$, and HCO$^+$ $J=4-3$  ALMA observations of our sample targets.

\subsection{Grid of Thermochemical Models}

\begin{deluxetable}{lCc}
\tabletypesize{\small}
\tablenum{4}
\tablecaption{Parameters of the Model Grid \label{tab:grid_params}}
\tablehead{ 
    \colhead{Parameter} &  \colhead{Range}      & \colhead{Unit}  
   }
\startdata 
($M_{*}$,  $T_{eff}$)\tablenotemark{a} &  (0.1, 3000),  \ (0.2, 3000)                               & ($M_{\odot}$, K)  \\
                                       &  (0.3, 3500), (0.4, 3500)                                  & (M$_{\odot}$, K)  \\ 
                                       &  (0.5, 3750),  \ (0.75, 4250)                              & (M$_{\odot}$, K)  \\ 
                                       &  (1.0, 4500), \ (1.5, 4750)                                & (M$_{\odot}$, K)  \\ 
                                       &  (2.0, 5000)                                               & (M$_{\odot}$, K)  \\ 
M$_{gas}$\tablenotemark{b}             &   3 \times 10^{-7}, \ 3 \times 10^{-6}, \ 3 \times 10^{-5},       & M$_{\odot}$        \\
    & \ 3 \times 10^{-4}, 3 \times 10^{-3},   & M$_{\odot}$        \\
             &  \ 3 \times 10^{-2}, \ 3 \times 10^{-1}       & M$_{\odot}$        \\
M$_{dust}$\tablenotemark{b}      &   3 \times  10^{-8}, \ 3 \times  10^{-7}, \ 3 \times 10^{-6},                           & M$_{\odot}$        \\ 
                                 &  \ 3 \times  10^{-5}, \ 3 \times  10^{-4}                             & M$_{\odot}$        \\ 
R$_{gas}$                              &  25, \ 50, \ 75, \ 100, \ 150, \ 250                       & AU                \\ 
$L_{UV}$\tablenotemark{c}                              &  10^{28}, \ 10^{29}, \ 10^{30}, \ 10^{31}, \ 10^{32}, \ 10^{33}       & ergs s$^{-1}$ \\
                                               & (1, \ 10, \ 10^{2}, \ 10^{3}, \ 10^{4}, \ 10^{5})                               & $G_0$  \\
$L_{XR}$                               &  10^{28}, \ 10^{29}, \ 10^{30}, \ 10^{31}, \ 10^{32}       & ergs s$^{-1}$ \\
$i$                                    &  5, \ 15, \ 30, \  45,   \ 60, \  75, \ 85      & $^{\circ}$ \\
$\theta$                               &  0, \ 30, \  60,   \ 90, \  120, \ 150,      & $^{\circ}$ \\
                                       &  180, \ 210, \  240,   \ 270, \  300, \ 330      & $^{\circ}$ \\
$v_{sys}$\tablenotemark{d}                              &              & km s$^{-1}$ \\
\enddata
\tablenotetext{a}{The stellar mass and effective temperature are varied in tandem.}
\tablenotetext{b}{Models with M$_{gas}$ $/$ M$_{dust}$ $<10$ and M$_{gas}$ $/$ M$_{dust}$ $>10^4$ are excluded from the grid.}
\tablenotetext{c}{Here we include an additional row showing the values of $L_{UV}$ in units of the Habing field, $G_0$, at $r \approx 25$ AU \citep[$G_0 = 1.6 \times 10^{-3}$ ergs cm$^{-2}$ s$^{-1}$, ][]{Habing68}.}
\tablenotetext{d}{Varied on a source-by-source basis, see Section \ref{sec:fitting_chisq}.}
\end{deluxetable}

Table \ref{tab:grid_params} lists the model parameters and parameter values used in our grid. There are a total of 9 free parameters: the stellar mass ($M_{*}$), the disk gas mass ($M_{gas}$), the disk dust mass ($M_{dust}$), the disk radius ($R_{gas}$), the FUV luminosity ($L_{UV}$), the X-ray luminosity ($L_{XR}$), the disk inclination ($^\circ$), the disk position angle ($\theta$), and the disk systemic velocity ($v_{sys}$). {All other thermochemical parameters that we explored in Section \ref{sec:modeling:pstudy} are fixed to the values listed in Table \ref{table:fiducial}, because they were found to have a smaller impact on the synthetic $350$ GHz continuum, CO $J=3-2$, and HCO$^+$ $J=4-3$ emission over realistic regions of parameter space, as we have shown in Figures \ref{fig:pstudy} - \ref{fig:cont_fluxes_pstudy} and discussed in  Appendix \ref{appendix:a1}.}

The stellar masses listed in Table \ref{tab:grid_params} cover the range expected for both M-dwarfs and Solar-type T-Tauri stars. Because solar-type stars are expected to be warmer than lower-mass M-dwarfs, we vary the effective temperature in tandem with the stellar mass, using the combination of values shown in Table \ref{tab:grid_params}. To determine the paired values of stellar mass and effective temperature, we used the pre-MS evolutionary tracks of \cite{Baraffe15} and \cite{Feiden16} and assumed a {stellar age of $\sim 1$ Myr, i.e, the canonical age of the ONC \citep[e.g.,][]{Hillenbrand97, DaRio12, DaRio16, Fang21}}. 

To incorporate the disk mass into the grid, we vary the gas mass from $3 \times 10^{-7}$ to $3 \times 10^{-1}$ $M_{\odot}$ and the dust mass from $3 \times 10^{-8}$ to $3 \times 10^{-4}$ $M_{\odot}$, comparable to the mass ranges explored in publicly-available grids of thermochemical models \citep[e.g.,][]{WB14, Miotello16} as well as the range of expected dust masses for our sample targets \citep[e.g.,][]{Mann14, Eisner18}. We enforce a minimum gas-to-dust mass ratio of $10$ and maximum gas-to-dust mass ratio of $10^4$ in our grid, and exclude mass pairings lying outside this range (see Appendix \ref{appendix:a1}). {Moreover, based on previous disk size measurements obtained by \cite{Boyden20}, who showed that disks in the ONC are smaller than the disks found in lower-mass star-forming of similar age \citep[e.g., ][]{Ansdell18}, we model disks with radii spanning $25 - 250$ AU. As we demonstrate in Section \ref{sec:fitting:results}, this range is sufficient to capture all disk sizes in our sample.}

The values of $L_{UV}$ and $L_{XR}$ that we consider in our grid are the same as the values covered in our parameter space exploration. Higher values of $L_{UV}$ represent represent disks that are exposed to the strong UV fields that are expected in clustered environments, whereas lower values of $L_{UV}$ represent cases where the disks are partially or completed shielded from strong UV fields due to extinction effects \citep[e.g.,][see also Table \ref{tab:grid_params}]{Winter19, Qiao22}. The chosen range of $L_{XR}$ values is consistent with the range of X-ray luminosities measured towards young stars in the ONC \citep[][see also Figure \ref{fig:COUP_ONC}]{Getman05}, 
as well as the typical luminosities expected from stochastic X-ray flaring events \citep[e.g.,][]{Waggoner22}, though we do not consider temporal variations in X-ray luminosity in our models. We have also verified that our grid renders HCO$^+$ abundance and column density profiles that more than cover the abundance enhancements expected from X-ray flaring  \citep[e.g.,][]{Waggoner22}. 
Finally, we include a broad range of possible viewing orientations for the disk inclination, position angle, and systemic velocity.

For each unique set of allowed parameter values, we generate a set of synthetic CO $J=3-2$ channel maps, synthetic HCO$^+$ $J=4-3$ channel maps, and a synthetic total dust continuum flux measurement at $350$ GHz, following the steps outlined in Section \ref{sec:modeling:thermochemical_calculations}. The thermochemical calculations are computationally intensive, and it can take anywhere from a few hours to a few days to run one model. We use the supercomputing resources provided by the University of Arizona Center for High Performance Computing to generate models in parallel over a large number of nodes and CPUs, enabling us to substantially reduce the total time needed to generate the grid. 
Altogether, we produce a total of 2,585,520  
sets of synthetic CO $J=3-2$, HCO$^+$ $J=4-3$, and $350$ GHz dust continuum observations (excluding the additional sets generated when varying $v_{sys}$, see Section \ref{sec:fitting_chisq}) that can be compared with our ALMA datasets. 
To match the resolution of the models and ALMA observations, we spectrally average the model channel maps down to $0.5$ km s$^{-1}$ resolution and convolve the model images with a $0\rlap{.}''09$ Gaussian beam. 

\subsection{ Fitting Procedure}\label{sec:fitting_chisq}

For each target in our sample, we adopt the following procedure for fitting our grids of synthetic $350$ GHz dust continuum fluxes, CO $J=3-2$ channel maps, and HCO$^+$ $J=4-3$ channel maps to the measured dust continuum fluxes and measured CO $J=3-2$ and/or HCO$^+$ $J=4-3$ channel map emission/absorption.

We first specify the $v_{sys}$ values that will be used in the fitting procedure. Previous kinematic modeling by \cite{Boyden20} has shown that the systematic velocities of our targets can usually be constrained down to $\lesssim 3$ velocity channels, such that varying  v$_{sys}$ over broad ranges of parameter space is computationally inefficient and unnecessary. We initially consider values within $\pm 1$ km s$^{-1}$ of the published measurements reported in \cite{Boyden20}, and only consider a broader range of values if it is needed to constrain the confidence interval. We utilize a step size of $0.5$ km s$^{-1}$, which matches the channel width of the synthetic observations and therefore enables us to vary $v_{sys}$ {\it a posteriori} \citep[see ][]{Boyden20}.

Next, we identify whether the target is detected in CO emission, HCO$^+$ emission, or CO absorption. If the target is detected in CO emission or HCO$^+$ emission, then we fit the grids of synthetic CO $J=3-2$ or HCO$^+$ $J=4-3$ observations directly to the ALMA line observations. If the target is a CO absorption detection, then we convert our grid of synthetic CO $J=3-2$ observations into a grid of simulated interferometric CO $J=3-2$ observations of a disk positioned in front of an extended molecular cloud via the equation
\begin{equation}\label{eq:simobs_main}
	I_{sim} = I_{d} - I_{c}(1 - {\it e}^{-\tau_{d}}), 
\end{equation}
and fit this converted quantity to the ALMA CO $J=3-2$ observations. Here, $I_{sim}$ denotes the corrected CO $J=3-2$ intensity, $I_{d}$ denotes the model disk CO $J=3-2$ intensity (i.e., the synthetic observations from the grid), $I_{c}$ denotes the CO $J=3-2$ background intensity of the Orion Molecular Cloud, and $\tau_{d}$ denotes the disk CO $J=3-2$ line optical depth. Equation \ref{eq:simobs_main} is derived from the equation of radiative transfer with the effects of spatial filtering included, and the details of this approach are discussed further in Appendix \ref{appendix:a}. To calculate the background cloud intensity, we extract a localized CO $J=1-0$ spectrum from the Carma-NRO Orion Survey \citep{Kong18, Kong21}. The CO $J=1-0$ intensity reasonably approximates the CO $J=3-2$ intensity under the expected physical conditions of the Orion Molecular Cloud (see Appendix \ref{appendix:a}). Because the resolution of the Carma-NRO Orion Survey ($8''$) exceeds the size of the sub-images created for each target in our sample ($1''$), we assume that the background intensity does not vary spatially across each sub-image.

We then perform joint fits to the channel map observations and free-free corrected $350$ GHz dust continuum fluxes via a $\chi^2$ minimization procedure. Here we simultaneously fit the two independent datasets (continuum flux measurements and line channel maps), and define the $\chi^2$ statistic as the weighted sum of the $\chi^2$ values obtained for each dataset:
\begin{equation}
    \chi^2 = w_{cont} \chi_{cont}^2 + w_{line} \chi_{line}^2.
\end{equation}
$\chi_{cont}^2$ is computed by comparing the model $350$ GHz dust continuum fluxes with the measured value, and is defined as 
\begin{equation}
    \chi_{cont}^2  = \bigg(\frac{F_{cont, mod} - F_{cont, obs}}{\sigma_{cont, obs}}\bigg)^2.
\end{equation}
The free-free corrected $350$ GHz dust continuum fluxes and uncertainties of each target are listed in Table \ref{tab:detection_info}. These measurements are taken from \cite{Eisner18}, who used cm-wavelength observations to constrain the free-free emission spectra towards each target and remove free-free contributions at $350$ GHz, which can be significant for cluster members with ionized proplyds \citep[e.g.,][]{Mann14, Sheehan16}.

$\chi_{line}^2$ is obtained by summing over the individual $\chi^2$ values computed at each pixel in the data cubes: 
\begin{equation}
    \chi_{line}^2  =  \text{\Large  $\sum_{\text{\small $x, y, v_z$} } $} \bigg(\frac{F_{line, mod}(x, y,  v_z)- F_{line, obs}(x, y,  v_z)}{\sigma_{line, obs}(v_z)}\bigg)^2, 
\end{equation}
with $\sigma_{line, obs}$ denoting the rms noise in each velocity channel (example rms values shown in Table \ref{tab:detection_info}). 
Since the velocity range of our model cubes ($\pm 100$ km s$^{-1}$) exceeds the velocity range of the channel map observations ($0.0 - 20.0$ km s$^{-1}$), we only consider pixels in the velocity range $0.0 - 20.0$ km s$^{-1}$ when calculating $\chi_{line}^2$. 
Moreover, because the pixel size of the channel map observations is smaller than the beam size, the flux measurements in each pixel are not spatially independent. We divide the channel map residuals by the number of pixels within a synthesized beam area\footnote{The number of square pixels within a synthesized beam area, $N_{pix}$, is related to the pixel size, $X_{pix}$, via: $N_{pix} = [\pi/4 \ln (2) ] (FWHM / X_{pix})^2$. For our adopted pixel size of $0.02''$ and synthesized beam FWHM of $0.09''$,  $N_{pix} \approx 24$.} so that the effective total number of residuals used to generate $\chi_{line}^2$ is equivalent to the number of independent channel map resolution elements.

Most of our targets are detected in either the CO $J=3-2$ or HCO$^+$ $J=4-3$ channel maps, so we typically use $\chi_{line}^2 = \chi_{CO}^2$ or $\chi_{line}^2 = \chi_{HCO^+}^2$ depending on the source. For 177-341W and HC482, which are detected in both the CO $J=3-2$ and HCO$^+$ $J=4-3$ channel maps, we use  $\chi_{line}^2  = \chi_{CO}^2 + \chi_{HCO^+}^2$ and sum over both sets of channel maps. We do, however, perform additional fits to 177-341W and HC482 with $\chi_{line}^2  = \chi_{CO}^2$  and $\chi_{line}^2  = \chi_{HCO^+}^2$ in order to quantify the effects of including both sets of channel maps in the fitting procedure.

We introduce the weight factors $w_{cont}$ and $w_{line}$ because the line channel maps have many more independent data points than the single continuum flux measurement, and we want to ensure that the best-fit models are consistent with both the measured continuum fluxes and the line channel maps. In general, we use $w_{cont} = 10$ and $w_{line} = 1$, which provides a good balance in fitting to the combined datasets. With this weighting scheme, the global $\chi^2$ surface is still predominantly influenced by the channel map emission, which is our focus in this paper, and the continuum fits are given enough weight such that they never dominate the combined fits but enable the best-fit models to always reproduce the measured continuum fluxes within $1\sigma$ or $2\sigma$.  The only exception is HC482, in which case we use $w_{cont} = w_{line} = 1$. We have found that the best-fit models for HC482 can already reproduce the measured continuum fluxes within $1\sigma$ without any additional weight to the continuum fits, and upweighting $w_{cont}$ results in a degradation of the fit to the line data. Hence, we elect not to upweight the continuum fits for HC482.

\section{\bf Results}\label{sec:fitting:results}

\begin{figure*}[ht!]
	\epsscale{1.15}
	\plotone{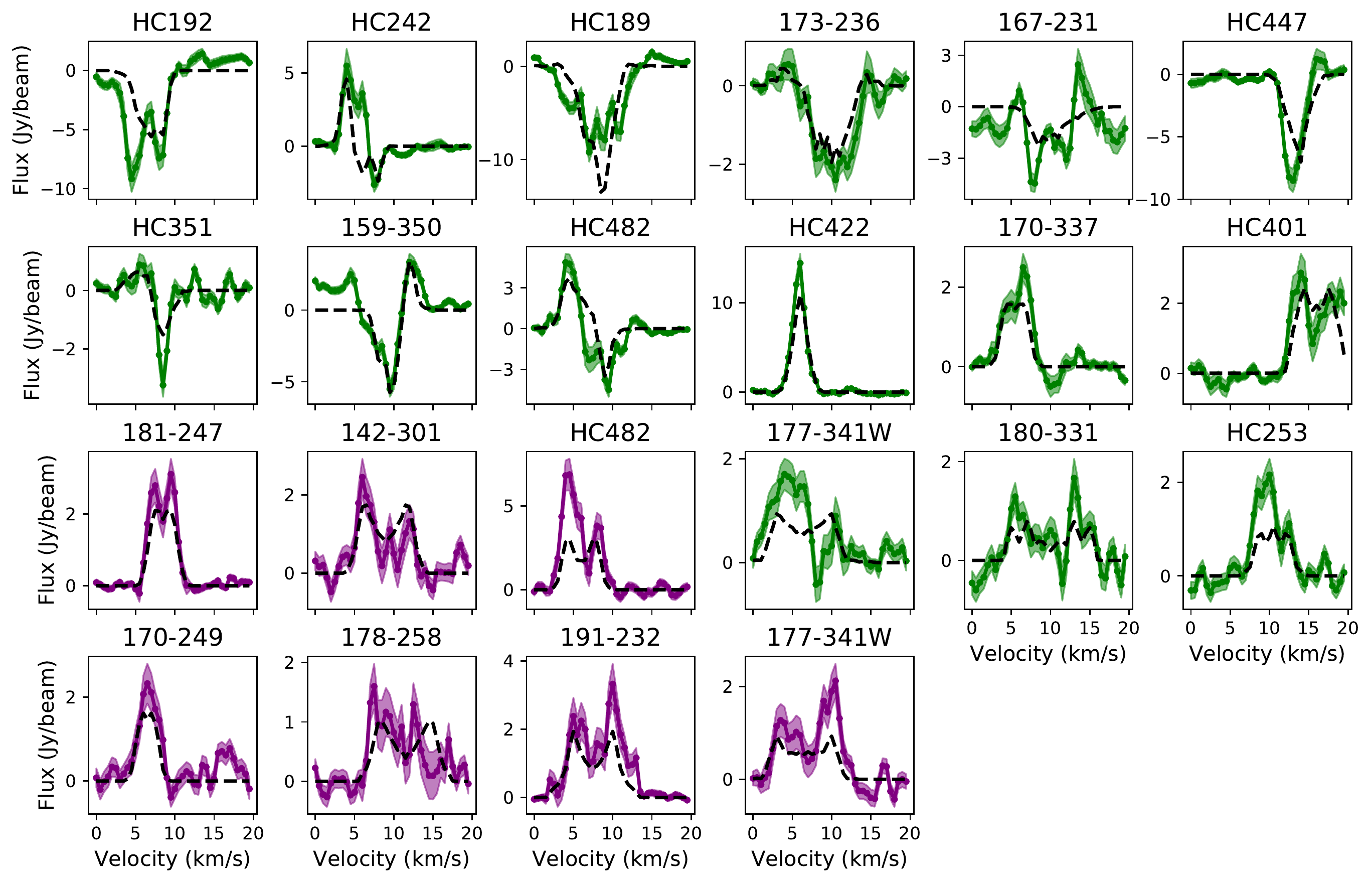}	
	\caption{Modeling results for all ONC targets in our sample. Each panel displays the integrated spectrum towards a target, with green corresponding to CO $J=3-2$ emission/absorption, and purple corresponding to HCO$^+$ $J=4-3$ emission. The best-fit model is plotted in each panel as a dashed black line.  All fits are performed on the channel map observations (see Appendix \ref{appendix:b}), but shown here as integrated spectra in order to summarize the modeling results for the entire sample.
	For 177-341W and HC482, we show the best-fit models obtained by fitting to both the CO $J=3-2$ and HCO$^+$ $J=4-3$ channel map observations (see Section \ref{sec:fitting_chisq}).
	\label{fig:modeling_results}}
\end{figure*}

\begin{deluxetable*}{llCCChChCCChCC}
\tablewidth{1000pt}
\tabletypesize{\footnotesize}
\tablenum{5}
\tablecaption{Best-Fit Model Parameters\tablenotemark{ }\label{tab:best_fit_params}}
\tablehead{ 
    \colhead{Source}          &      \colhead{Tracer}       &     \colhead{$M_{*}$}         &      \colhead{$\log{(M_{gas})}$}   & 
    \colhead{$\log{(M_{dust})}$} &      \nocolhead{$gdr$}        &      \colhead{$R_{gas}$}       &      \nocolhead{$\log{(L_{UV})}$}  &      \colhead{$\log{(L_{UV})}$}     &     \colhead{$\log{(L_{XR})}$}  &      \colhead{$i$}          &     \nocolhead{$\theta$}   &     \colhead{$\theta$}        &      \colhead{$v_{sys}$}    \\       
    \colhead{}                &      \colhead{}              &  \colhead{($M_{\odot}$)}   &     \colhead{($M_{\odot}$)}  & \colhead{($M_{\odot}$)}   &   \colhead{}                 & 
    \colhead{(AU)}            & \nocolhead{($G_{0}$)}          &  \colhead{(ergs s$^{-1}$)}        & 
    \colhead{(ergs s$^{-1}$)} & \colhead{(deg)} & \nocolhead{(deg)}       &
    \colhead{(deg)}           & \colhead{(km s$^{-1}$)}                    
}
\decimalcolnumbers
\startdata 
181-247    &  HCO$^+$  &  < 0.2  &  -2.5_{1.0}^{1.0}  &  -5.5_{1.0}^{1.0}  &  3.0_{1.0}^{1.0}  &  < 50.0  &  3.0_{1.0}^{1.0} &  31.0_{1.0}^{1.0}  &  > 31.0  &  75.0_{15.0}^{10.0}  &  150.0_{30.0}^{30.0}  &  120.0_{30.0}^{30.0}  &   8.5_{0.5}^{0.5} \\
142-301    &  HCO$^+$  &  0.75_{0.25}^{0.75}  &  -4.5_{1.0}^{2.0}  &  -6.5_{2.0}^{1.0}  &  2.0_{1.0}^{2.0}  &  75.0_{25.0}^{25.0}  &  < 2.0  & < 30.0  & > 29.0  &  85.0_{25.0}^{5.0}  &  270.0_{30.0}^{30.0}  &  0.0_{30.0}^{30.0}  &  9.0_{0.5}^{1.0} \\
170-249    &  HCO$^+$  &  < 0.2  &  > -3.5  &  > -5.5  &  3.0_{2.0}^{1.0}  &  < 50.0  &  > 2.0  & > 30.0  &  > 30.0  &  45.0_{15.0}^{15.0}  &  180.0_{30.0}^{60.0}  & 90.0_{60.0}^{30.0}  &  6.5_{0.5}^{1.0} \\
178-258   &  HCO$^+$  &  0.75_{0.55}^{0.25}  &   > -4.5  & -5.5_{1.0}^{1.0}  &  3.0_{1.0}^{1.0}  &   50.0_{25.0}^{25.0}   &   > 0.0  & > 28.0   &  ...  &  85.0_{10.0}^{5.0}  &  -120.0_{60.0}^{30.0}  &  30.0_{60.0}^{30.0}  &  11.5_{1.5}^{1.0} \\
191-232     &  HCO$^+$  &   0.75_{0.25}^{0.5}  &  -4.5_{1.0}^{2.0}  &  -7.5_{1.0}^{1.0}  &  > 2.0  &  100.0_{25.0}^{50.0}  &  < 3.0  & < 31.0  &   31.0_{2.0}^{1.0}  &  85.0_{10.0}^{5.0}  &  60.0_{60.0}^{60.0} &  210.0_{60.0}^{60.0} &  7.5_{0.5}^{1.0} \\
177-341W  &  HCO$^+$  &  > 0.4  &  -2.5_{3.0}^{1.0}  &  -6.5_{2.0}^{1.0}  &  > 1.0  &  > 25.0  &  > 0.0  &  > 28.0  &  > 29.0  &  75.0_{30.0}^{15.0}  &  330.0_{60.0}^{30.0}  & 300.0_{30.0}^{60.0}  &  7.0_{3.5}^{0.5} \\
HC482      &  HCO$^+$  &  0.4_{0.1}^{0.35}  &  > -4.5  &  -5.5_{1.0}^{1.0}  &  > 1.0  &  75.0_{25.0}^{75.0}  &  > 1.0  &  > 29.0  &  > 30.0  &  60.0_{15.0}^{25.0}    &  330.0_{30.0}^{30.0}  & 300.0_{30.0}^{30.0}  &  6.0_{0.5}^{1.0} \\
HC422      &  CO  &  < 0.2  &  > -2.5  &  -5.5_{1.0}^{1.0}  &  > 3.0  &  < 50.0  &  3.0_{1.0}^{2.0}  & 31.0_{1.0}^{2.0}  &  ...              &  45.0_{15.0}^{15.0}  &  90.0_{30.0}^{60.0}  & 180.0_{60.0}^{30.0}  &  6.0_{0.5}^{0.5} \\
170-337    &  CO  &  0.75_{0.35}^{0.75}  &  > -2.5  &  > -5.5  &  3.0_{1.0}^{1.0}  &  < 50.0  &  3.0_{2.0}^{1.0}  & 31.0_{2.0}^{1.0}  &  31.0_{1.0}^{1.0}  &  15.0_{10.0}^{15.0}  &  240.0_{60.0}^{60.0}  & 240.0_{60.0}^{60.0}  &  5.5_{0.5}^{1.0} \\
HC401      &  CO  &  0.2_{0.1}^{0.1}  &  -2.5_{2.0}^{1.0}  &  -6.5_{1.0}^{1.0}  &  > 2.0  &  < 50.0  &  > 1.0  & > 29.0  &  ...              &  60.0_{15.0}^{15.0}  &  0.0_{60.0}^{30.0}  &  270.0_{30.0}^{60.0}  &  16.0_{1.0}^{0.5} \\
177-341W &  CO  &   > 0.1  &   > -6.5   &  > -8.5   &  > 1.0  &  < 150              &  ...   &  ...              &  ...              &   85.0_{80.0}^{5.0}             &  360.0_{60.0}^{30.0}  & 270.0_{30.0}^{60.0}  &  4.5_{0.5}^{1.5} \\
HC253      &  CO  &  0.4_{0.3}^{0.6}  &  > -4.5  &  -5.5_{1.0}^{1.0}  &  > 1.0  &  < 50.0  &  > 0.0  & > 28.0  &  ...              &  30.0_{15.0}^{15.0}  &  60.0_{90.0}^{150.0}  & 210.0_{150.0}^{90.0}  &  10.5_{1.0}^{1.0} \\
180-331    &  CO  &  0.4_{0.3}^{0.6} &   > -4.5  &    > -7.5   &  > 2.0  &  < 50.0  &  ...     &  ...              &  ...              &   > 75.0_{30.0}^{15.0}  &  30.0_{120.0}^{120.0}  & 240.0_{120.0}^{120.0}  &  10.0_{4.0}^{1.0} \\
HC192      &  CO\tablenotemark{a}  &  0.3_{0.2}^{0.7}  &  > -5.5  &  > -6.5  &  < 4.0  &  50.0_{25.0}^{100.0}  &  ...   &  ...              &  ...              &  75.0_{45.0}^{15.0}  &  120.0_{60.0}^{60.0}  &  150.0_{60.0}^{60.0}  &  6.0_{1.5}^{1.0} \\
HC242      &  CO\tablenotemark{a}  &  > 0.75  &  > -2.5  &  > -5.5  &  3.0_{1.0}^{1.0}  &  50.0_{25.0}^{25.0}  &  3.0_{1.0}^{1.0}  & 31.0_{1.0}^{1.0}  &  > 31.0  &  15.0_{10.0}^{15.0}  &  ...              &      ...         &  5.5_{1.0}^{2.0} \\
HC189      &  CO\tablenotemark{a}  &  > 1.5  &  -2.5_{1.0}^{1.0}  &  > -5.5  &  2.0_{1.0}^{1.0}  &  75.0_{25.0}^{25.0}  &  > 3.0  & > 31.0  &  < 32.0  &  30.0_{15.0}^{15.0}  &  240.0_{30.0}^{30.0}  &   30.0_{30.0}^{30.0}  &   7.5_{0.5}^{0.5} \\
173-236    &  CO\tablenotemark{a}  &  > 1.5  &  -2.5_{2.0}^{1.0}  &  > -5.5  &  < 3.0  &  50.0_{25.0}^{25.0}  &  ...              &  ...    &  ...              &  60.0_{15.0}^{15.0}  &  180.0_{60.0}^{60.0}  & 90.0_{60.0}^{60.0}  &  9.0_{1.0}^{1.5} \\
167-231    &  CO\tablenotemark{a}  &  > 0.3  &  > -6.5  &  -5.5_{2.0}^{1.0}  &  ...              &  < 250.0  &  ...              &  ...      &  ...           &  75.0_{30.0}^{10.0}  &  ...         &         ...          &  11.0_{4.0}^{1.5} \\
HC447      &  CO\tablenotemark{a}  &  < 0.2  &  -2.5_{2.0}^{1.0}  &  -6.5_{1.0}^{1.0}  &  > 2.0  &  < 50.0  &  < 3.0  &  < 31.0  &  < 32.0  &  60.0_{15.0}^{15.0}  &  -90.0_{30.0}^{30.0}  &   0.0_{30.0}^{30.0}  &  14.0_{1.0}^{0.5} \\ 
HC351      &  CO\tablenotemark{a}  &  < 2.0  &  > -6.5  &  > -7.5  &  ...              &  < 250.0  &  ...        &  ...             &  ...              &  75.0_{30.0}^{15.0}  &  ...              &   ...              &   6.5_{2.5}^{6.5} \\
159-350    &  CO\tablenotemark{a}  &  > 1.5  &  -2.5_{1.0}^{1.0}  &  > -5.5  &  2.0_{1.0}^{1.0}  &  50.0_{25.0}^{25.0}  &  > 1.0  & > 29.0  &  < 32.0  &  15.0_{10.0}^{45.0}  &  ...              &    ...              &   10.5_{0.5}^{1.5} \\
HC482      &  CO\tablenotemark{a}  &  0.5_{0.1}^{0.25}  &  -3.5_{1.0}^{1.0}  &  -5.5_{1.0}^{1.0}  &  2.0_{1.0}^{1.0}  &  50.0_{25.0}^{25.0}  &  > 2.0  &  > 30.0  &  > 31.0  &  45.0_{15.0}^{15.0}  &  330.0_{60.0}^{30.0}  &  300.0_{30.0}^{60.0}  &  6.5_{1.0}^{0.5} \\
177-341W\tablenotemark{b}  &   &  0.75_{0.25}^{0.75}  &  -4.5_{2.0}^{1.0}  &  -6.5_{2.0}^{1.0}  &  2.0_{1.0}^{1.0}  &  50.0_{25.0}^{50.0}  &  < 3.0  & < 31.0  &  > 29.0  &  75.0_{30.0}^{15.0}  &  300.0_{30.0}^{60.0}  & 330.0_{60.0}^{30.0}  &  6.5_{1.0}^{0.5} \\
HC482\tablenotemark{b}       &    &  0.5_{0.3}^{0.25}  &  -2.5_{2.0}^{1.0}  &  -5.5_{1.0}^{1.0}  &  3.0_{2.0}^{1.0}  &  50.0_{25.0}^{25.0}  &  2.0_{1.0}^{2.0}  &  30.0_{1.0}^{2.0}  &  > 31.0  &  45.0_{15.0}^{15.0}  &  330.0_{60.0}^{30.0}  &   300.0_{30.0}^{60.0}  &  6.0_{0.5}^{1.0} \\
\enddata
\tablenotetext{ }{{\bf Notes.} Column (1): source name; Column (2): molecular tracer; Columns (3) - (11): best-fit model parameters derived from our modeling of $350$ GHz dust continuum fluxes and CO $J=3-2$ and/or HCO$^+$ $J=4-3$ 
channel map emission/absorption, with uncertainties spanning a 1$\sigma$ confidence interval. Upper or lower limits are given when we are unable to derive full confidence intervals with our grid. Ellipses indicate cases where we are unable to place any limit on a model parameter. 
}
\tablenotetext{a}{ Denotes a CO $J = 3-2$ absorption detection.}
\tablenotetext{b}{ Best-fit model parameters obtained when fitting to both the CO $J = 3-2$ and HCO$^+$ $J = 4-3$ observations.}
\end{deluxetable*}

\begin{figure*}[ht!]
	\epsscale{1.1}
	\plotone{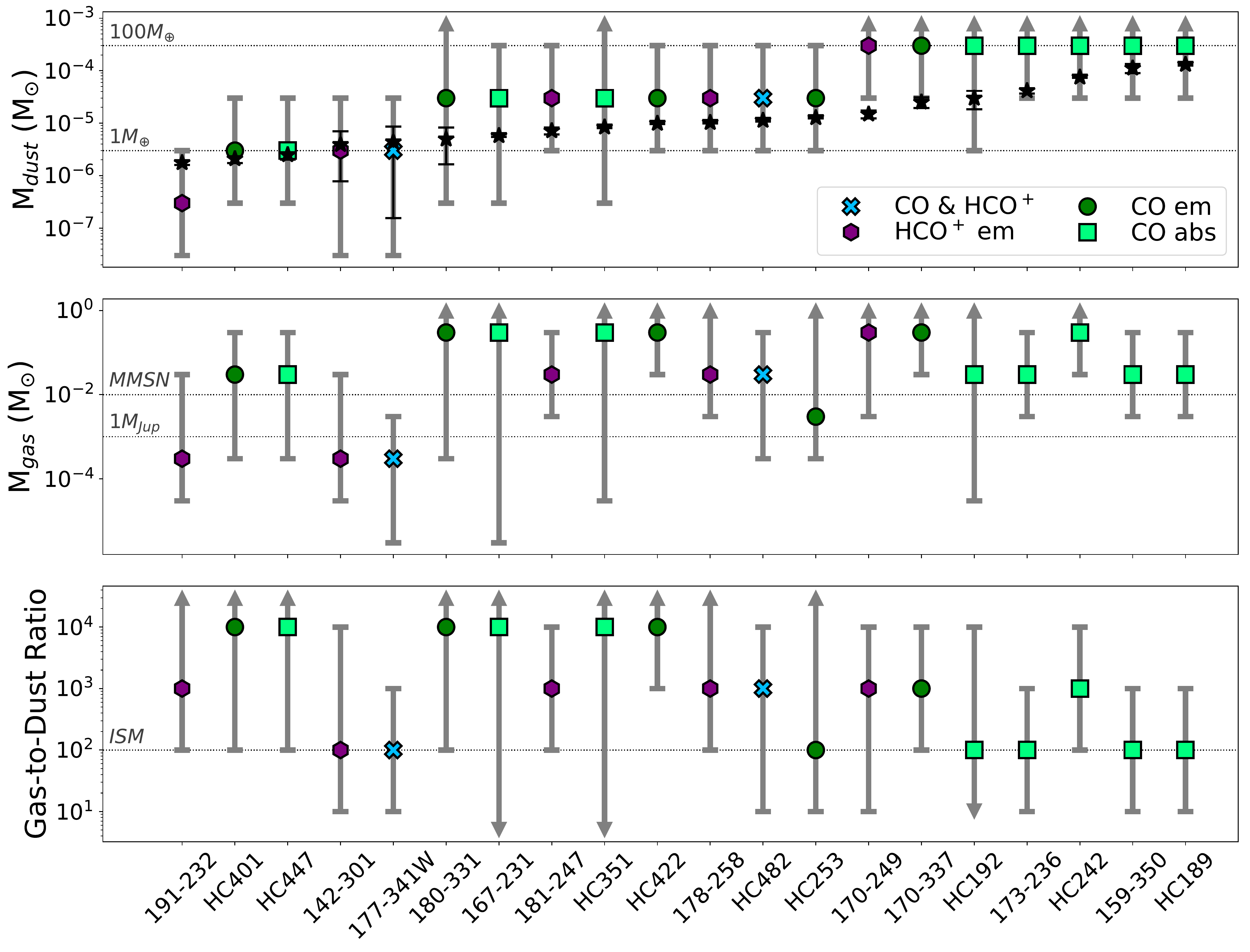}
	\caption{Disk dust masses (top), gas masses (middle), and gas-to-dust ratios (bottom) inferred for all ONC targets in our sample. Best-fit values are specified using dark green circles for the CO emission detections, light green squares for the CO absorption detections, and purple hexagons for the HCO$^+$ emission detections. For ONC members 177-341W and HC482, which are modeled in both CO and HCO$^+$, we use a cyan X-symbol.  Arrows indicate upper or lower limits. In the top panel, we include black stars that show the dust masses derived from the measured $350$ GHz continuum fluxes assuming that the dust is optically thin, via Equation \ref{eq:dust}. 
	\label{fig:Mgas_dust_ratio}}
\end{figure*}

Figure \ref{fig:modeling_results} shows the best-fit models obtained for each ONC member in our sample. 
While the fitting is performed on the channel maps, we plot the ALMA observations and best-fit models as integrated spectra in Figure \ref{fig:modeling_results}, in order to illustrate the results for the entire sample. We provide the full sets of observed channel maps for the sources, best-fit models, and residuals of the fits in Appendix \ref{appendix:b}. All of the fits reproduce the observed continuum emission within $<3\sigma$ of the measured values, and also reproduce the observed line emission/absorption with few or no residuals $>3\sigma$.

Our modeling of the CO absorption detections yields realistic fits to the observed negative emission. Strong CO $J=1-0$ cloud emission, as measured from the CARMA-NRO Orion Survey, is typically found over the same regions of position-velocity space as the  CO $J=3-2$ absorption that we measure with ALMA. Hence, when we convert our synthetic CO $J=3-2$ observations into simulated interferometric CO $J=3-2$  observations of a disk positioned in front of a background cloud (via Equation \ref{eq:simobs_main}), we can obtain models with negative emission comparable to what is seen in the ALMA observations. There are a few cases where the absorption detections show strong negative emission that our models cannot reproduce (e.g., HC192). These difference arise when the measured background CO $J=1-0$ intensity is too faint to yield strong negative model emission along those particular regions of position-velocity space, and could be reconciled if the  background CO $J=3-2$ intensity is stronger than implied by our simple extrapolation 
from CO $J=1-0$, which is possible for cloud conditions different from what we have assumed (see Appendix \ref{appendix:a}).

Since the intensity of the background cloud varies as a function of velocity, we can also obtain models that emit positive CO emission in some velocity channels and negative CO emission in other channels. This feature has enabled us to reproduce the asymmetric CO absorption profile of 173-236 as well as the CO emission and CO absorption detected towards HC482 and 159-350, although for 159-350 we find that the models do not prefer fitting to the emission observed at low-velocity channels (see Appendix \ref{appendix:b}). As such, the emission in those channels, which follows a different velocity gradient than the absorption and emission seen in higher-velocity channels, could be tracing an infall or other non-disk structure, as suggested previously by \cite{Boyden20}.

\subsection{Best-fit model parameters}

Table \ref{tab:best_fit_params} lists the best-fit model parameters derived from the fits. Our grid spans broad enough regions of parameter space such that we are able to derive full $1\sigma$ confidence intervals or place upper or lower limits on the majority of model parameters for all sample targets. The fits tend to prefer compact gas disks with radii $<100$ AU, and a range of stellar masses and viewing geometries are found across the sample.

In the top panel of Figure \ref{fig:Mgas_dust_ratio}, we plot the best-fit dust masses inferred from our modeling. We also show the dust masses derived from the measured $350$ GHz dust continuum fluxes assuming that the dust is optically thin, via the equation
\begin{equation}\label{eq:dust}
M_{dust} = \frac{S_{v} \  d^2}{\kappa_{\nu} B_{\nu}(T_{dust})}.
\end{equation}
To use this equation, we assume that $d = 400$ pc \citep[the canonical distance to Orion; e.g.,][]{Kounkel17, Kounkel18, Grob18} and $\kappa_{\nu} = \kappa_{0} (\nu/\nu_{0})^{\beta}$ with $\beta = 1$ and $\kappa_0 = 10$ cm$^2$ g$^{-1}$ at $1000$ GHz \citep{Beckwith90}. While many studies often assume an average dust temperature of $T_{dust} = 20$ K when calculating the dust mass from Equation \ref{eq:dust} \citep[e.g.,][]{Ansdell16, Pascucci16}, we instead use the average dust temperatures obtained from our best-fit thermochemical models, as our targets typically have warmer average dust temperatures $\sim 30-40$ K due to their compact sizes, consistent with previous radiative transfer modeling of compact disks  \citep[e.g.,][]{Hendler17, Eisner18, Ballering19}.

For many targets, we constrain the dust mass at an order-of-magnitude level of precision, with best-fit values spanning $\sim 1-100$ $M_{\oplus}$ (see Figure \ref{fig:Mgas_dust_ratio}). A subset of cluster members, however, prefers the largest dust mass in our grid, in which case we can only place lower limits on the dust mass. In general, we find that the dust masses derived from our modeling are consistent with the values obtained under the assumption of optically thin dust when the model dust mass is $<10$ $M_{\oplus}$. For dust masses $\geq10$ $M_{\oplus}$ and radii $\leq 50$ AU, the dust disks become optically thick at $350$ GHz and the masses computed from Equation \ref{eq:dust} underestimate the model dust mass. This is especially evident for sources 170-249 and 170-337, whose massive dust disks are underpredicted by more than $1\sigma$ from Equation \ref{eq:dust} due to the high optical depths arising from their large dust masses and very compact sizes (see Table \ref{tab:best_fit_params}).

The second panel of Figure \ref{fig:Mgas_dust_ratio} shows the best-fit gas masses obtained for each source. For approximately half of the targets, the derived gas masses are lower limits because the largest gas mass in our grid is within the derived confidence intervals of the fits.  When we are able to fully constrain the gas mass, the precision is typically one or two orders of magnitude. Despite these somewhat broad constraints on the gas mass, we find that all of the fits prefer massive gas disks with gas masses that are consistent with the ``Minimum Mass Solar Nebula'' \citep[$\sim 0.01$ M$_{\odot}$; ][]{Weidenschilling77, Desch07}. This includes the subset of targets in which the derived gas masses are lower limits, as those limits still allow us to rule out low gas masses. Even for the cluster members with relatively low dust masses ($\sim 1$ $M_{\oplus}$), we find that the combined line and continuum observations are best reproduced by gas rich model disks that have similar gas masses as the cluster members with more massive dust disks.

Because the model CO $J=3-2$ and HCO$^+$ $J=4-3$ emission are found to be sensitive to the gas-to-dust mass ratio (see Figure \ref{fig:checker_Mgas}), we also track the model gas-to-dust ratios during the fitting procedure, and include a third panel in Figure \ref{fig:Mgas_dust_ratio} that shows the best-fit gas-to-dust ratios and their uncertainties obtained by treating the gas-to-dust ratio as an independent parameter. With this implementation, we find that all of the derived gas-to-dust ratios are either consistent with or, in some cases, larger than the canonical ISM value of 100.  The tendency of the fits to prefer ISM-level gas-to-dust ratios is consistent with the results of our parameter space exploration, which demonstrate that gas-rich model disks with gas-to-dust ratios $\gtrsim 100$ are needed to produce bright CO and/or HCO$^+$ emission at the level that is seen towards our sample targets (c.f., Figure \ref{fig:checker_Mgas} with Appendix \ref{appendix:b}). The only sources in which we are unable to constrain the gas-to-dust ratios are HC351, 167-231,  and HC192, as the observations towards those sources are too noisy for us to derive precise confidence intervals on a number of parameters, including the gas-to-dust ratio. 

While our joint fits to the continuum and CO/HCO$^+$ observations have enabled us to constrain the gas masses at a higher level of precision than from fitting to the molecular line observations alone, the gas masses obtained through this approach are not independent, and instead depend on the dust mass and by extension, the gas-to-dust ratio. Our derived confidence intervals may therefore underestimate the uncertainties on the gas mass when they include models with the minimum or maximum dust mass in our grid and/or the minimum or maximum allowed gas-to-dust ratio. For cluster members HC351, 167-231, and HC192, whose confidence limits include models with the minimum allowed gas-to-dust ratio (see Figure \ref{fig:Mgas_dust_ratio}), this would imply that we have not constrained the gas masses from our modeling. For all other cluster members, however, the fits universally reject models with the lowest considered dust mass and/or lowest gas-to-dust ratio. We therefore expect to have constrained the lower end of the best-fit gas masses for the vast majority of sample targets.

We are also able to constrain {the UV and X-ray luminosities for the majority of targets} detected in HCO$^+$. As shown in Table \ref{tab:best_fit_params}, those fits converge to intermediate/high values of $L_{UV}$ and $L_{XR}$. This follows from the behavior shown in Figure \ref{fig:checker_GO_LXR}, and demonstrates that specific combinations of $L_{UV}$ and $L_{XR}$ are needed to match the observed HCO$^+$ $J=4-3$ intensity of our targets. For the CO emission and CO absorption detections, the derived confidence intervals for $L_{UV}$ and $L_{XR}$ are usually broader, and for a subset of these sources we are unable to constrain either quantity. This is again consistent with our parameter space exploration, where we show that broader combinations of $L_{UV}$ and $L_{XR}$ can reproduce similar levels of CO emission or CO absorption (see Figure \ref{fig:checker_GO_LXR}). Interestingly, none of the fits prefer regions of parameter space where the X-ray radiation field is strong enough relative to the UV field to initiate significant chemical reprocessing of CO into atomic carbon and other neutral carbon carriers (see Appendix \ref{appendix:a1}).

Finally, our modeling of 177-341W and HC482 demonstrates the benefits of fitting to both CO and HCO$^+$ channel map observations. Since CO and HCO$^+$ react differently to changes in the disk radius, UV and X-ray radiation fields, and gas-to-dust ratio, including both sets of channel maps in the fitting procedure enables us to constrain those parameters at a higher level of precision than from separate fits to each tracer. For example, we are unable to place an upper limit on the disk radius of 177-341W when we fit to just the continuum and HCO$^+$ observations (see Table \ref{tab:best_fit_params}), as HCO$^+$ is not particularly sensitive to changes in R$_{gas}$ at larger radii (see Figures \ref{fig:pstudy2} and \ref{fig:pstudy_spec2}). When we include the CO observations along with the continuum and HCO$^+$ observations, the precision on the derived radius improves significantly, as expected from Figures \ref{fig:pstudy} and \ref{fig:pstudy_spec1}. Models with gas-to-dust ratios $>100$ are also rejected in the combined CO and HCO$^+$ fits since they are too bright in CO to match the CO observations of 177-341W. The combined fits can therefore produce better constraints (i.e., narrower confidence intervals) on the gas-to-dust ratio and gas mass. The combined fits also produce better constraints on derived systemic velocity and stellar mass, since they reject models with both low systemic velocities ($\sim 4$ km s$^{-1}$) and high stellar masses ($2$ M$_{\odot}$), which were allowed in the separate fits to the CO and HCO$^+$ channel maps (see Table \ref{tab:best_fit_params}).

\section{\bf Discussion}\label{sec:discussion}

\subsection{Gas-to-dust ratios}\label{sec:discussion:GDR}

The gas-to-dust ratios that we derive for our sample of ONC disks are systemically larger than the values derived for disks in lower-density regions from CO and/or CO isotopologue observations. ALMA surveys in low-mass star-forming regions reveal that many nearby disks emit faint CO isotopologue emission, 
such that comparisons of the measured line fluxes with physical-chemical models often yield gas masses $<1$ Jupiter mass and gas-to-dust ratios of $\sim 1-10$ \citep[e.g.,][]{WB14, Ansdell16, Long17, Miotello17}. A growing body of work suggests that the weak CO line emission is attributed to subinterstellar CO abundances rather than a rapid dispersal of the gas disk. In this framework, gas-phase CO is gradually removed from the disk via physical and chemical reprocessing, and if the reprocessing is not completely accounted for when inferring gas masses from CO and/or CO isotopologue observations, then the masses will be underestimated.  Direct evidence of depleted gas-phase CO has been found in the subset of disks with measured HD fluxes, which provide independent constraints on the gas mass that are typically larger than the gas masses derived from CO isotopologue observations assuming an ISM-like CO abundance \citep[e.g., ][]{Bergin13, Favre13, McClure16, Schwarz16, Calahan21b, Schwarz21}. Recent surveys targeting other molecular species, such as C$_2$H and N$_2$H$^+$, also find that the measured fluxes are better reproduced by models with low CO abundances than by models with low gas-to-dust ratios  \citep[e.g.,][]{Anderson19, Anderson22, Miotello19, Bosman21, Trapman22}. 

The ISM-like CO- and HCO$^+$-based gas-to-dust ratios that we derive suggest that disks in the ONC are less prone to CO depletion than the disks commonly studied in nearby star-forming regions. In fact, our models require particular balances of UV and X-ray radiation that enable gas-phase CO to retain an abundance close to the initial, ISM-like value in the emitting molecular layers, as this ingredient is essential for reproducing bright CO and HCO$^+$ emission at the level that is observed towards our targets (c.f., Figure \ref{fig:checker_GO_LXR} with Appendix \ref{appendix:b}). 
These irradiation properties are precisely the ones expected in clusters like the ONC (see Figure \ref{fig:COUP_ONC}), 
and as illustrated in Figure  \ref{fig:checker_Mgas}, the gas-to-dust ratio plays a central role in regulating the propagation of the radiation fields into the disk. If all models in our grid were strongly depleted in gas-phase CO, none of them would resemble what is seen in the HCO$^+$ observations, and even higher gas masses and gas-to-dust ratios would be required to match the CO observations.

Most targets in our sample also have disk temperatures $\gtrsim 30$ K, a trend directly related to the compact sizes derived from our fitting in addition to the irradiation properties of the ONC (e.g., see Figure \ref{fig:fiducial}). These temperatures are larger than the values in which gas-phase CO freezes out onto dust grains \citep[$\sim 15 - 30$ K; e.g., ][]{Bisschop06}. CO freeze-out is thought to be one of the main avenues for reprocessing gas-phase CO into less volatile species that can effectively lock up the available carbon and oxygen in circumstellar disks  \citep[e.g.,][]{Eistrup16, Bosman18, Schwarz18, Krijt20}. Compact disks may simply lack the cold gas and dust reservoirs needed to efficiently deplete gas-phase CO, and instead harbor warm environments that enable CO and HCO$^+$ to remain in gas phase at relative high abundances, similar to what is seen in the warm disks surrounding Herbig Ae/Be stars \citep[e.g.,][]{Kama20, vdMarel21a, Sturm22}.

It will be worth investigating whether compact disks in other star-forming regions are less prone to CO depletion than the massive and extended disks that are commonly used to study disk chemistry but not necessarily representative of all disks found in nearby star-forming regions. In Lupus, $\sim 50\%$ of disks with dust masses $>1$ $M_{\oplus}$ have CO-based gas-to-dust ratios $\leq10$ and $\sim 75\%$ have CO-based gas-to-dust ratios $<100$ \citep[see Figure 10 of][]{Miotello17}, a strong contrast with what is found in our sample. However, the region also contains a subset of disks towards the lower end of the disk mass distribution ($\lesssim1$ $M_{\oplus}$) that were undetected in the initial ALMA surveys targeting their CO isotopologue emission but have upper limits to the isotopologue fluxes that imply CO-based gas-to-dust ratios of $\lesssim 500$ \citep{Ansdell16, Miotello17}. A small number of low-dust-mass disks were detected, though, and interestingly, these disks tend to have CO-based gas-to-dust ratios that are consistent with ISM values and thus, more consistent with the gas-to-dust ratios found in our sample than with the gas-to-dust ratios found towards higher-dust-mass Lupus disks \citep[see Figure 10 of][]{Miotello17}.

\cite{Miotello21} recently analyzed a subset of the lower-dust mass Lupus disks that were undetected in $^{13}$CO and C$^{18}$O but detected in $^{12}$CO, and found that the unresolved $^{12}$CO line fluxes and $^{13}$CO upper limits can be reproduced entirely by compact disks with sizes that are similar to those found in the ONC and smaller than the extended sizes commonly found towards higher-dust-mass disks in Lupus \citep[$R \gtrsim 200$ AU; e.g., ][]{Ansdell18}.  If deeper, higher-resolution ALMA observations of Lupus were to confirm that the majority of undetected Lupus gas disks are compact and have ISM-like CO-based gas-to-dust ratios, similar to what is seen in the ONC, this would indicate that weaker CO depletion is a universal feature of compact disks.

Follow-up observations targeting more optically thin tracers are also needed to confirm that our sample of ONC targets are warm and CO-rich in the deeper, midplane layers of their disks. If confirmed, this would indicate that CO depletion is less of an uncertainty for disk mass measurements in clustered star-formation environments than it is for disk mass measurements in lower density regions. Such a finding would be advantageous for ALMA surveys of nearby clusters, which are too distant to observe large samples of cluster members using some of the faint molecular tracers that have been proposed for constraining CO depletion \citep[e.g.,][]{Anderson19, Anderson22, Trapman22}, but close enough for brighter molecules to be detected over large samples \citep[e.g,][]{Boyden20}. Observations targeting the bright CO isotopologues and/or other bright, commonly-detected molecules, such as HCO$^+$, may be all that is needed to constrain the disk gas mass distributions of nearby clusters at high precision if CO depletion is a non-factor. We will explore such observing strategies in future work.

\subsection{External photovaporation in the ONC}\label{sec:discussion:photoevap}

The presence of massive, gas-rich disks in a high-density, high-ionization environment like the ONC is intriguing. For the expected $\sim 1$ Myr age of the ONC \citep{Hillenbrand97, DaRio12, DaRio16, Fang21}, the high photevaporation rates inferred for many proplyds in the region \citep[$10^{-8} - 10^{-6}$ M$_{\odot}$ yr$^{-1}$;][]{Henney99}, including a number of targets in our sample (e.g., 170-337, 177-341W), 
imply that our detected gas disks should have dispersed well before the present time in which we are observing them. 
Such a rapid dispersal is also predicted by simulations of external photoevporation, which find typical disk lifetimes of $\lesssim 1$ Myr under the extreme UV  conditions found in the ONC \citep[e.g.,][]{Winter18, ConchaRamirez19, Nicholson19,Winter20a,Parker21b}. Our derived gas masses, as shown in Figure \ref{fig:Mgas_dust_ratio}, suggest that while environmental effects such as external photoevaporation are capable of rapidly depleting disks of gas and dust, the ONC still contains a subset of disks with enough material to form giant planets.

The massive disks found in our sample may represent a subset of disks that have more recently begun to externally photoevaporate than other disks in the region. The positions of the ONC members on the HR diagram indicate a large spread in luminosity \citep[e.g.,][]{Hillenbrand97}, and comparisons with pre-MS evolutionary tracks yield stellar ages ranging from $<$ 0.5 Myr to $\sim$3 Myr, with the youngest stars concentrated towards the center of the ONC  \citep{Hillenbrand97, DaRio12, DaRio16, Fang21}. \cite{Boyden20} showed that many of our sample targets lie well above the $1$-Myr isochrones of a variety of pre-MS evolutionary models \citep[e.g.,][]{Baraffe15, Feiden16}, suggesting that they are indeed younger than other disks in the region. 

Recent spectroscopic surveys argue that the star-formation activity in the ONC peaked closer to $\sim$2 Myr, with only a minority of stars forming $\lesssim$ 1 Myr ago \citep[e.g.,][]{DaRio10, Reggiani11, Beccari17, Kroupa18, Jerabkova19}. If our targets are young and in the minority, then we would expect them to have undergone external photoevaporation for a shorter duration than the more evolved disks that make up the majority of cluster members in the ONC \citep[e.g., ][]{Winter19}, which could explain why we are still detecting massive gas disks in the ONC in spite of the anticipated environmental impact. It is worth noting, however, that the star-formation history and median age of the ONC remain active subjects of debate, as a more recent spectroscopic analysis by \cite{Fang21} indicates that the median age of the ONC is closer to $\sim 1$ Myr and that an age spread is the not the main cause of the observed stellar luminosity spread, implying that an extended, $\gtrsim 1$ Myr period of star-formation is unlikely to have occurred in the ONC.

External photoevporation may have recently begun operating in all of the ONC. Stellar clusters form out of dense clouds of molecular gas, and while the gas gradually dissipates as more stars are born, the remaining gas can shield cluster members from external ionizing radiation. Recent simulations by \cite{Qiao22} show that newly-formed cluster members can be shielded from a strong external UV source for as long as $\sim 0.5$ Myr, and thus evolve for an extended period of time before being externally irradiated. This behavior underscores how the stellar ages are likely overestimates of the actual time in which a disk-bearing cluster member has been undergoing external photoevaporation. If the majority of cluster members in the ONC are young and were once subject to strong shielding, then it would be more likely for them to retain large disk masses after $\sim 1$ Myr of cluster evolution rather than after $\sim 1$ Myr of evolution with no shielding \citep[e.g.,][]{Winter19, Qiao22}. Given that the ONC is surrounded by dense molecular gas \citep[for recent examples, see][]{Kong18, Hacar20b}, we might expect shielding to have played a central role in the early evolution of most, if not all, cluster members.

We might also expect a number of disks in the ONC to still be shielded from the external UV field if $\theta^1$ Ori C formed $<1$ Myr ago \citep[e.g.,][]{Qiao22}. This scenario may account for the high gas masses inferred for HC192 and HC189, as these sources are undetected as proplyds but detected in HCO$^+$ absorption \citep{Boyden20}, implying that they are positioned near very dense molecular gas \citep[$\gtrsim 10^6$ cm$^{-3}$;][]{Tielens05}.

A deeper ALMA survey is needed to constrain the disk gas masses of additional cluster members and examine whether other cluster members are more or less photoevaporated than those found in our small sample, and/or whether any cluster members are still shielded from the external UV field. An increased number of lower-mass gas disks would, for example, point to more significant photoevaporation in the region than what is suggested from our sample.

\subsection{Disk kinematics and stellar masses}\label{sec:discussion:kinematics}

\begin{figure*}[ht!]
	\epsscale{1.15}
	\plotone{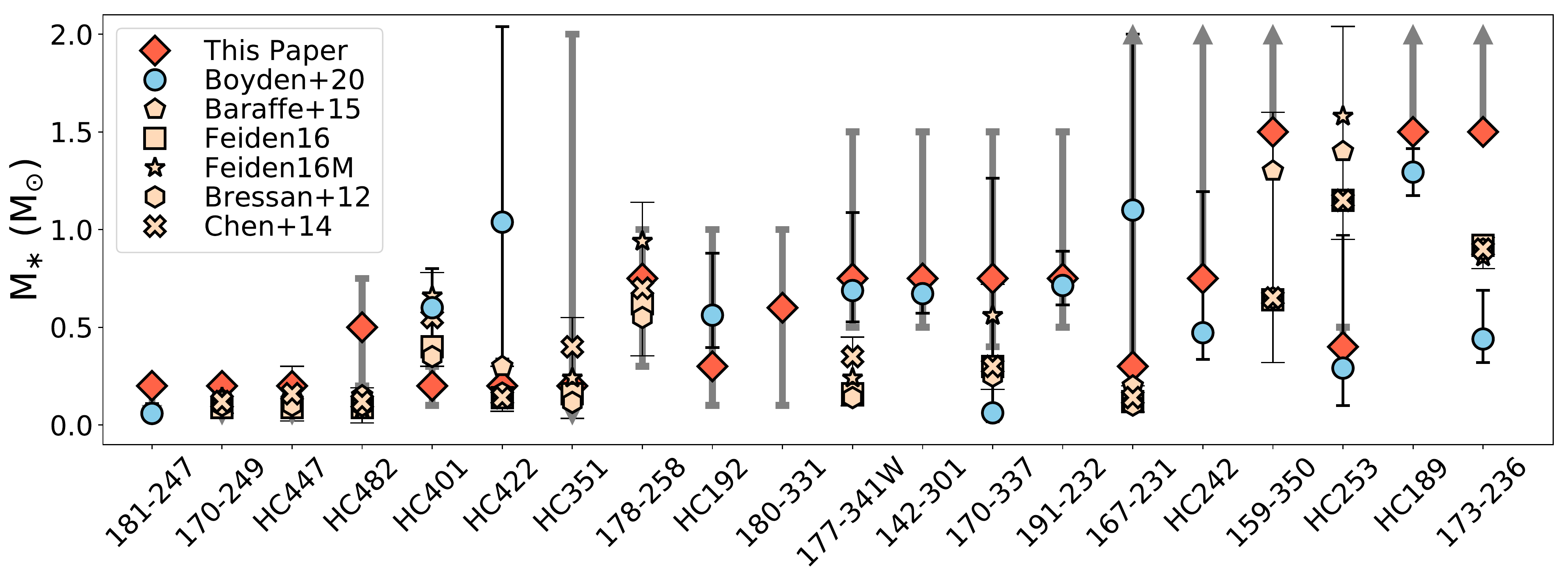}
	\caption{Comparison of the stellar masses derived from this paper (red diamonds), the stellar masses derived from \cite{Boyden20} (blue circles), and the stellar masses derived spectroscopically using a variety of pre-MS evolutionary tracks {(orange markers). The spectroscopic stellar masses are computed from the following sets of evolutionary tracks: the models of \citet[][Baraffe$+15$]{Baraffe15},  the magnetic and nonmagnetic tracks of \citet[][Feiden$+16$M and Feiden$+16$, respectively]{Feiden16}, the PARSEC models of \citet[][Bressan$+12$]{Bressan12}, and the updated PARSEC models of \citet[][Chen$+14$]{Chen14}. }The errorbars for the \cite{Boyden20} mass measurements have been recomputed in this figure (see text). The spectroscopic stellar masses are only shown for the sources with available measurements of bolometric luminosity and effective temperature. 
	\label{fig:mdyn_new} }
\end{figure*}

The best-fit channel maps produced from our modeling tend to resemble those generated by \cite{Boyden20}, who modeled the observed kinematics of our sample targets using a simpler approach assuming a geometrically thin Keplerian disk with a power-law emission profile. Figure \ref{fig:mdyn_new} compares the stellar masses derived in this paper to the stellar masses derived by \cite{Boyden20}. 
The uncertainties plotted for the \cite{Boyden20} measurements are different from the values reported in \cite{Boyden20}, who did not correct for the number of pixels within a synthesized beam area, $N_{pix}$, in their $\chi^2$ statistic, as we have done in our new study. Excluding a correction factor for $N_{pix}$ can result in narrower confidence intervals and smaller uncertainties on the derived best-fit model parameters. 
We re-examined the $\chi^2$ surfaces generated by \cite{Boyden20} and divided the $\chi^2$ values by $N_{pix}$ to obtain new $1\sigma$ confidence intervals on the best-fit stellar masses. We find that the new uncertainties are larger than the previously-derived values, although the degree varies on a source-by-source basis.

With these newly-obtained uncertainties, we see a broad agreement between the two dynamically-derived stellar mass measurements for the majority of cluster members, underscoring the similarity of the modeled kinematics obtained from the two approaches. Minor differences arise for some of the best-fit values, but the measurements are all within either $1\sigma$ or  $2\sigma$ of each other, with the exception of 173-236, in which case we find a $>3\sigma$ disagreement between the two dynamical mass measurements.

The strong discrepancy seen towards the dynamical mass measurements of 173-236 can be attributed to our new treatment of CO absorption, which has enabled us to more accurately model the observed kinematics of this target. As shown in Figures \ref{fig:data_mom0} and \ref{fig:data_mom1}, 173-236 has a well resolved, inclined dust disk, and strong CO absorption is seen only one side of the dust disk. This asymmetry can be naturally reproduced in our models when we account for the velocity-dependent intensity of the background cloud, as illustrated in Figure \ref{fig:modeling_results} (see also Appendix \ref{appendix:b}). Since \cite{Boyden20} did not consider the background cloud in their modeling, they were unable to reproduce the observed asymmetry with their model, and their kinematic fits were influenced by extended absorption not associated with the compact absorption seen towards the disk.

Moreover, while our derived stellar mass for HC242 is in agreement with the previously-derived value, we derive a significantly different systemic velocity due to our treatment of absorption, unlike what is seen with any other source in our sample. The ALMA CO $J=3-2$ observations towards HC242 show bright extended emission at velocity channels $\sim 4-6$ km s$^{-1}$ but compact absorption at velocity channels $\sim 6-10$ km s$^{-1}$ (see Figure \ref{fig:modeling_results} and Appendix \ref{appendix:b}). Because the CO $J=1-0$ background intensity towards HC242 is brighter at $>6$ km s$^{-1}$ than at $\leq 6$ km s$^{-1}$, our model pipeline can produce synthetic channel maps of HC242 that are seen in emission at $\leq 6$ km s$^{-1}$ and in absorption at $>6$ km s$^{-1}$. The synthetic channel maps can thus fit to the extended emission and compact absorption present in the ALMA observations, and fitting to both of these features renders the smallest $\chi^2$ values. This results in a best-fit model with a significantly lower systemic velocity than the velocity derived \cite{Boyden20}, whose fits were only sensitive to the compact absorption due to their assumed  intensity profile. 

The similarly of the kinematic fits obtained for the CO emission and HCO$^+$ emission detections, as illustrated in Figure \ref{fig:mdyn_new}, supports the assumption of \cite{Boyden20} that including radiative transfer in the model pipeline is not essential for constraining the kinematics of those sources \citep[see also][]{Rosenfeld12}. 
However, the different kinematic fits obtained for sources 173-236 and HC242 indicate that radiative transfer effects associated with the background cloud can influence the observed kinematics of the absorption detections, especially when the absorption is asymmetric. 
Future modeling efforts should therefore include a treatment of the background the cloud when modeling detections in absorption, 
even if the model is simpler than what we have adopted in this current study. 

Finally, we note that the inclination and position angle that we obtain for 170-337 are consistent with the orientation of the optical jet known to originate from this system \citep[e.g.,][]{Bally98}. As discussed in \cite{MendezDelgado22}, the kinematics of the jet imply a disk inclination of  $\sim 18^{\circ}$ and a position angle aligned east-west, both of which are not in agreement with the inclination and position angle reported in \cite{Boyden20}. With new our kinematic fit to 170-337, we find that inclinations $\sim 18^{\circ}$ and east-west-aligned position angles are within the $1\sigma$ confidence intervals of the best-fit values (e.g., see Table \ref{tab:best_fit_params}).

\subsubsection{Comparisons with spectroscopic stellar masses}

Since we have recomputed the uncertainties on the dynamical mass measurements reported in \cite{Boyden20}, and since we have also obtained new dynamical mass measurements in this paper, we find it useful to compare the dynamically-derived stellar mass measurements with the stellar masses derived spectroscopically from pre-MS evolutionary tracks, as was done previously in \cite{Boyden20}. In Figure \ref{fig:mdyn_new}, we include estimates of the spectroscopic stellar masses for the subset of cluster members with available spectroscopic measurements of bolometric luminosity and effective temperature. The spectroscopic data for 177-341W are taken from \cite{DaRio12}, and for all other cluster members we use the spectroscopic data from \cite{Fang21}. 
Following the same procedure outlined in \cite{Boyden20}, we compute the spectroscopic stellar masses using the following sets of evolutionary tracks: the models of \citet{Baraffe15}, the magnetic and nonmagnetic tracks of \citet{Feiden16}, the PARSEC models of \citet{Bressan12}, and the updated PARSEC models of \citet{Chen14}. All evolutionary tracks are compiled using the code {\tt pdspy} \citep{Sheehan18b}.  

With our new dynamical mass measurements, we find strong agreement between the dynamically- and spectroscopically-derived stellar masses. Of the 13 cluster members with available spectroscopic mass measurements, 9 have dynamical mass measurements that are within $1\sigma$ of the spectroscopically-derived values, and 12 have dynamical mass measurements that are within $2\sigma$ of the spectroscopically-derived values. These levels of agreement are consistent with the statistical deviations expected from a Gaussian noise distribution: for two independent sets of stellar mass measurements and a sample size of $13$, we would expect under a Gaussian noise distribution that ~$\sim 4$ of the $13$ measurements ($\sim 33\%$) differ by $>1\sigma$ and ~$\sim 1$ of the $13$ measurements ($\sim 5\%$) differ $>2\sigma$, which is exactly what we see in our comparisons. 
This also indicates that the dynamical mass measurements from \cite{Boyden20}, after correcting for the number of independent measurements in the uncertainty estimates (see Section \ref{sec:discussion:kinematics}), are consistent with the spectroscopic mass measurements. 

The cluster member in our sample that exhibits most discrepant dynamical and spectroscopic mass measurements is 173-236, and the level of discrepancy is $>3\sigma$. In comparison with the entire spectroscopic catalog of \cite{Fang21}, 173-236 happens to also be a statistical outlier on the HR diagram. Namely, it is part of a small group of cluster members that appear subluminous compared with other cluster members with the same spectral type \citep[see Figure 9 of][]{Fang21}. This subpopulation make up less than $\sim 1\%$ of the spectroposcopically-characterized ONC members, and comparisons with pre-MS evolutionary tracks yield typical ages of $>10$ Myr for those cluster members, i.e., ages that are substantially larger than the expected age of the ONC \citep[$\sim 1$ Myr; see][]{Fang21}. 173-236 and other subluminous cluster members could be relics of an extended period of star formation in the ONC (see Section \ref{sec:discussion:photoevap}).  However, the fact that 173-236 is a statistical outlier in both our sample and the sample from \cite{Fang21} suggests that the mass measurements for 173-236 need to be re-examined.

\cite{Fang21} argue that the bolometric luminosities of the subluminous cluster members are likely to have been underestimated due to edge-on circumstellar disks. This scenario seems likely to explain why 173-236, which possesses an edge-on disk (see Figures \ref{fig:data_mom0} and \ref{fig:data_mom1}), is an outlier. Based on the current location of 173-236 on the HR-diagram \citep[see Figure 17 of ][]{Boyden20}, an increased luminosity would result in not only larger spectroscopically-derived stellar masses that more closely resemble our dynamically-derived value, but also younger stellar ages that are more similar to the canoncial $\sim 1$ Myr age of the ONC. 

\subsubsection{Importance of dynamical mass measurements in the ONC}

Our sample of dynamical mass measurements demonstrates that the ONC is an ideal region for measuring the dynamical masses of low-mass ($\leq 0.5 M_{\odot}$) M-dwarfs, which are not only the most common types of stars in the Galaxy \citep[e.g., ][]{Reid97, Bochanski10}, but also common hosts of planetary systems \cite[e.g.,][]{Dressing15, Mulders15}. In lower-density star-forming regions, disk-based dynamical mass measurements have been challenging to obtain for M-dwarfs due to their typically smaller disk sizes and fainter line emission, and only a small number of high-precision dynamical mass measurements have been obtained for these objects \citep[e.g.,][]{Sheehan19, Simon19, Pegues21}. The ONC, on the other hand, appears to contain many low-mass M-dwarfs whose disks are either 1) bright in CO and/or HCO$^+$, a trend we attribute to the irradiation properties of the cluster (see Section \ref{sec:discussion:GDR}); or 2) positioned in localized regions of the cluster that enable them to be seen in absorption. 

The distinct disk properties exhibited by the M-dwarf cluster members in our sample illustrate how the rich cluster environment can alleviate some of the main challenges associated with detecting the disks surrounding M-dwarfs, even though nearby clusters like the ONC are further away than Taurus, Lupus, and other low-density regions. With currently-available ALMA observations of the ONC, we have already more than doubled the sample size of available dynamical mass measurements in the mass range $\leq 0.2 M_{\odot}$, and we have increased the number of dynamical mass measurements in the mass range $0.2 - 0.5 M_{\odot}$ by $\sim 30\%$. With deeper observations of the ONC, we can expect to obtain additional dynamical mass measurements for other low-mass M-dwarfs, and we can also improve the precision on our current dynamical mass measurements, which is essential for calibrating models of pre-MS stellar evolution in the low-mass regime \cite[e.g.,][]{Pegues21}.

Dynamical mass measurements also offer an alternative way to constrain the central masses of cluster members for which it is challenging to obtain spectroscopic measurements of bolometric luminosity and effective temperature due to, for example, ionized proplyds that can obscure emission from the central star \cite[e.g.,][]{Mann14}, massive edge-on circumstellar disks that can influence the positions of cluster members on the HR diagram \cite[][see above]{Fang21}, or localized enhancements of the stellar density, nebulosity, and extinction \citep[e.g., ][]{Hillenbrand97, Hillenbrand00, Scandariato11, DaRio12}. An alternative method is especially needed in the ONC, as $\sim 30\%$ of the dust disks detected by \cite{Eisner18} lack the spectroscopic data needed to derive their stellar masses.

Of the 7 targets in our sample that lack spectroscopic estimates of stellar mass, 3 have bright proplyds  (181-247, 180-331, and 142-301), 5 are located in the more extincted regions of the ONC (HC192, 142-301, 191-232, HC242, and HC189), and 3 have large dust masses and dynamical-derived stellar masses (173-236, HC242, and HC189, see Table \ref{tab:best_fit_params}). The latter trend raises the possibility that the weak dependence of disk mass/size as a function of stellar mass\textemdash a behavior found by \cite{Eisner18} that differs strongly from the correlations seen in lower-density regions \citep[e.g.,][]{Pascucci16} and from the behavior expected from external photoevoporation \citep[e.g.,][]{Haworth18, Winter19b}\textemdash is contaminated by missing spectroscopic stellar mass measurements at the upper end of the stellar mass distribution. Settling this, however, requires an increased sample of dynamical mass measurement towards the remaining cluster members with unavailable spectroscopic data.

\subsection{Properties of non-detected gas disks}\label{sec:discussion:stat}

\begin{figure*}[ht!]
	\epsscale{1.1}
	\plotone{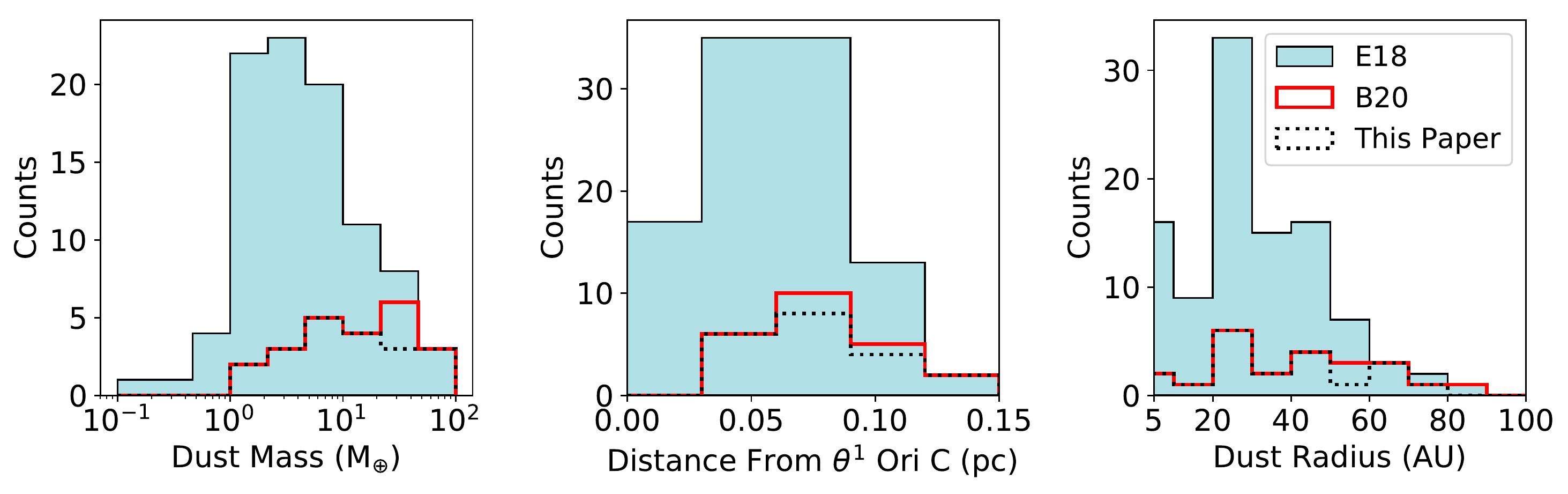}
	\caption{Distributions of dust-disk mass (left), projected distance from $\theta^1$ Ori C (center), and dust radius (right) for the  ALMA-detected ONC members. The blue histograms are generated from all the disks detected at $350$ GHz ($0.87$ mm) by \cite{Eisner18}. The red histograms include only the subset of dust disks that are also detected in gas by \cite{Boyden20}, and the dotted black histograms are produced from the gas disks selected in this current study. Here, the dust radius is measured as the half-width at tenth-maximum (HWTM) major axis of a Gaussian fit to the submillimeter image of a source \citep[see][]{Boyden20}.
	\label{fig:histograms_samples}}
\end{figure*}

\begin{figure}[ht!]
	\epsscale{1.1}
	\plotone{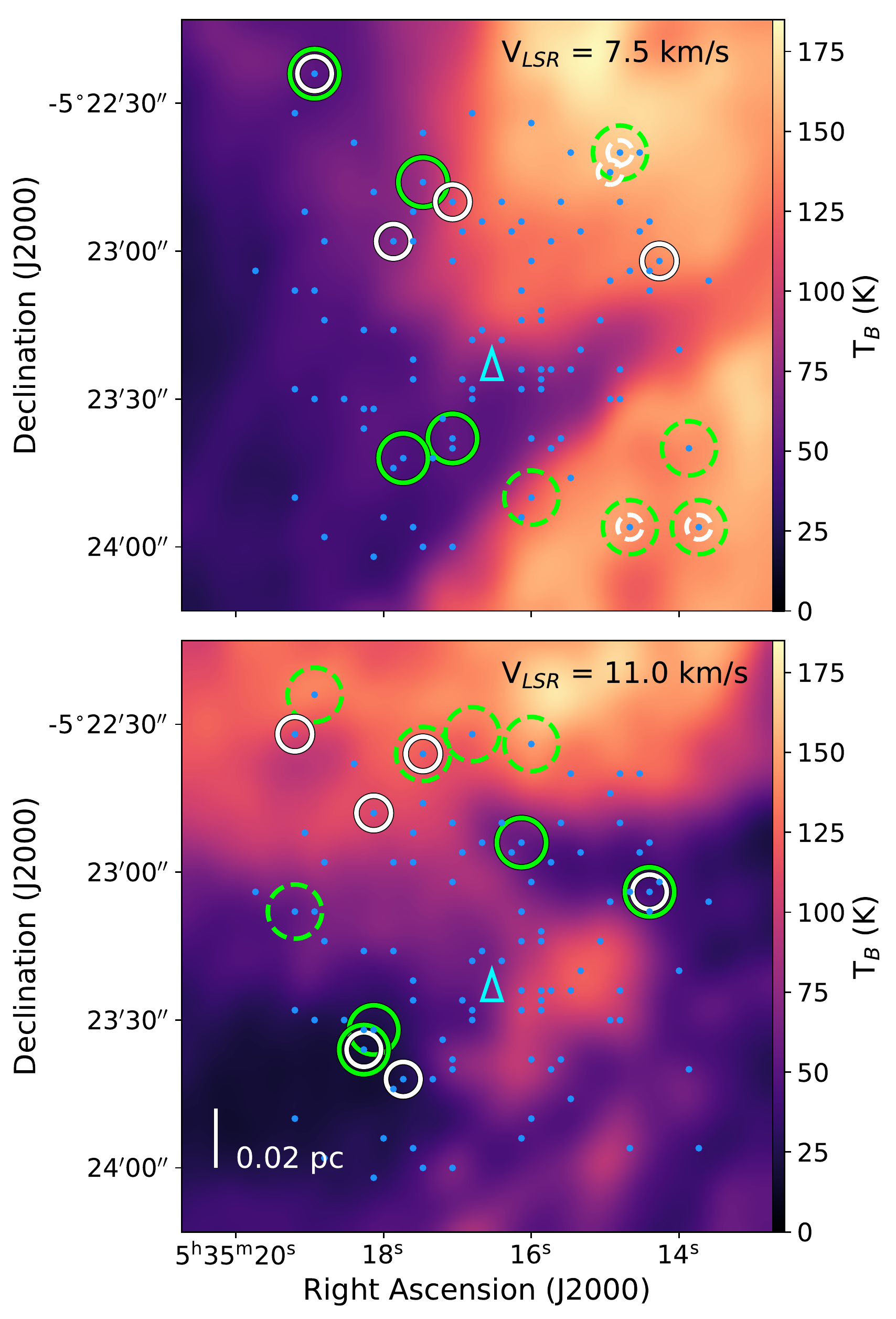}
	\caption{Large-scale CO $J=1-0$ emission seen towards the central $\sim$ $2'$ $\times$ $2'$ region of the ONC at $\sim 7.5$ km s$^{-1}$ (top) and $\sim 11$ km s$^{-1}$ (bottom). The data are taken from the CARMA-NRO Orion survey \citep{Kong21} and have an angular resolution of $8''$. Filled small circles indicate the position of circumstellar disks detected at $350$ GHz ($0.87$ mm) by \cite{Eisner18}. Larger, open circles indicate the positions of disks also detected in CO $J=3-2$ (green) and/or HCO$^+$ $J=4-3$ (white) by \cite{Boyden20}, with solid open circles corresponding to detections in emission, and dashed open circles denoting detections in absorption against the warm background. Gas detections with peak signals near $\sim 7.5$ km s$^{-1}$ are plotted in the top panel, while gas detections with peak signals closer to $\sim 11$ km s$^{-1}$ are plotted in the bottom panel. The cyan triangle indicates the position of the massive star $\theta^1$ Ori C.
	\label{fig:CARMA_ONC_main}}
\end{figure}

The targets that we have modeled are selected from the subset of circumstellar disks that were detected in CO and/or HCO$^+$ by \cite{Boyden20}. The majority of ALMA-detected dust disks in the ONC, however, were undetected in gas. In Figure \ref{fig:histograms_samples}, we plot histograms of the dust masses, projected distances from $\theta^1$ Ori C, and dust radii for the entire sample of ALMA-detected dust disks, and we compare these with the distributions generated from the subset of dust disks that are also detected in gas. These histograms show that the non-detected gas-disks tend to have lower dust masses than the detected gas disks, with $\sim80\%$ of the gas non-detections falling in the mass range $\sim 1 - 10$  $M_{\oplus}$. Furthermore, no gas disks are detected in the most central regions of the ONC, i.e.,  within  $\lesssim 0.03$ pc of $\theta^1$ Ori C.

To quantify the statistical significance of the differences illustrated in Figure \ref{fig:histograms_samples}, we apply a two-sided Kolmogorov-Smirnov test to each histogram pair and derive p-values of 0.003, 0.05, and 0.2 for the distributions of dust mass, projected distance from $\theta^1$ Ori C, and dust radius, respectively.  These p-values confirm that our sample of detected gas disks is biased towards higher dust masses, somewhat biased towards larger projected distances, and not biased towards any particular dust-disk size. In the following paragraphs, we discuss the possible theoretical mechanisms responsible for reducing the CO/HCO+ detection rate at the lower end of the dust mass and projected  distance distributions, and we examine how the conclusions drawn from our modeling of the detected gas disks, in particular the ones about gas-to-dust ratios, may be affected by any biases in our sample selection. 

In dense, clustered environments like the ONC, we might expect a gas non-detection to arise entirely from more significant external photoevaporation. When disks are exposed to the extreme levels of UV irradiation found in the ONC, they eventually truncate to very compact sizes \citep[$\lesssim 15$ AU; e.g., ][]{Adams04} where optically thick molecular line emission would fall well below the sensitivity limit of our observations regardless of the excitation conditions. Disk truncation due to external photoevaporation seems to explain why no gas disks are detected within $\sim 0.03$ pc of $\theta^1$ Ori C  (see Figure \ref{fig:histograms_samples}), as these innermost regions of the ONC are where the external UV field is the strongest and where the disk lifetime is expected to be the shortest \citep[e.g.,][]{Johnstone98, Haworth18}. It also reinforces the notion that an age spread exists amongst the individual cluster members of the ONC (see Section \ref{sec:discussion:photoevap}). Namely, if the majority of gas non-detections are older than the detected gas disks, then they should also be more strongly evaporated, less massive, smaller, and thus, harder to detect in CO and/or HCO$^+$ than the younger gas disks in our sample. These expected properties for an older, more evaporated disk population match up well with the reduced gas detection rate observed towards lower disk masses (see Figure \ref{fig:histograms_samples}), 
{although some of these non-detections may simply be disks that initially formed with compact sizes \citep[e.g., ][]{Kuffmeier20}.}

Since the UV and X-ray radiation fields are known to vary spatially, temporally, and on a source-by-source basis in clustered regions, it is also possible that chemical variations across the sample account for a fraction of gas non-detections. Our thermochemical modeling indicates that particular balances of UV and X-ray irradiation are needed to reproduce the bright HCO$^+$ emission of the detected gas disks. If the non-detected cluster members have low X-ray luminosities relative to the local UV field strength, due to intrinsically fainter X-ray emission or to a lack of flaring activity \citep[e.g.,][]{Waggoner22}, then their HCO$^+$ emission may simply fall below the detection threshold of our observations. Moreover, in the limit of weak UV fields and strong X-ray fields, X-rays dominate the photochemistry and reprocess CO into neutral carbon \citep[see also][]{Schwarz18}, causing the models become depleted in gas-phase CO and produce faint CO and HCO$^+$ emission (see Figure \ref{fig:checker_GO_LXR}). X-ray-active cluster members that are also shielded from strong external (or internal) UV fields may therefore be unaccounted for in our sample, more prone to CO depletion, and thus, chemically different from our sample targets. We do not expect this scenario to apply to all gas non-detections, though, as many of them have optically-identified proplyds and are therefore not shielded from strong UV fields \cite[e.g.,][]{Odell94, Bally98, Ricci08}.

Even if a cluster member is not shielded from the external UV field, closer proximity to the background cloud can result in a gas non-detection.  As shown in Equation \ref{eq:simobs_main}, when the disk emission and spatially-filtered background cloud emission are of similar magnitudes, they will effectively cancel each other out in the interferometric observations, and produce faint compact emission/absorption towards the disk. We attribute this behavior as a contributing reason to why cluster members 170-249, 178-258, 181-247, 191-232, and 142-301 are all detected in HCO$^+$ but undetected in CO. Those cluster members are all located in a similar region of the ONC, as illustrated in Figure \ref{fig:CARMA_ONC_main}, and they are all detected at velocities where the CARMA-NRO Orion Survey indicates strong CO cloud emission. The compact CO emission/absorption towards those sources is likely to have been obscured by spatially-filtered cloud emission, considering that we find no regions of parameter space where the model disk alone can produce undetectable CO emission and detectable HCO$^+$ emission (e.g., see Figures \ref{fig:pstudy} - \ref{fig:pstudy_spec2}). Other cluster members may be undetected in CO for similar reasons, although this scenario requires that the disk and cloud brightness temperatures cancel out perfectly over multiple velocity channels, which seems unlikely to happen over large samples.

An alternative and perhaps more tantalizing explanation for why we detect some disks in gas but not in others is that gas non-detections in the ONC are a result of ongoing planet formation.  As disks evolve under the influence of external photoevaporation, the gas and dust undergo distinct evolutionary paths that can result in a wide range of possible gas-to-dust ratios. The gas content is primarily removed from the disk via the thermally-induced photoevaporative wind \citep[e.g.,][]{Johnstone98, Storzer99}, and while some dust is initially lost to the wind, the dust ultimately grows to mm-sizes and decouples from the gas, such that it is no longer entrained in the wind and is instead subject to inward radial drift \cite[e.g.,][]{Facchini16}.  Recent simulations by \cite{Sellek20} have shown that when the effects of radial drift are accounted for in simulations of external photoevaporation, the overall gas-to-dust ratio tends to increase or remain near $\sim 100$ as the gas photoevaporates outward and the mm-grains migrate radially inward. Thus, subinterstellar gas-to-dust ratios in externally-irradiated disks should only be expected if the inward migration of the dust has been halted by other mechanisms, such as pressure traps \citep[e.g.,][]{Pinilla12} or the piling up of grains at the water snowline \citep[e.g.,][]{Draz16, Schoonenberg17}, both of which are commonly associated with planet formation.

The thermochemical models that we have produced in this study show that over the expected range of dust masses for the ALMA-detected ONC dust disks, gas-to-dust ratios lower than $100$ can result in faint CO and HCO$^+$ emission that is below the sensitivity limit of our observations (see Figures \ref{fig:checker_Mgas} and \ref{fig:Mgas_dust_ratio} for examples). The current sample of ALMA-detected gas-disks may therefore be drawn from the population of cluster members that have so far evolved in the absence of pressure traps or other drift-suppressing mechanisms, allowing them to retain gas-to-dust ratios $\geq 100$ despite the overall loss of gas and dust due to external photoevaporation and radial drift.  The non-detected gas disks, on the other hand, might have lower gas-to-dust ratios as a result of the physical mechanisms that suppress radial drift and induce planet formation, causing them to emit fainter CO/HCO$^+$ emission that we are currently unable to detect with available ALMA observations. While the pressure trap scenario is more likely to apply to gas non-detections towards the upper end of the disk mass and size distributions, given that disks with strong pressure traps tend to retain large dust masses and sizes \citep[e.g.,][]{Pinilla20, vdMarel21}, the grain piling scenario is still expected to operate in lower-dust-mass disks \citep[e.g.,][]{Ormel17}, including the ones that are exposed to strong UV fields \citep[e.g.,][]{ Haworth18b}.

{We suggest that the majority of dust disks in the ONC are undetected in CO and/or HCO$^+$ due to a combination of smaller disk sizes and lower gas-to-dust ratios resulting from compact disk formation, external photoevaporation, 
and/or ongoing planet formation.} Most non-detections should therefore be as warm as the detected gas disks, and lack the cool gas and dust that can facilitate gas-phase CO depletion and lead to a reduction of the CO/HCO$^+$ intensity  \citep[see also][]{Walsh13, Tsamis13, Champion17, Haworth21a}. However, if the majority of non-detectections have gas-to-dust ratios $<100$, then the ISM-like gas ratios derived for the disks in our sample would not be representative of the typical gas-to-dust ratios found in the region. Deeper, higher-resolution ALMA observations are needed to constrain the disk sizes and gas-to-dust ratios towards additional cluster members and test whether the average disk in the ONC is gas-rich or gas-poor.

\subsection{Importance of mapping the background cloud}\label{sec:discussion:future}

The prevalence of ONC gas disks seen in absorption with ALMA \citep[e.g.,][]{Bally15, Boyden20} indicates that absorption detections should be common in surveys of young stellar clusters, unlike what is expected from surveys targeting lower-density regions \citep[e.g., ][]{Ansdell16, Ansdell18}. It is therefore essential that future surveys in clustered environments are not only optimized for detecting gas disks both in absorption and in emission, but also equipped with the supplemental information necessary for interpreting the observations and facilitating accurate comparisons with physical-chemical models. 

Our modeling of CO absorption demonstrates that the background molecular cloud plays a central role in setting the depth of the observed absorption, even when extended ALMA configurations filter out much of the large-scale cloud emission. The CO absorption detections that we have examined in this paper are all positioned in front of dense molecular gas, allowing us to reasonably extrapolate the CO $J=3-2$ cloud intensity from publicly-available observations targeting the $J=1-0$ transition (see Appendix \ref{appendix:a}). In other clusters, the cloud densities may be lower such that the CO $J=3-2$ and CO $J=1-0$ background intensities are more likely to differ significantly. Furthermore, no information on extended HCO$^+$ emission in the ONC is currently available, hindering our ability to model sources that are detected in HCO$^+$ absorption \citep[e.g., HC192, HC189, HC771; see][]{Boyden20}. In future surveys, we plan to include observations that map extended cloud structures over the same molecular transition(s) that are used to probe the compact gas disks, unless prior knowledge of the cloud intensity at those transitions is readily available \citep[e.g., ][]{Hacar18, Kong18, Kong21}. 

Observations targeting large-scale emission can also shed light on whether any disk-bearing cluster members are actively accreting material from their parent cloud reservoirs. Recent ALMA observations have revealed the presence of large-scale spirals, remnant envelopes, and/or filaments towards multiple evolved class II disks in lower-density star-forming regions \citep[e.g., ][]{Alves20, Huang20, Huang21, Ginski21}, hinting that the parent cloud can still regulate the transport and distribution of material in disks at late evolutionary stages. With the limited spatial resolution of our current ALMA observations \citep[which are sensitive to compact, $<800$ AU scales;][]{Eisner18, Boyden20}, we are unable to detect large-scale spiral, envelope, and/or filamentary structures. With deeper observations probing larger ($>800$ AU) spatial scales, we might expect to detect these structures around a subset of our targets.

\section{\bf Conclusions}\label{sec:conclusions}

We employed a detailed thermochemical disk framework to analyze $350$ GHz continuum, CO $J=3-2$, and HCO$^+$ $J=4-3$ ALMA observations of $20$ gas disks in the ONC that were previously detected by \cite{Boyden20}. We constructed an azimuthally symmetric model of a hydrostatically-supported, viscously-evolving disk, and used the code {\tt RAC2D} to simulate the physical-chemical evolution of the gas and dust and generate synthetic observations of $350$ GHz continuum, CO $J=3-2$, and HCO$^+$ $J=4-3$ emission. We produced a large grid of models as a function of disk, stellar, and environmental parameters, and then developed a procedure for fitting these models to the measured $350$ GHz dust continuum fluxes and CO $J=3-2$ and/or  HCO$^+$ $J=4-3$ channel map observations. For the cluster members detected in absorption, we included a proper radiative transfer treatment for converting the synthetic observations into simulated interferometric observations of a disk positioned in front of a bright background cloud.

From our modeling we are able to constrain the disk masses, disk sizes, UV and X-ray radiation fields, viewing geometries, and central stellar masses of all ONC sources in our sample. We find that the ONC sources all have compact gas disks ($<100$ AU in radius), relatively large gas masses ($\gtrsim 10^{-3} - 10^{-2} M_{\odot}$), and gas-to-dust ratios $\sim 100-1000$. Our CO- and HCO$^+$-based gas-to-dust ratios are consistent with expected ISM values but also systematically larger than the values derived for disks in lower-density star-forming regions from CO and/or CO isotopologue observations, signifying strong differences in the underlying disk composition. We suggest that disks in the ONC are warmer and less prone to gas-phase CO depletion than the massive, extended gas disks that are commonly detected in lower-density regions due to a combination of compact disk sizes and the irradiation properties of the ONC.

The gas masses that we derive for many of the cluster members are larger than the typical values predicted from external photoevaporation models after 1 Myr of evolution under the extreme UV conditions found in the ONC. This discrepancy can be reconciled if the disks in our sample are younger than the canonical 1 Myr age of the ONC and/or younger than  other cluster members. It can also be reconciled if external photoevaporation has more recently begun to operate in the ONC due to strong intracluster extinction, a behavior recently identified in simulations of young stellar clusters.

Our dynamically-derived measurements of stellar mass are broadly consistent with the dynamical masses derived by \cite{Boyden20}. However, we show that the uncertainties on the \cite{Boyden20}  measurements are likely to have been systematically underestimated. After recomputing the uncertainties on the \cite{Boyden20} dynamical mass measurements, we find that the dynamical masses derived in this study and in \cite{Boyden20} are all in excellent agreement with the stellar masses derived spectropically from pre-MS evolutionary tracks. Our study has significantly increased the number of disk-based dynamical mass measurements in the mass range $\leq 0.5 M_{\odot}$. As such, we suggest that the ONC is an ideal region for obtaining large samples of dynamical mass measurements towards young M-dwarf systems.

The thermochemical models that we have presented in this study demonstrate that disks in the ONC are likely to have higher gas-to-dust ratios and warmer gas and dust temperatures than disks in lower-density regions. However, deeper observations are needed to detect additional gas disks in the region, confirm similar physical-chemical properties over larger samples, and constrain disk, stellar, and environmental parameters at a higher precision than what we have obtained in this study. HCO$^+$ in particular appears well suited for deeper surveys of the ONC and other clusters, given that it is bright in clustered regions, less likely to be influenced by the background cloud than CO, and provides constraints on the disk mass along with other physical/chemical properties. Since ALMA is readily capable of mapping nearby clusters at the sensitivity and resolution needed to detect and resolve gaseous circumstellar disks, it is only a matter of time until larger samples of gas disks in these regions are obtained.

\

{\it \noindent  Acknowledgements:} We are grateful to K. Schwarz, who provided useful ideas for some of the analysis presented in this work. We also thank F. Du for proving assistance on using {\tt RAC2D}, and D. Marrone for useful discussions on interferometric observations of CO absorption. {Finally, we thank the anonymous referee for their detailed comments, which helped improve the clarity and quality of this manuscript.} 
This work was supported by NSF AAG grant 1811290. R. Boyden also acknowledges support from the University of Arizona's College of Science Fellowship. This paper makes use of the following ALMA data: ADS/JAO.ALMA \#2015.1.00534.S. ALMA is a partnership of ESO (representing its member states), NSF (USA) and NINS (Japan), together with NRC (Canada), MOST and ASIAA (Taiwan), and KASI (Republic of Korea), in cooperation with the Republic of Chile. The Joint ALMA Observatory is operated by ESO, AUI/NRAO and NAOJ. The National Radio Astronomy Observatory is a facility of the National Science Foundation operated under cooperative agreement by Associated Universities, Inc. 
{This material is based upon work supported by the National Aeronautics and Space Administration under Agreement No. 80NSSC21K0593 for the program ``Alien Earths.'' The results reported herein benefitted from collaborations and/or information exchange within NASA's Nexus for Exoplanet System Science (NExSS) research coordination network sponsored by NASA's Science Mission Directorate.}

\

{\it Facility: ALMA}

{\it Software:} {\tt RAC2D} \citep{Du14}, {\tt Astropy} \citep{astropy13}, matplotlib \citep{Hunter07},   {\tt pdspy} \citep{Sheehan18b}, {\tt CASA} \citep{McMullin07}, {\tt RADEX} \citep{vdTak07}



\appendix


\twocolumngrid

\section{Treatment of Background Cloud}\label{appendix:a}

\twocolumngrid

As discussed in Section \ref{sec:data}, a subset of our sample targets are detected in CO absorption against the warm background, meaning that the ALMA-detected emission is negative (see Table \ref{tab:detection_info}). In order to compare the observed negative emission to the positive emission produced from our thermochemical models, we must first understand how the negative emission observed with ALMA relates to the emission from the disk. We constructed a toy model intensity profile of a compact disk positioned in front of a bright molecular cloud, following the equation of radiative transfer:
\begin{equation}
	I_{model} = I_{d} + I_{c} {\it e}^{-\tau_{d}},
\end{equation} 
where $I_{d}$ denotes the intensity profile of the compact disk, $I_{c}$ denotes the intensity profile of the extended molecular cloud, and $\tau_{d}$ denotes the optical depth of the disk. We define the intensity profile of the disk as a Gaussian with a full-width-at-half-maximum (FWHM) of $0\rlap{.}''1$, and the cloud as an elliptical Gaussian with major and minor axes of $6''$ and  $3''$, respectively. Here we choose a cloud size that is larger than the maximum recoverable scale (MRS) of our ALMA observations ($\sim 1''$), as cloud structures exceeding the MRS are large enough to be resolved out in interferometric observations. Finally, we assume that the compact disk is significantly optically thick ($\tau_{d} >> 1$), as is often the case with CO and HCO$^+$ line emission.

\begin{figure}[ht!]
	\epsscale{1.1}
	\plotone{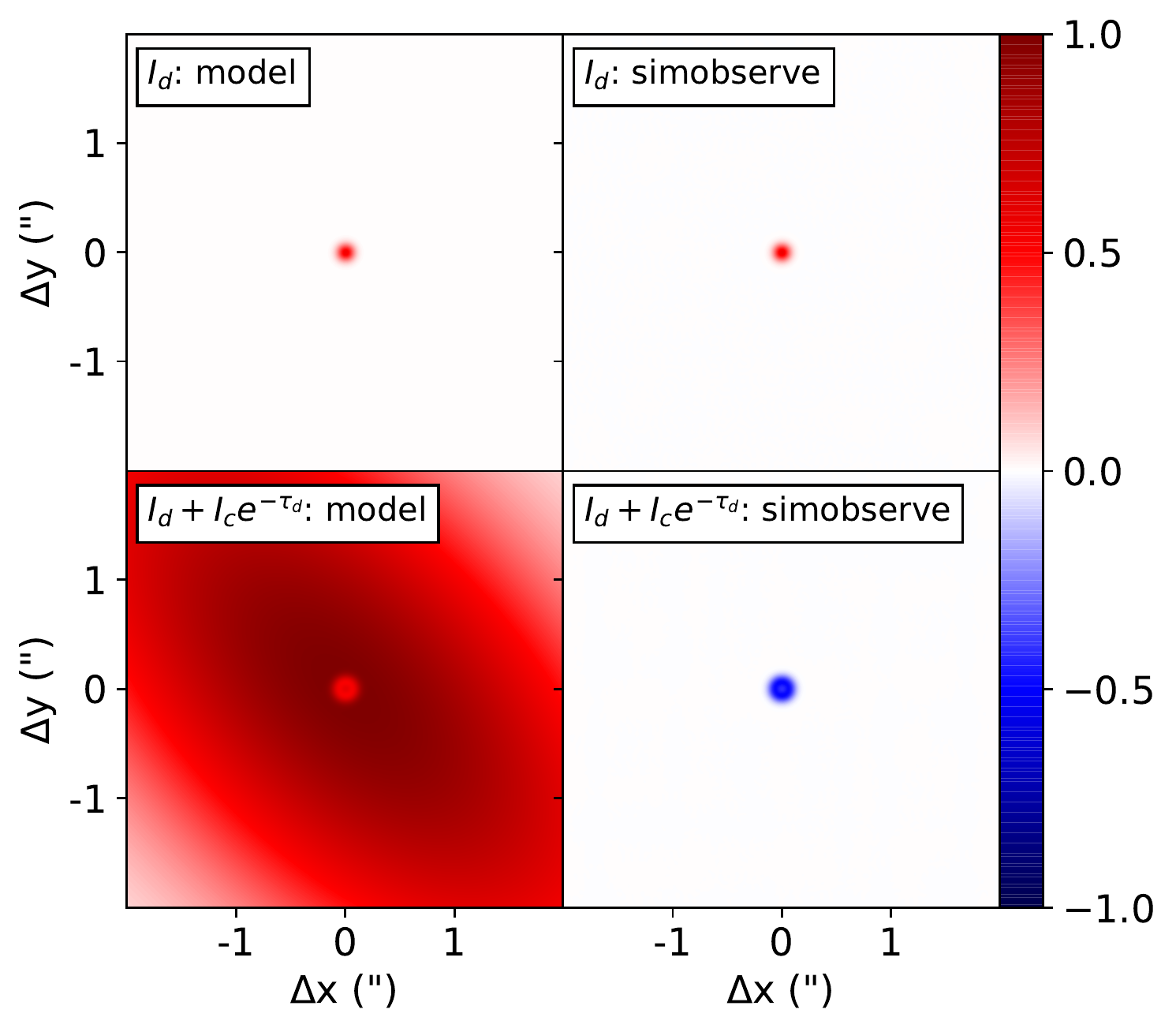}
	\caption{Normalized simulated ALMA observations of a compact, optically thick disk positioned in front of a bright background cloud. We model the disk as a compact Gaussian with a full-width-at-half-maximum of $0\rlap{.}''1$, and the background cloud as an extended Gaussian with major and minor axes of $6''$ and  $3''$, respectively. The left column shows the model emission profiles of just the disk (top) as well as the combined profile of the disk and extended background cloud (bottom). The right column shows synthetic interferometric observations of the emission profiles, generated using CASA's {\tt simobserve} task.
	\label{fig:simobserve}}
\end{figure}

Figure \ref{fig:simobserve} shows synthetic ALMA observations of our toy model, {generated using the \tt simobserve} task from {\tt CASA} version 5.0.0 {\citep{McMullin07}.} We design the synthetic observations to be representative of our ALMA mosaic, adopting an observing frequency of $345$ GHz (i.e., the approximate frequency of the CO $J=3-2$ transition) and employing an extended $12$m array configuration that yields an angular resolution of $\sim 0\rlap{.}''08$ and an MRS of $\sim 1''$. With this setup, the background cloud is mostly resolved out in the synthetic observations. However, along the positions where the cloud emission is attenuated by the compact disk, the synthetic emission of the cloud is negative. The total observed emission towards the disk is equivalent to the sum of the compact disk emission and the negative emission from the disk-attenuated background cloud. If the disk is bright enough to overcome the negative deficit, then we recover positive emission in the synthetic observations. Otherwise, the observed emission is negative.

\begin{figure}[ht!]
	\epsscale{1.0}
	\plotone{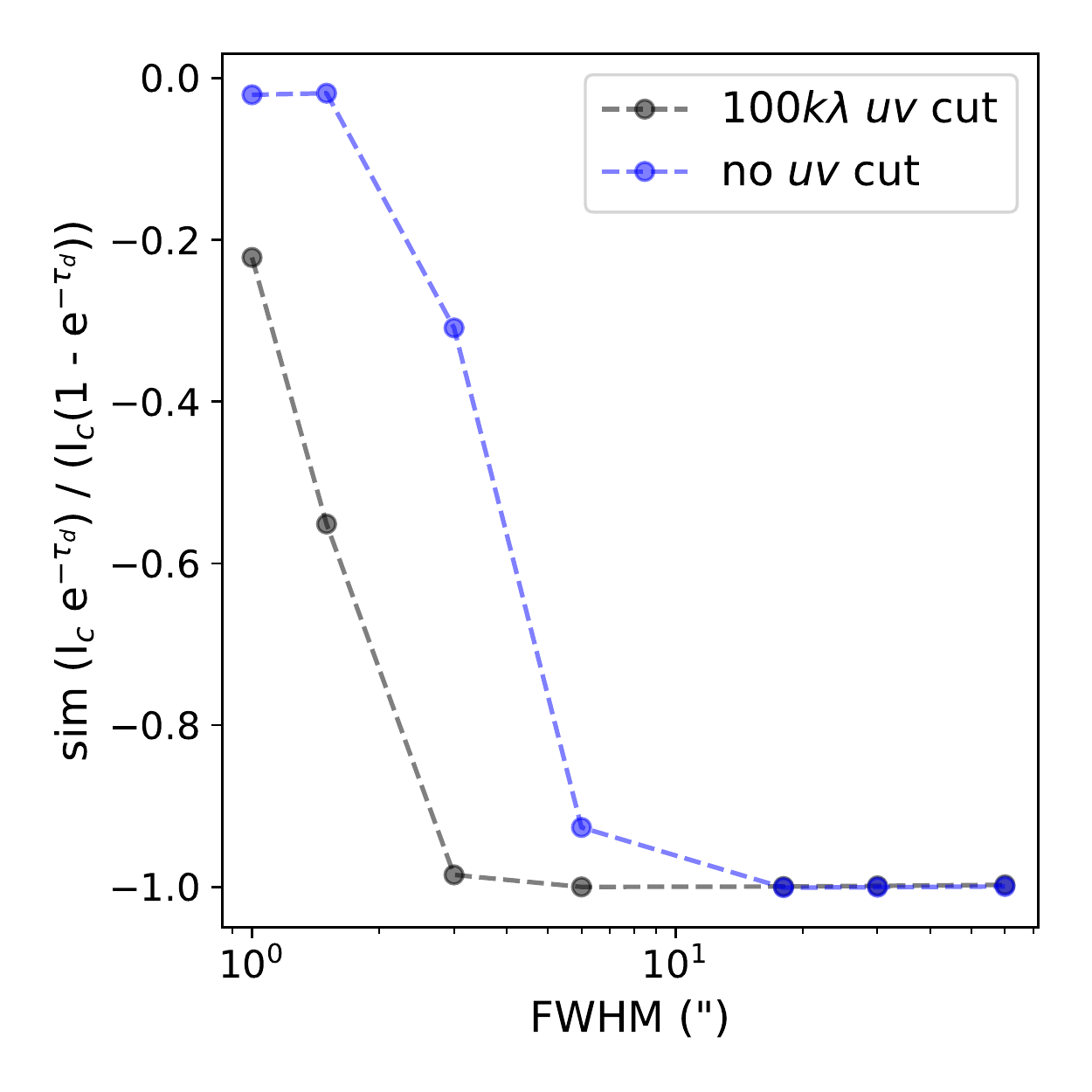}
	\caption{Simulated central peak intensity 
	of the background cloud component of our toy model ($I_{c} {\it e}^{-\tau_{d}}$), 
	plotted as a function of the cloud size. 
	We model the background as a Gaussian and define the background cloud size as the Gaussian full-width-at-half-maximum (FWHM). 
	Here the simulated peak intensities are normalized by the difference between the unattenuated and disk-attenuated background cloud intensities: $I_{c} (1 - {\it e}^{-\tau_{d}})$. The dotted gray line shows the simulated intensities for images generated with a $100$k$\lambda$ {\it uv} cut, and the dotted blue line shows the intensities when all baselines are included in the simulated images. 
	\label{fig:simobserve_peaks}}
\end{figure}

In Figure \ref{fig:simobserve_peaks}, we plot the observed relationship between the size of the background cloud and the peak intensity of the simulated background component ($I_{c} {\it e}^{-\tau_{d}}$). For backgrounds with FWHMs $\gtrsim 6''$, the synthetic observations resolve out the large-scale structure, and the magnitude of the compact negative deficit closely matches the difference between the unattenuated and disk-attenuated background cloud: $I_{c} (1 - {\it e}^{-\tau_{d}})$. As the  size of the background decreases, more structure is retained in the interferometric observations and the negative deficit decreases in magnitude. When the majority of the cloud structure is retained in the observations, the synthetic emission becomes positive and converges to the model value: $I_{c} {\it e}^{-\tau_{d}}$. 

For our modeling of the sources detected in CO absorption, we assume that the background cloud is large enough such that the extended structure is completely resolved out in the ALMA observations. Thus, the observed emission towards each absorption detection is assumed to be 
\begin{equation}\label{eq:simobs}
	I_{obs} = I_{d} - I_{c}(1 - {\it e}^{-\tau_{d}}),
\end{equation} 
where the second term represents the negative contribution due to the resolving out of the background cloud. Our assumptions are motivated by the CARMA-NRO Orion survey \citep{Kong18, Kong21}, which mapped the large-scale CO $J=1-0$ emission in the ONC down to a spatial resolution of $8''$ and spectral resolution of $0.25$ km s$^{-1}$ (shown in Figure \ref{fig:CARMA_ONC_main}), and revealed that the dominant large-scale structures in the central ONC are on spatial scales of $\sim 10-100''$. Our toy modeling demonstrates that molecular cloud structures $\gtrsim 10''$ should be resolved out in our ALMA observations, particularly when we employ a $100$k$\lambda$ {\it uv} cut during the imaging procedure (see Figure \ref{fig:simobserve_peaks}).

Additionally, we assume that the CO $J=3-2$ cloud intensity towards each absorption detection is equivalent to the localized CO $J=3-2$ emission measured from the CARMA-NRO Orion survey. Although the CARMA-NRO Orion survey targets a different transition of CO than our ALMA mosaic, we have found that the large-scale CO $J=1-0$ emission reasonably estimates the CO $J=3-2$ cloud intensity. 
The absorption detections are typically located near the OMC-1 and BN/KL regions, where the cloud is expected to be dense ($n \sim 10^6$ cm$^{-3}$), CO-rich ($N_{^{12}CO} \sim 10^{19}$ cm$^{-2}$),  and warm ($T \sim 100-200$ K), as demonstrated in previous radiative transfer modeling of these regions \citep[e.g., ][]{Peng12}. Using the non-LTE radiative transfer program {\tt RADEX} \citep{vdTak07}, we find that under the expected cloud conditions of the OMC-1 and BN/KL regions, the CO $J=1-0$ and CO $J=3-2$ intensities are within a factor of $< 2$. In particular, when we vary the cloud parameters in order to match the bright CO $J=1-0$ intensities observed towards our absorption detections ($T_B \sim 150$ K), we find that the predicted CO $J=1-0$ and CO $J=3-2$ intensities are nearly equal.

For the sources detected in CO and/or HCO$^+$ emission, we assume that $I_c <<< I_d$, and therefore neglect the effects of the background cloud when modeling their ALMA-observed emission. The CO emission detections are located away from the OMC-1 and BN/KL regions, and towards the positions of the CO emission detections, the CO $J=1-0$ cloud intensity is typically $\lesssim 20$ K over the velocity ranges where we detect significant CO $J=3-2$ emission with ALMA (e.g., see top panel of Figure \ref{fig:CARMA_ONC_main}). In the instances where the localized CO $J=1-0$ cloud intensity exceeds $\gtrsim 50$ K, we find only a small overlap in the velocity channels where the disks are detected and where the strong cloud emission is seen. The lack of bright CO $J=1-0$ cloud emission in the majority of position-velocity space where we detect the disks in CO emission suggests that the cloud does not significantly influence the observed morpohology of the CO emission detections. 

Because there are no datasets probing extended HCO$^+$ emission in the ONC at a resolution comparable to that of the CARMA-NRO Orion Survey, the spatial distribution of large-scale HCO$^+$ emission in the ONC remains unknown. Nevertheless, we argue that the background cloud plays an insignificant role in the observed morphology of the disks detected in HCO$^+$ emission, such that it can be neglected in our modeling. Similar to the CO emission detections, the HCO$^+$ emission detections are also found away from the OMC-1 and BN/KL regions. The gas density of the cloud is expected to be $<10^6$ cm$^{-3}$ in those regions \citep[e.g.,][]{Peng12}, i.e., well below the critical density of the HCO$^+$ $J=4-3$ transition \citep[$\sim 10^7$ cm$^{-3}$,][]{Tielens05}. We therefore do not expect significant extended HCO$^+$  emission outside of the dense OMC-1 and BN/KL regions. This is in  contrast with the extended CO $J=3-2$ emission, which is expected to be widespread over the diffuse regions of the ONC, given the low critical density of the CO $J=3-2$ transition \citep[$\sim 10^3$ cm$^{-3}$,][]{Tielens05}.


\section{Internal vs. External Ionization Sources}\label{appendix:ionization}

\twocolumngrid

\begin{figure}[ht!]
	\epsscale{1.25}
	\hspace{-0.4in}
	\plotone{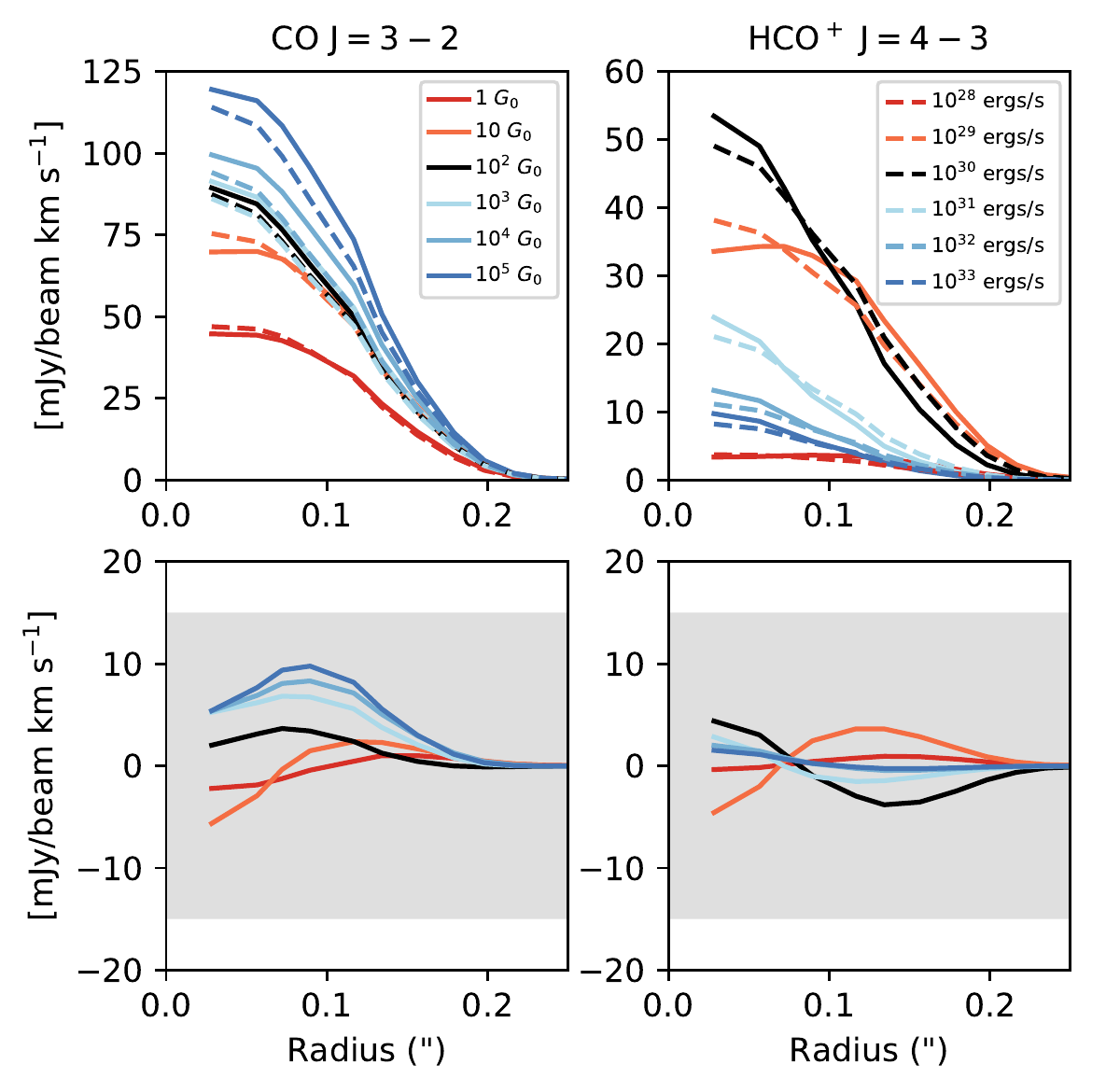}
	\caption{Top row: simulated integrated CO $J=3-2$ (left) and HCO$^+$ $J=4-3$ (right) intensity profiles of thermochemical models with externally-sourced FUV radiation from an ambient interstellar radiation field (solid lines) or internally-sourced  FUV radiation from accretion-UV emission (dashed lines). The intensity profiles cover FUV fluxes from $1$ $G_0$ to $10^5$ $G_0$ (solid lines) and FUV luminosities from $10^{28}$ ergs s$^{-1}$ to $10^{33}$ ergs s$^{-1}$ (dashed lines), and are normalized and color-coded such at  $r \approx 25$ AU, the FUV flux of a model with internally-sourced FUV radiation matches the FUV flux of a model with externally-sourced FUV radiation and the same plotted color. Bottom row: CO $J=3-2$ (left) and HCO$^+$ $J=4-3$ (right) intensity difference between models with similarly-valued internal/external FUV radiation fields. 
	The shaded grey region represents the typical rms levels seen towards our targets.
	\label{fig:FUV_test}}
\end{figure}

Here we examine the impact of including different FUV ionization sources in our thermochemical modeling. Figure \ref{fig:FUV_test} shows the integrated CO $J=3-2$ and HCO$^+$ $J=4-3$ intensity profiles of models with UV emission from nearby massive stars as the source of ionizing FUV photons, and models with accretion-UV emission from the central star as the source of ionizing FUV photons. Models with externally-sourced FUV radiation are generated by varying the interstellar FUV flux parameter in {\tt RAC2D}.  We consider interstellar FUV fluxes ranging from $1$ $G_0$ to $10^5$ $G_0$ in order-of-magnitude increments, where  $G_0 = 1.6 \times 10^{-3}$ ergs cm$^{-2}$ s$^{-1}$ and is defined as the local ISM FUV field strength over the wavelength range $930-2000$ \r{A} \citep{Habing68}. Models with internally-sourced FUV radiation are generated using an input stellar UV spectrum. We use the observed spectrum of TW Hya \citep{Herczeg02, Herczeg04}, which has a total UV luminosity of $\sim 10^{31}$ ergs s$^{-1}$ between $930$ and $2000$ \r{A}, as a reference input UV spectrum. To include a range of internally-sourced FUV radiation fields, we rescale the reference spectrum to cover FUV luminosities ranging from $10^{28}$ to $10^{33}$ ergs s$^{-1}$ in order-of-magnitude increments. At $r \approx 25$ AU, these luminosities correspond to FUV fluxes that are equivalent to the FUV fluxes used to generate the set of models with externally-sourced FUV radiation. For both model sets, we fix all other parameters to the fiducial values listed in Table \ref{table:fiducial}.

Figure \ref{fig:FUV_test} demonstrates that internally- and externally-sourced FUV radiation have a similar effect on the CO $J=3-2$ and HCO$^+$ $J=4-3$ emission of our baseline model. For both ionization sources, stronger FUV radiation results in brighter CO $J=3-2$ emission, brighter HCO$^+$ $J=4-3$ emission at low/intermediate FUV fields ($\lesssim 10^{30}$ ergs s$^{-1}$), and fainter HCO$^+$ $J=4-3$ emission at intermediate/high FUV fields ($\gtrsim 10^{30}$ ergs s$^{-1}$, see Section \ref{sec:modeling:ionization3} for a physical explanation of these trends).  Moreover, when we compare the intensity profiles of models that have different FUV ionization sources but similar FUV fluxes at $r \approx  25$ AU, we find that the intensity profiles are nearly identical, with only minor deviations due to the radial variations in the internally-sourced FUV radiation field strength. These deviations result in $<10\%$ differences in the integrated fluxes, and they are well below the typical sensitivity limit of our ALMA observations ($>10$ mJy beam$^{-1}$ km s$^{-1}$). 

While we might expect greater variations in the intensity profiles at larger disk radii \citep[e.g., $R = 200$ AU, ][]{Cleeves13}, our targets are compact enough to lack these outer regions where the accretion-UV emission from the central star is significantly weakened. Therefore, we conclude that with our current ALMA observations and sample targets, our modeling is insensitive to whether we treat the FUV radiation as internal or external, such that we can use a single ionization parameter to encapsulate contributions from both the central star and external environment.

\section{Parameter Space Exploration Supplemental Text}\label{appendix:a1}

\twocolumngrid

\subsection{Ionization Environment}\label{sec:modeling:ionization3}

We first consider the subset of parameters that directly influence our model photon sources: $L_{UV}$, $L_{XR}$, $T_{eff}$, $R_*$, and $\zeta_{CR}$. The CO $J=3-2$ intensity increases as we raise $L_{UV}$ (see Figure \ref{fig:pstudy}), due the increased gas temperature near the optically thick CO emission surface. Since the primary effect of $L_{UV}$ is to change the normalization of the CO emission, no significant effect is seen in the modeled kinematics, as shown in Figure \ref{fig:pstudy_spec1}. 

While strong UV fields in general lead to the photodissociation of molecules, we find that the self-shielding of CO enables a large majority of the disk to remain CO-abundant at higher UV fields. Compared with its sensitivity to $L_{UV}$, the sensitivity of the CO emission to $L_{XR}$ is much weaker. Only for $L_{XR} =  10^{32}$ ergs s$^{-1}$ is the X-ray irradiation strong enough to dissociate the CO gas.  

Since the UV and X-ray ionization balance can influence the entire chemical landscape of circumstellar disks, we also examined how the CO abundance and intensity vary as a function of both $L_{UV}$ and $L_{XR}$. Figure \ref{fig:checker_GO_LXR} indicates that at lower UV fields, X-rays are more effective at dissociating gas-phase CO and reducing the CO intensity. When the X-ray field is enhanced relative to the UV field, gas-phase CO preferentially reprocesses into CI, and the CI then reacts with other species to form CN, HCN, C$_2$H, and other neutral carbon carriers. Since many of the neutral carbon carriers are dissociated by FUV radiation (rather than by X-rays), they essentially reprocess back into CO when we increase $L_{UV}$. Gas-phase CO is therefore able to remain bright and intact at high UV fields regardless of the X-ray radiation field strength, at least over the ranges of $L_{UV}$ and $L_{XR}$  that we have considered.

With HCO$^+$, we find that the model emission exhibits  similar responses to changes in  L$_{UV}$ and $L_{XR}$ (see Figures \ref{fig:pstudy2} and \ref{fig:pstudy_spec2}). We initially see a strong enhancement of the HCO$^+$ $J=4-3$ emission as we increase $L_{UV}$ or $L_{XR}$, but eventually a maximum intensity is reached, and the emission declines at the highest explored values of $L_{UV}$ or $L_{XR}$. In the elevated layers of our model disks where the HCO$^+$ abundance peaks (see Figure \ref{fig:fiducial}), HCO$^+$ formation is dominated by the reaction \citep{Dalgarno84}:
\begin{equation}\label{eq:HCO_form}
    CO + H_3^+ \rightarrow HCO^+ + H_2.
\end{equation}
The initial increase of the HCO$^+$ intensity with respect to $L_{UV}$ reflects the increased CO abundance due to the dissociation of neutral carbon carriers, as noted previously. The subsequent decline in intensity at higher $L_{UV}$ reflects the reduced H$_3^+$ abundance, which is inversely proportional to the UV radiation field strength. If we instead increase $L_{XR}$, the H$_3^+$ abundance increases due to the increased levels of ionization, and so the HCO$^+$ intensity  also increases. But at higher $L_{XR}$, the CO starts to dissociate, causing the  HCO$^+$ emission to decline.
 
Since L$_{UV}$ and $L_{XR}$ can produce correlated effects on the CO abundance and intensity, we also searched for correlated effects on the  HCO$^+$ intensity. Figure \ref{fig:checker_GO_LXR} confirms that bright HCO$^+$ emission is favorable under a correlated range of L$_{UV}$ and $L_{XR}$. If too much $H_3^+$ is removed due to the increased UV field, then the HCO$^+$ abundance declines and the emission becomes faint. If too much CO is reprocessed into neutral carbon due to the increased X-ray radiation field relative to the UV field, then the HCO$^+$ abundance also declines. In between these limiting cases are ``sweet spots'' in the UV and X-ray ionization balance where HCO$^+$ can exist in relative high abundance and be nearly as bright as CO. 

As we increase $T_{eff}$ or $R_{*}$, the photon output increases at wavelengths comparable to the sizes of the small dust grain population, leading to efficient photon absorption, a warmer dust disk, and brighter dust emission. Because $T_{eff}$ influences the stellar UV photon output, it has a similar effect on the synthetic CO and HCO$^+$ emission as L$_{UV}$, but to a smaller degree over the explored range of temperatures. Higher values of $T_{eff}$ result in an increased UV intensity and thus, brighter CO and HCO$^+$ emission, except for $T_{eff} = 5000$ K (the highest value considered), in which case the HCO$^+$ emission declines due to the reduced H$_3^+$ abundance. 

While changes in $R_*$ can also impact the UV photon output, no strong sensitivity to $R_*$ is seen over the explored values for both CO and HCO$^+$. Since the CO and HCO$^+$ emission are unaffected by small changes in $R_{*}$, and since $T_{eff}$ and $R_{*}$ have a similar effect on the $350$ GHz continuum flux, we fix $R_{*}$ for the remainder this paper while treating $T_{eff}$ as variable. 

Finally, we note that changes in the cosmic ray ionization rate have a weak effect on the CO, HCO$^+$, and $350$ GHz continuum emission. As shown in Figures \ref{fig:pstudy} - \ref{fig:cont_fluxes_pstudy}, the model emission is nearly identical for ionization rates spanning 10$^{-21}$ to 10$^{-15}$ s$^{-1}$. Cosmic rays mainly effect the gas ionization and chemistry in the disk midplane, and since the CO $J=3-2$ and HCO$^+$ $J=4-3$ lines are usually optically thick, changes in midplane composition are not strongly imprinted on the synthetic emission. We therefore fix $\zeta_{CR}$ to $\sim 10^{-19}$ s$^{-1}$ \citep[i.e., the expected value for an accreting T-Tauri system; ][]{Cleeves13, Cleeves15}  for the remainder of this paper.

\subsection{Disk Mass}\label{sec:modeling:diskmass}


Next, we explore sensitivities to changes in the disk dust and gas masses. The $350$ GHz continuum flux exhibits a strong sensitivity to the dust mass (see Figure \ref{fig:cont_fluxes_pstudy}), and no sensitivity to the gas mass, as expected. The flux scales sub-linearly with $M_{dust}$, due to the increased dust temperature found towards lower-mass dust disks and the increased optical depth found towards massive dust disks. Nevertheless, out of all the parameters that we consider in our parameter space exploration, the disk dust mass has the strongest influence on the model  $350$ GHz continuum flux (see Figure \ref{fig:cont_fluxes_pstudy}), consistent with previous studies \citep[e.g.,][]{Ballering19}.

The CO $J=3-2$ emission exhibits a strong sensitivity to both the gas and dust masses (see Figure \ref{fig:pstudy}). As we increase the gas mass, the CO self shields at a higher elevation where the temperatures are warmer, and this results in brighter CO emission. For lower gas masses, the CO self shields in a deeper, cooler layer of the disk, and so the emission is fainter. When we instead raise the dust mass, the dust optical depth increases, and this in turn reduces the FUV intensity, the gas temperature near the CO surface, and thus, the intensity of the CO emission. Increasing the dust mass therefore has the opposite effect on CO emission as increasing the gas mass, and the same goes for reducing the dust/gas masses (see Figure \ref{fig:pstudy}).

The HCO$^+$ $J=4-3$ emission also exhibits strong, opposing sensitivities to the gas and dust masses. As shown in Figure \ref{fig:pstudy2}, an order-of-magnitude increase in the dust mass causes enough of a decline in the FUV intensity such that the relative strengths of the UV and X-ray radiation fields are no longer suitable for HCO$^+$ formation. Hence, we see a sharp decline in the HCO$^+$ emission. Similarly, an order-of-magnitude decrease in the gas mass causes enough of a reduction of the gas opacity to increase the X-ray intensity within the disk to the point where the UV and X-ray ionization balance is again unsuitable for HCO$^+$ formation, causing a decline in the HCO$^+$ intensity. 

The opposing behaviors of the dust and gas masses on the model CO and HCO$^+$ emission suggest that both molecular lines are sensitive to the gas-to-dust mass ratio. In Figure \ref{fig:checker_Mgas}, we plot the peak CO $J=3-2$ and HCO$^+$ $J=4-3$ intensities as a function of both the dust and gas masses. For gas masses $\gtrsim 10^{-5} M_{\odot}$, the opposing sensitivities of the CO emission to the dust and gas masses are preserved, and when we vary the dust and gas masses over a fixed gas-to-dust ratio, the opposing behaviors of the gas and dust effectively cancel one another out. Below $\sim 10^{-5} M_{\odot}$, we see less of a canceling out,  as the emission becomes less optically thick and more sensitive to changes in gas column density. 

Models with gas-to-dust ratios of $1$ or $10$ also do not yield the ideal balance of UV and X-ray irradiation that is suitable for HCO$^+$ formation, so they produce fainter emission HCO$^+$ emission than  gas-rich model disks with gas-to-dust ratios $\geq 100$, which strike a more favorable UV/X-ray ionization balance (see Figure \ref{fig:checker_Mgas}). In particular, gas-to-dust ratios of $100$ or $1000$ appear to provide the ``sweet spot'' in the ionization balance for the brightest HCO$^+$ emission, as shown in Figure \ref{fig:checker_Mgas}. For gas masses up to $\sim 10^{-3} M_{\odot}$, the HCO$^+$ emission increases as we increase the dust and gas masses over a fixed gas-to-dust ratio, as the emission is less optically thick than the CO emission and is therefore sensitive to changes in gas column density at higher gas masses.

Figure \ref{fig:checker_Mgas} demonstrates that while the CO and/or HCO$^+$ emission alone can be used to constrain the disk gas mass, the precision on the gas mass measurements can be improved significantly with an independent constraint on the disk dust mass, especially for CO-based mass estimates. The model $350$ GHz continuum flux is capable of providing such a constraint, as illustrated in Figure \ref{fig:cont_fluxes_pstudy}. Modeling the continuum and CO and/or HCO$^+$ emission together can therefore help overcome the high optical depths associated with the CO and HCO$^+$ lines, and provide more refined measurements of the gas (and dust) masses for the ONC targets in our sample. 

Finally, we examine the impact of $\gamma$ on the model emission. Small changes in $\gamma$ do not influence where the optically thick CO and HCO$^+$ emission originate vertically, and so the abundances, emission surface temperatures, and line intensities are all unaffected. Since the dust is less optically thick than the molecular lines, the $350$ GHz continuum flux is marginally sensitive to changes in the  radial distribution of material. Lower values of $\gamma$ result in an enhanced continuum flux due to the increased concentration of material in outer, cooler regions of the disk, and vice-versa for higher values of $\gamma$. Since the effects of $\gamma$ on the $350$ GHz continuum flux are small compared with other model parameters, such as the dust mass, we fix $\gamma$ to $1$ \citep[e.g., ][]{Andrews11, Tazzari16, Ballering19} for the remainder of this study (see Table \ref{table:fiducial}).

\subsection{Disk Morphology}

In this section, we examine the effects of varying the stellar mass ($M_*$), disk inclination ($i$), and disk inner and outer radii ($R_{in}$ and $R_{out}$) on the synthetic CO $J=3-2$, HCO$^+$ $J=4-3$, and $350$ GHz continuum emission. Changes in $M_*$ and $i$ have a direct effect on the observed kinematics, with higher values resulting in broader line profiles  (see Figures \ref{fig:pstudy_spec1} and \ref{fig:pstudy_spec2}). Higher-inclination disks also emit fainter continuum, CO, and HCO$^+$ emission than lower-inclination disks due to a reduction of the emission surface area and increased line-of-sight optical depth. 

The intensities of the CO $J=3-2$, HCO$^+$ $J=4-3$,  and $350$ GHz continuum emission are also sensitive to changes in the stellar mass, although the degree of sensitivity varies for each quantity. As we reduce the stellar mass, the disk scale height increases as a result of our enforcement of vertical hydrostatic equilibrium, and this leads to a subsequent increase in the gas and dust column densities as well as an increase in the emitting surface area. Since the CO $J=3-2$ line is optically thick, the intensity of the CO $J=3-2$ emission is sensitive to stellar-mass-driven changes in the emitting surface area, and so we find slightly brighter CO emission in models with lower-mass stars. Since the dust and the HCO$^+$ $J=4-3$ line are less optically thick than the CO $J=3-2$ line, their emission is also sensitive to stellar-mass-driven changes in the gas and dust column densities, which can explain why the continuum and HCO$^+$ emission undergo a stronger increase in intensity than the CO emission when we reduce the stellar mass. 

The CO, HCO$^+$, and continuum emission all exhibit strong sensitivities to changes in the outer radius. As illustrated in Figures \ref{fig:pstudy} - \ref{fig:cont_fluxes_pstudy}, the fluxes increase significantly for models with larger radii as a result of the increased emission surface area. 
We note that the sensitivity of the HCO$^+$ emission to the disk size begins to saturate for radii $\gtrsim 100$ AU, since the outer regions of the disk are less suitable for HCO$^+$ formation due to the reduced X-ray radiation field. The continuum flux also exhibits a similar decline in sensitivity, although here it is because of the reduced dust temperature in the outer regions of the disk. 

Changes in the inner radius do not impact the model CO and HCO$^+$ emission, as the emission mainly originates in outer, cooler regions of the disk. The $350$ continuum flux increases as we raise $R_{in}$, due the increased concentration of material in cooler regions of the disk away from the central star. However, the effect is small compared with the effects of other model parameters. We therefore fix $R_{in}$ to the fiducial value listed in Table \ref{table:fiducial} for the remainder of this study.

\subsection{Other parameters}\label{sec:modeling:other}

We explore the subset of model parameters that influence the mass distribution of the small and large dust grain populations, namely, $f_{mm}$, $a_{max, \mu m}$, and $a_{max, mm}$. As we increase $f_{mm}$ or $a_{max, \mu m}$, the disk becomes more opaque to UV radiation due to the increased amount of small, micron-sized dust grains, and vice-versa for reducing  $f_{mm}$ or $a_{max, \mu m}$. As shown in Figures \ref{fig:pstudy} -  \ref{fig:pstudy_spec2}, these changes in the localized UV intensity are not strong enough to significantly alter the CO/HCO$^+$ chemistry and line intensity.

If more of the dust mass is found in mm-sized grains (which depends on choices of $f_{mm}$, $a_{max, \mu m}$, and $a_{max, mm}$), then the $350$ GHz continuum flux will increase.  However, as shown in Figure \ref{fig:cont_fluxes_pstudy} \citep[and also in previous studies; e.g.,][]{Ballering19}, the total mass of the large and small grains, set by $M_{dust}$, has a stronger effect on the continuum flux  than the relative masses of the small and large grains. Since we are computationally limited in the the number of parameters we can treat as variable, and since the model CO and HCO$^+$ emission are not sensitive to $f_{mm}$, $a_{max, \mu m}$, and $a_{max, mm}$, we fix $f_{mm}$, $a_{max, \mu m}$, and $a_{max, mm}$ to the fiducial values listed in Table \ref{table:fiducial}.

We also examine how changes in the initial CO abundance impact our molecular lines of interest. We consider initial abundances within a factor of $\sim 2$ of our fiducial value, reflective of differences in initial cloud compositions \citep[e.g.,][]{Bolatto13, Lacy17}. As shown in Figures \ref{fig:pstudy} and \ref{fig:pstudy_spec1}, these variations have a small effect on the model CO $J=3-2$ emission, with higher abundances resulting in brighter emission, and vice-versa for lower abundances. While a reduction of the initial CO abundance also leads to fainter HCO$^+$ emission (see Figures \ref{fig:pstudy2} and \ref{fig:pstudy_spec2}), an enhancement does not lead to brighter HCO$^+$ emission. This is because HCO$^+$ abundances that are larger than the maximum values found in our models with bright HCO$^+$ emission ($\sim 10^{-7}$, see Figure \ref{fig:fiducial}) are not chemically stable, and tend to reprocess down to lower abundances on short timescales even when there is additional initial CO gas. Since the effect of the initial CO abundance on the model emission is small over the explored values, we assume a fixed initial CO abundance for the remainder of this study (see Table \ref{tab:init}).

Finally, we consider whether including additional reactions involving vibrationally-excited molecular hydrogen, henceforth denoted as ``H$_2^*$'', can impact the model HCO$^+$ emission. Recent modeling by \cite{Kamp17} has shown that including H$_2^*$ chemistry in the reaction network can enhance the abundances of HCO$^+$ in disks, although they do not discuss how this impacts the HCO$^+$ line fluxes. We regenerated our fiducial model with a set of H$_2^*$ reactions in the chemical network, following the procedure outlined in \cite{Kamp17}, and found no significant difference in the CO or HCO$^+$ line fluxes when the H$_2^*$ network was included (see Figures \ref{fig:pstudy} and \ref{fig:pstudy2}). Given the lack of a difference, and given that many of the H$_2^*$ reactions are uncertain, we opt to exclude an H$_2^*$ network in this study.

\onecolumngrid

\vspace{0.5in}

\onecolumngrid

\section{Best-Fit Channel Maps}\label{appendix:b}

Here we provide the full results of our modeling, namely, the sets of plots showing the observed channel maps, best-fit models, and residuals for each ONC member in our sample.



\begin{figure*}[ht!] 
	\epsscale{1.2}
	\vspace{-1pt}
	\centering
	\plotone{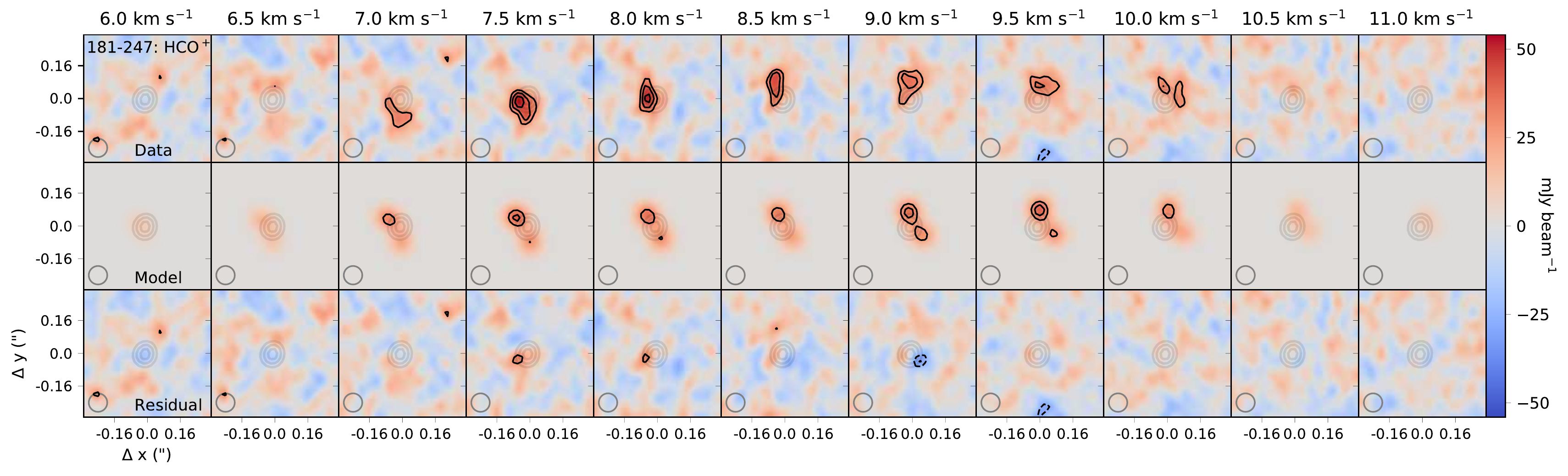}
	\caption{
	Modeling results for ONC member 181-247. The top row shows ALMA HCO$^+$ $J=4-3$ channel map emission; the middle row shows the best-fit thermochemical model channel maps obtained from our fitting procedure (see Section \ref{sec:fitting}); and the bottom row shows the residuals of the fit.  Solid black contours show 3, 4, and 5$\sigma$ emission, whereas dashed black contours show  -3, -4, and -5$\sigma$ absorption.  Solid gray contours show the 350 GHz continuum emission. The channel velocities are shown at the top of each column.
	\label{fig:appendix:fit_80}}
\end{figure*}


\begin{figure*}[ht!] 
	\epsscale{1.2}
	\vspace{-1pt}
	\centering
	\plotone{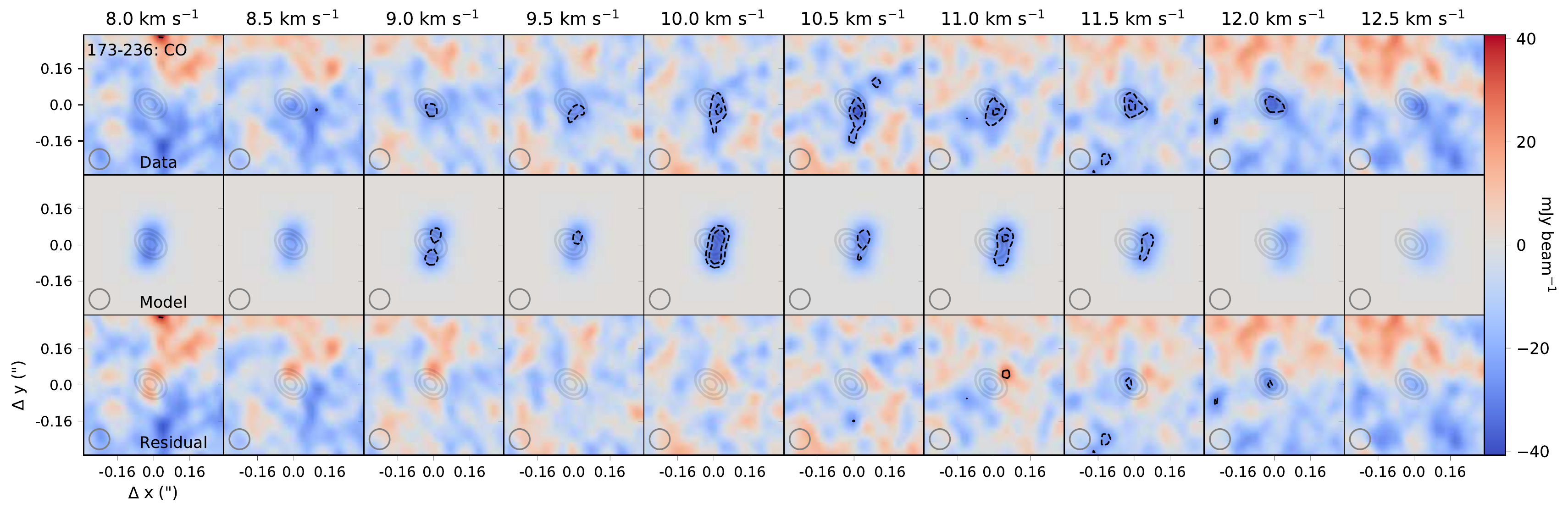}
	\caption{Modeling results for ONC member 173-236. The layout of this plot is identical to that of Figure \ref{fig:appendix:fit_80}, except here we show CO $J=3-2$ observations rather than HCO$^+$ $J=4-3$ observations.
	\label{fig:appendix:fit_65}}
\end{figure*}

\begin{figure*}[ht!] 
	\epsscale{1.2}
	\vspace{-1pt}
	\centering
	\plotone{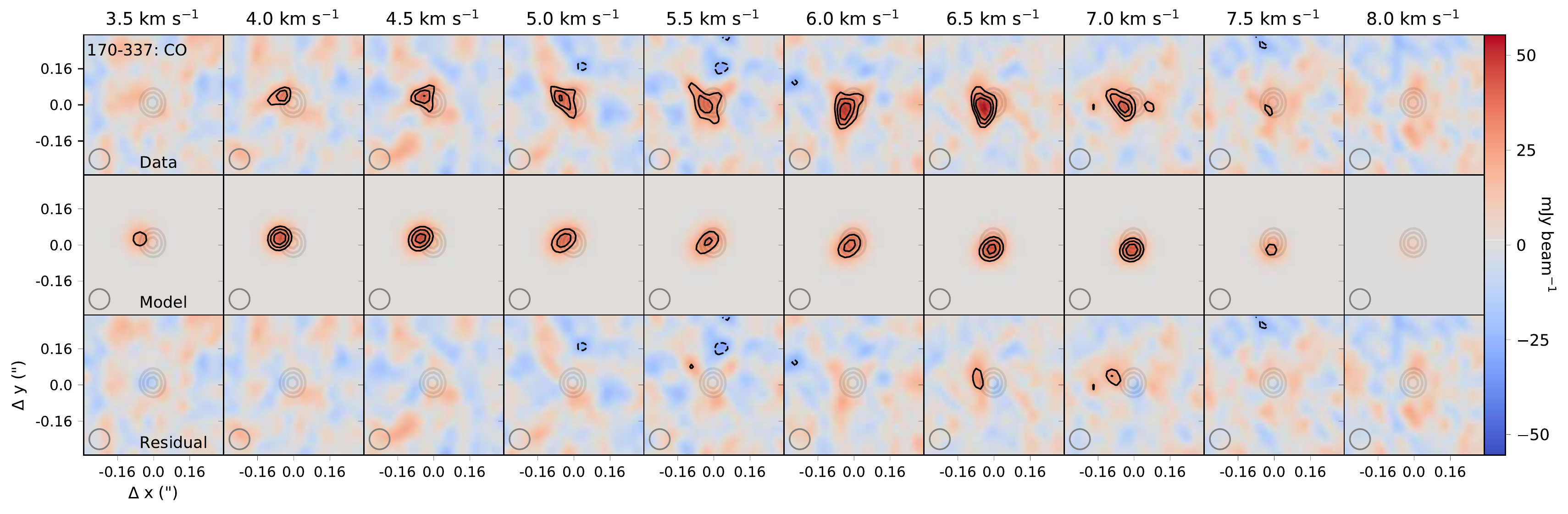}
	\caption{Modeling results for ONC member 170-337. The layout of this plot is identical to that of Figure \ref{fig:appendix:fit_65}.
	\label{fig:appendix:fit_61}}
\end{figure*}

\begin{figure*}[ht!] 
	\epsscale{1.2}
	\vspace{-1pt}
	\centering
	\plotone{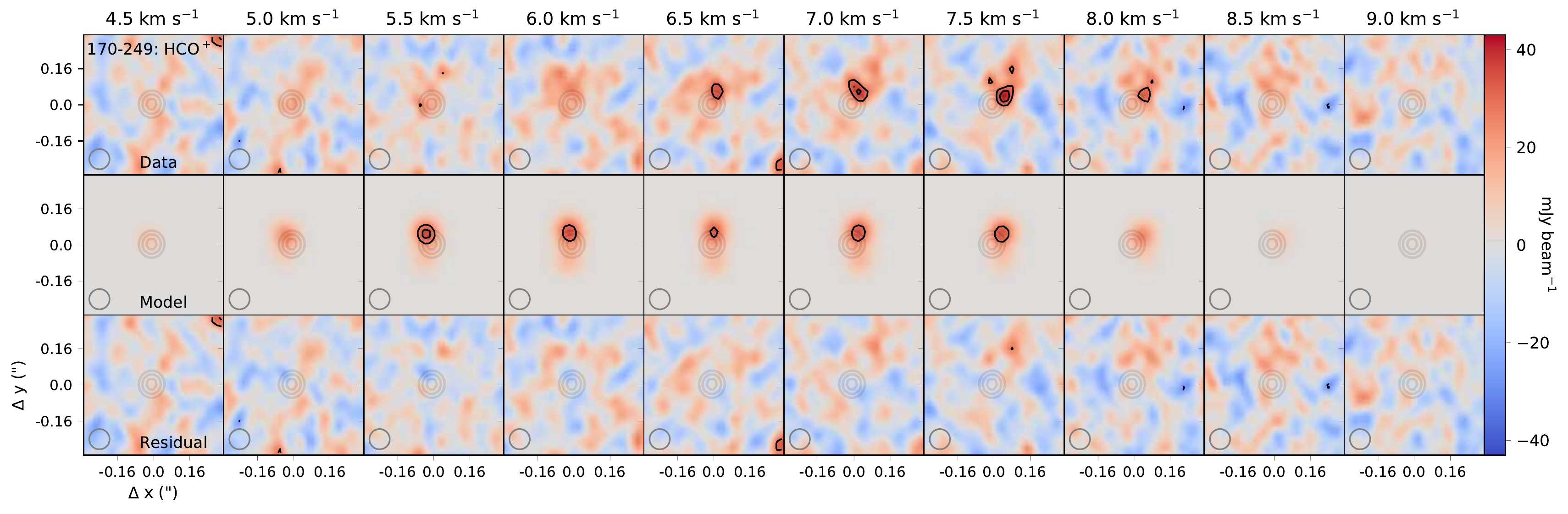}
	\caption{Modeling results for ONC member 170-249. The layout of this plot is identical to that of Figure \ref{fig:appendix:fit_80}.
	\label{fig:appendix:fit_60}}
\end{figure*}

\begin{figure*}[ht!] 
	\epsscale{1.2}
	\vspace{-1pt}
	\centering
	\plotone{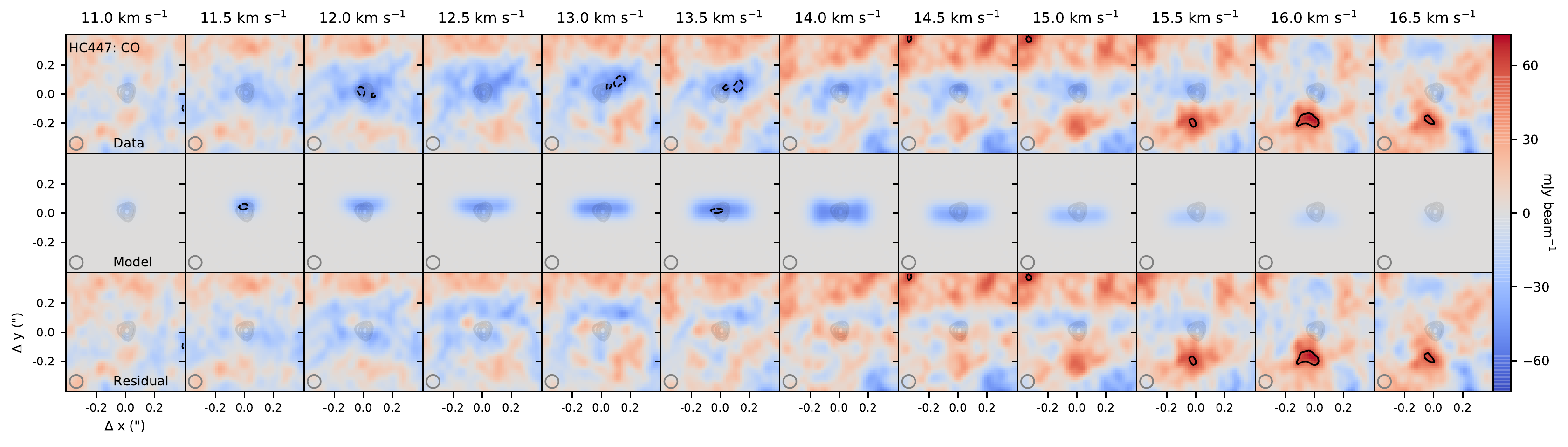}
	\caption{Modeling results for ONC member HC447. The layout of this plot is identical to that of Figure \ref{fig:appendix:fit_65}.
	\label{fig:appendix:fit_37}}
\end{figure*}

\begin{figure*}[ht!] 
	\epsscale{1.2}
	\vspace{-1pt}
	\centering
	\plotone{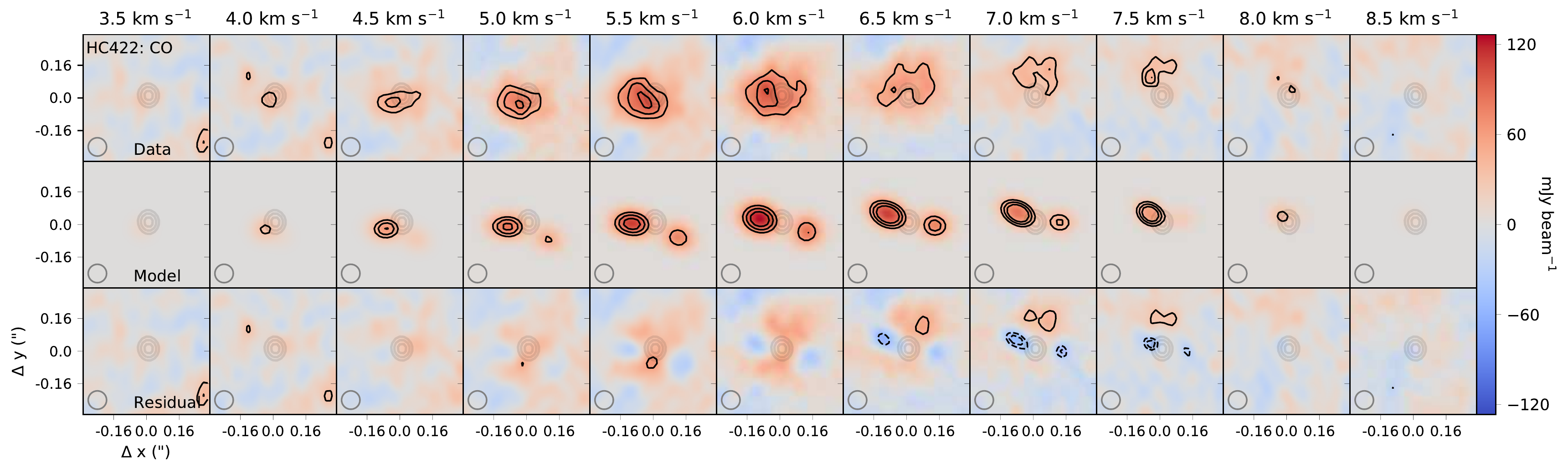}
	\caption{Modeling results for ONC member HC422. The layout of this plot is identical to that of Figure \ref{fig:appendix:fit_65}.
	\label{fig:appendix:fit_66}}
\end{figure*}

\begin{figure*}[ht!] 
	\epsscale{1.2}
	\vspace{-1pt}
	\centering
	\plotone{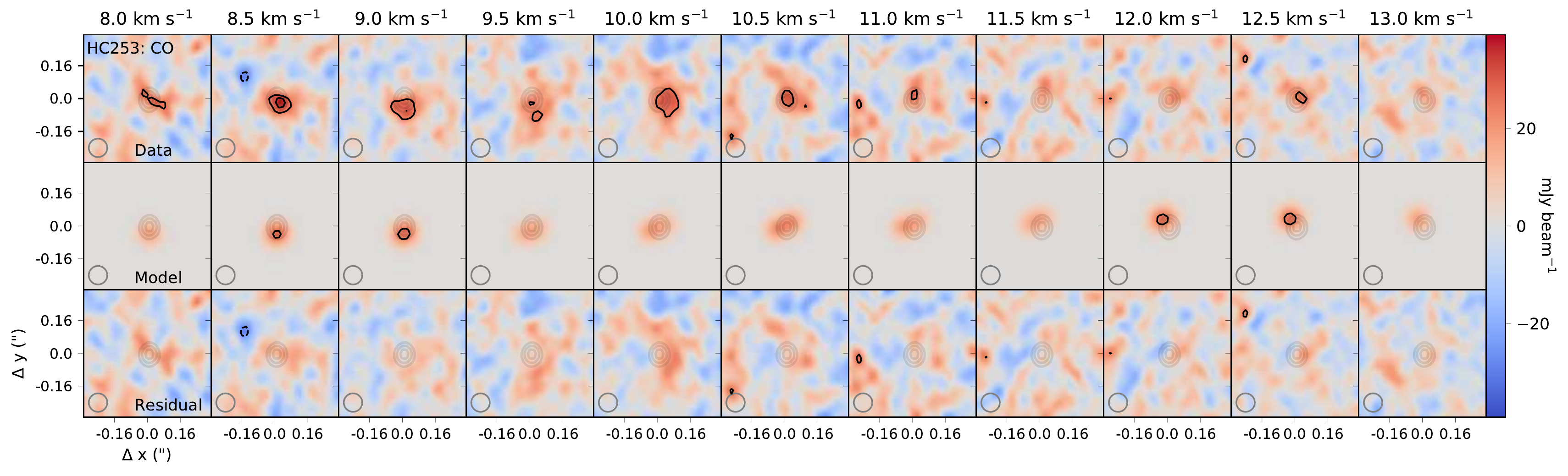}
	\caption{Modeling results for ONC member HC253. The layout of this plot is identical to that of Figure \ref{fig:appendix:fit_65}.
	\label{fig:appendix:fit_82}}
\end{figure*}

\begin{figure*}[ht!] 
	\epsscale{1.2}
	\vspace{-1pt}
	\centering
	\plotone{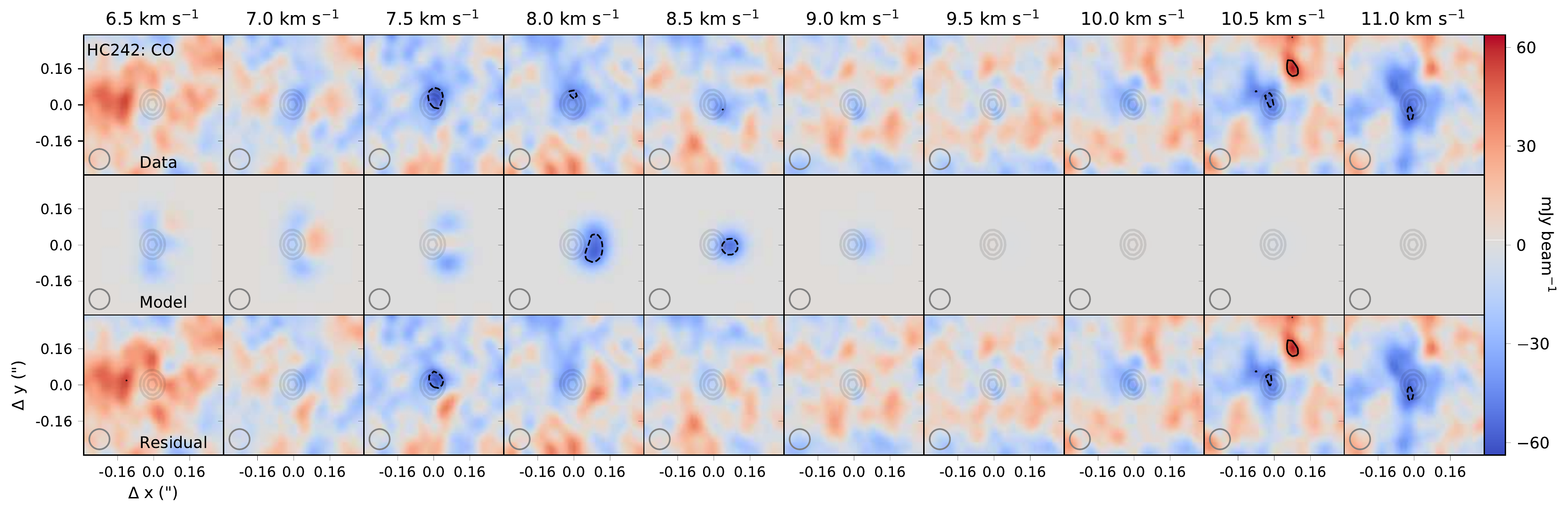}
	\caption{Modeling results for ONC member HC242. The layout of this plot is identical to that of Figure \ref{fig:appendix:fit_65}.
	\label{fig:appendix:fit_2}}
\end{figure*}

\begin{figure*}[ht!] 
	\epsscale{1.2}
	\vspace{-1pt}
	\centering
	\plotone{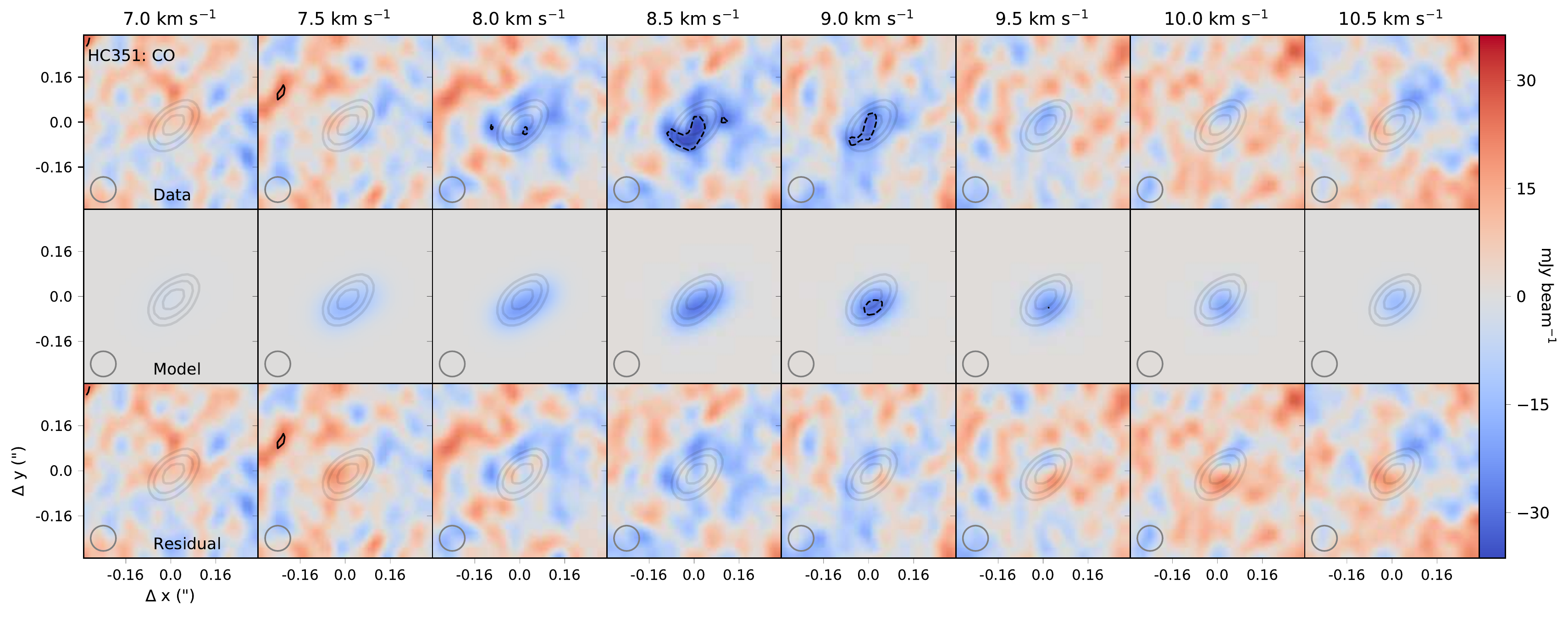}
	\caption{Modeling results for ONC member HC351. The layout of this plot is identical to that of Figure \ref{fig:appendix:fit_65}.
	\label{fig:appendix:fit_94}}
\end{figure*}

\begin{figure*}[ht!] 
	\epsscale{1.2}
	\centering	
	\gridline{\fig{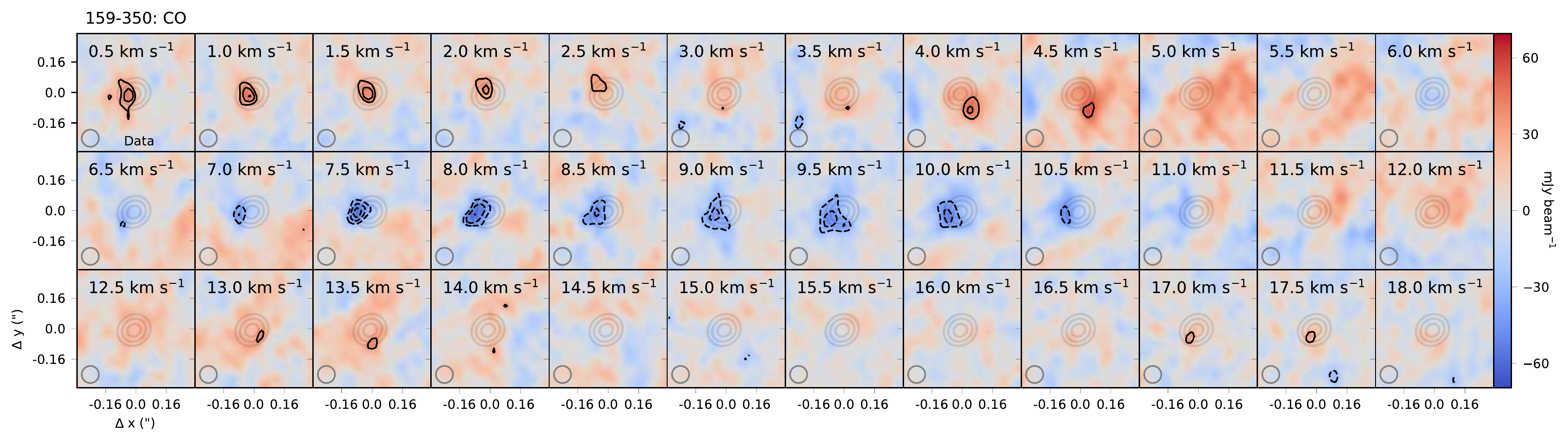}{1.0\textwidth}{}}
	\gridline{\fig{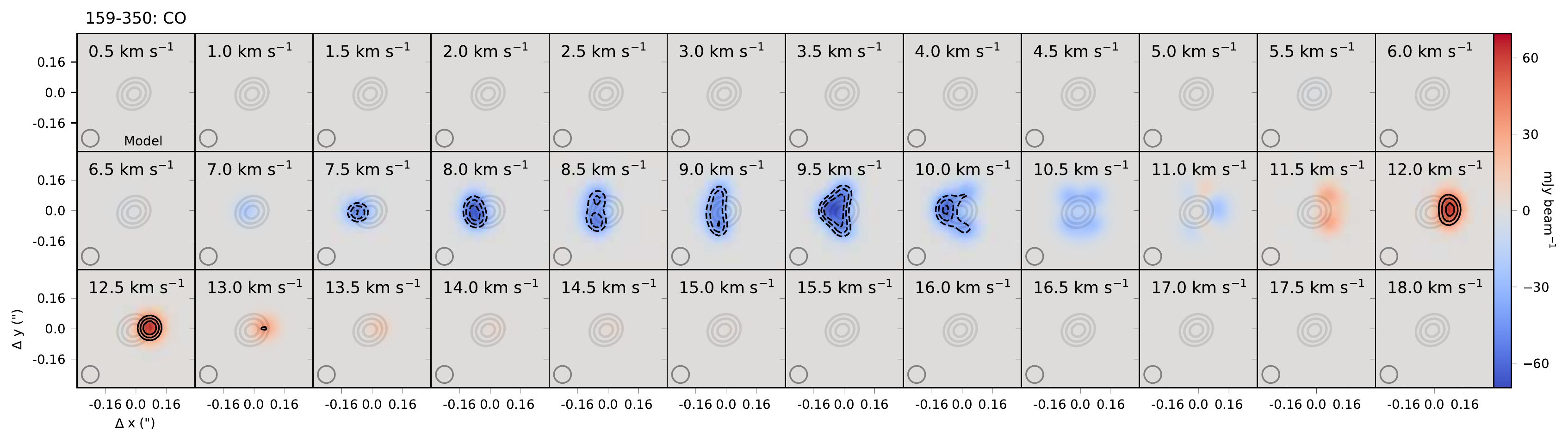}{1.0\textwidth}{}}
	\gridline{\fig{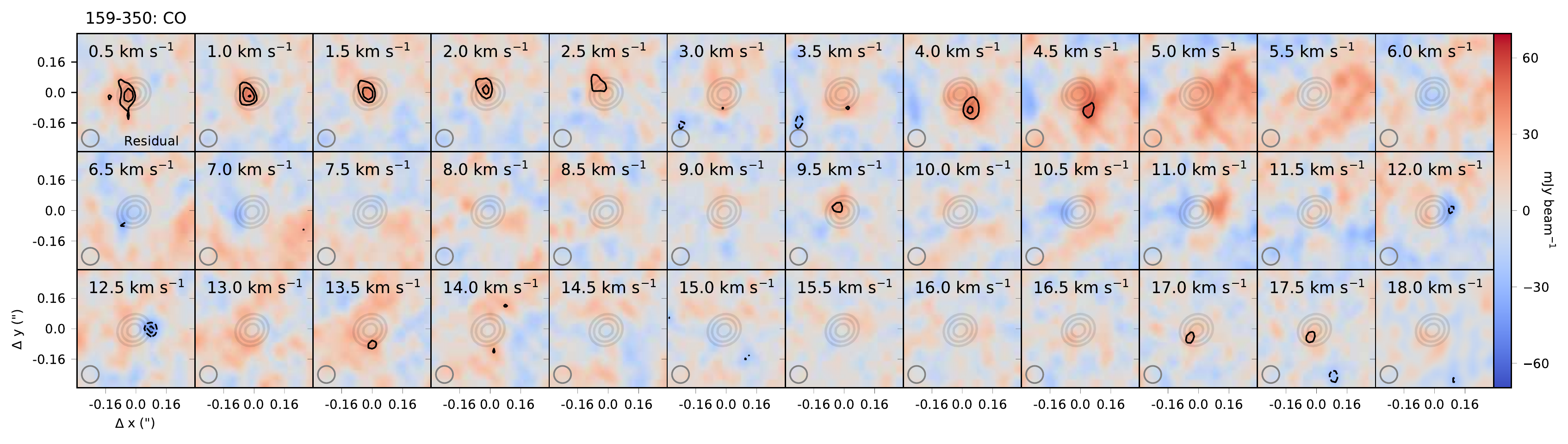}{1.0\textwidth}{}}
	\caption{Modeling results for ONC member 159-350. The layout of this plot is similar to that of Figure \ref{fig:appendix:fit_65}. 
	\label{fig:appendix:fit_39}}
\end{figure*}

\begin{figure*}[ht!] 
	\epsscale{1.1}
	\centering	
	\gridline{\fig{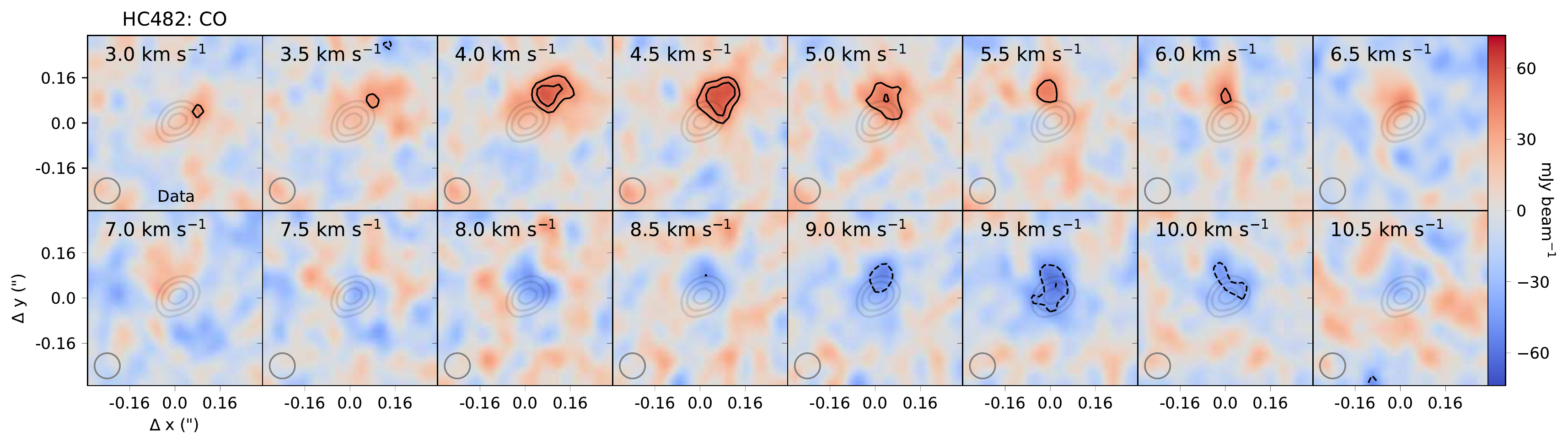}{1.0\textwidth}{}}
	\gridline{\fig{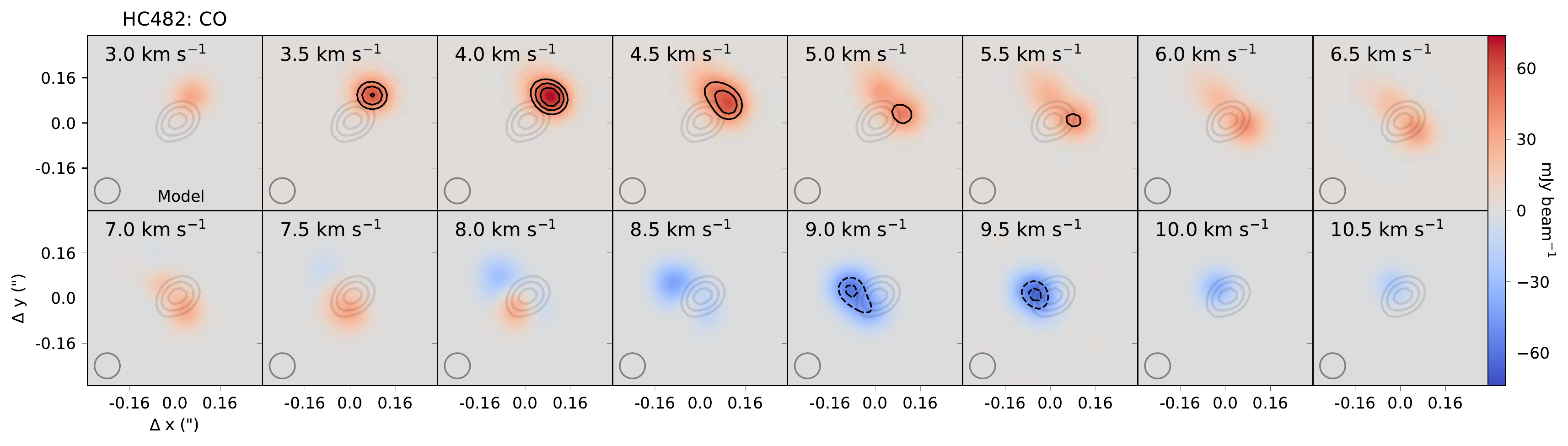}{1.0\textwidth}{}}
	\gridline{\fig{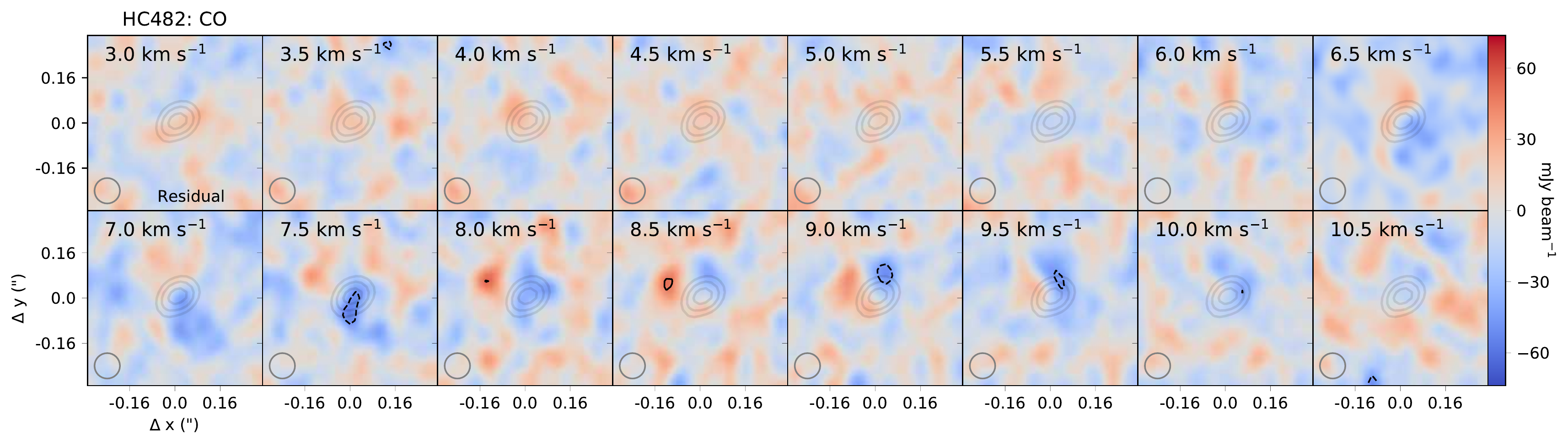}{1.0\textwidth}{}}
	\caption{Modeling results for ONC member HC482. The layout of this plot is similar to that of Figure \ref{fig:appendix:fit_65}. 
	\label{fig:appendix:fit_89CO}}
\end{figure*}

\begin{figure*}[ht!] 
	\epsscale{1.1}
	\centering	
	\gridline{\fig{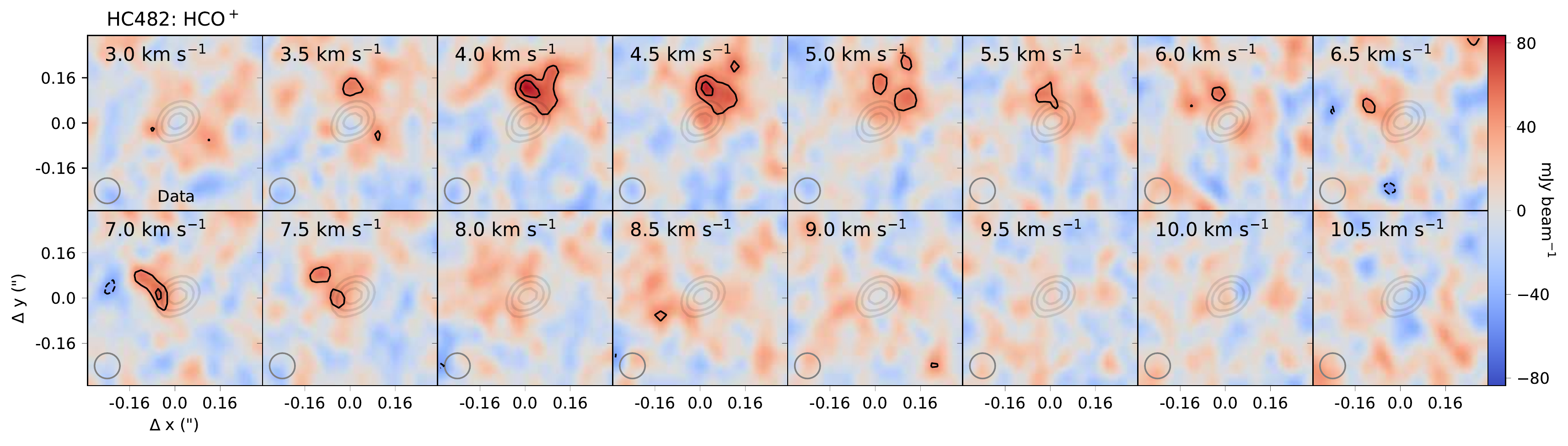}{1.0\textwidth}{}}
	\gridline{\fig{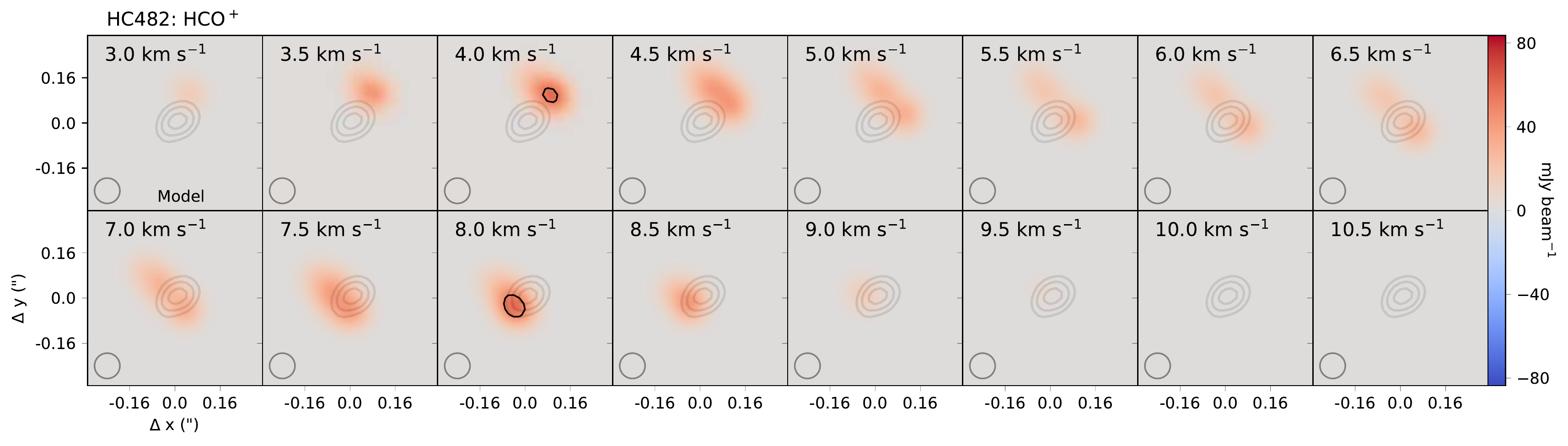}{1.0\textwidth}{}}
	\gridline{\fig{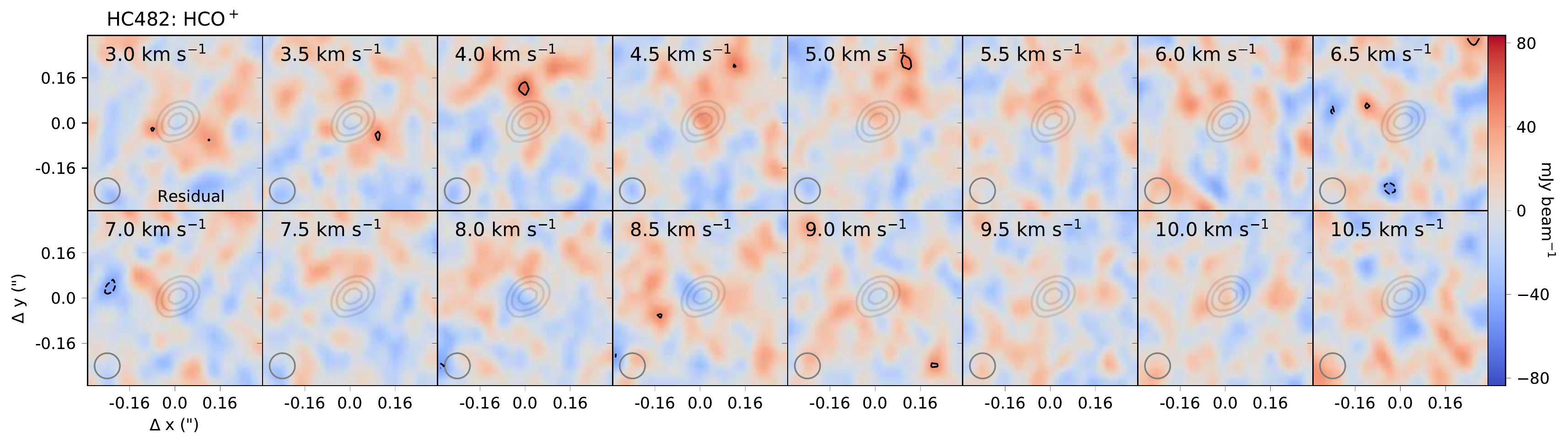}{1.0\textwidth}{}}
	\caption{Modeling results for ONC member HC482. The layout of this plot is similar to that of Figure \ref{fig:appendix:fit_80}. 
	\label{fig:appendix:fit_89HCO}}
\end{figure*}

\begin{figure*}[ht!] 
	\epsscale{1.1}
	\centering	
	\gridline{\fig{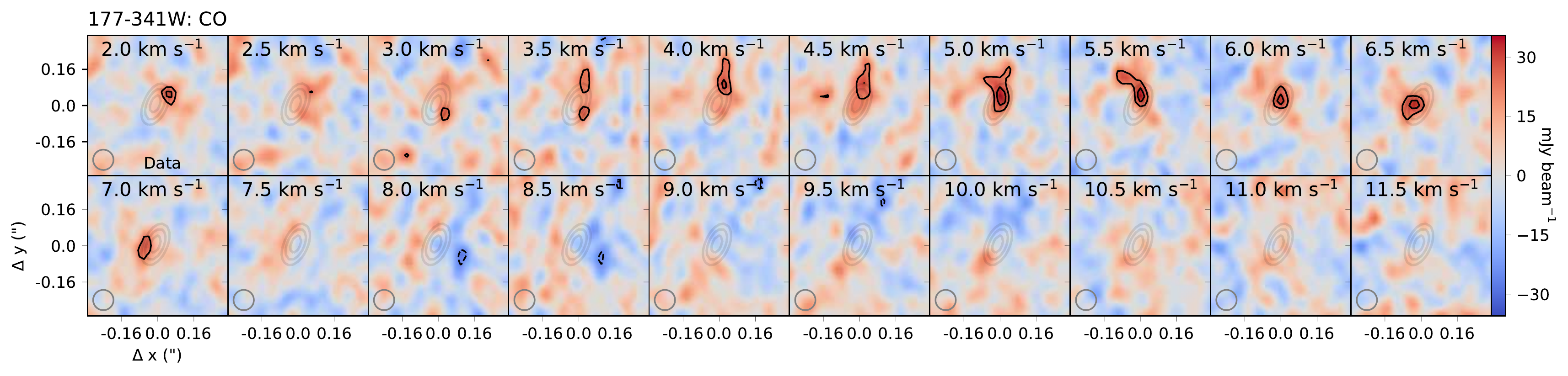}{1.0\textwidth}{}}
	\gridline{\fig{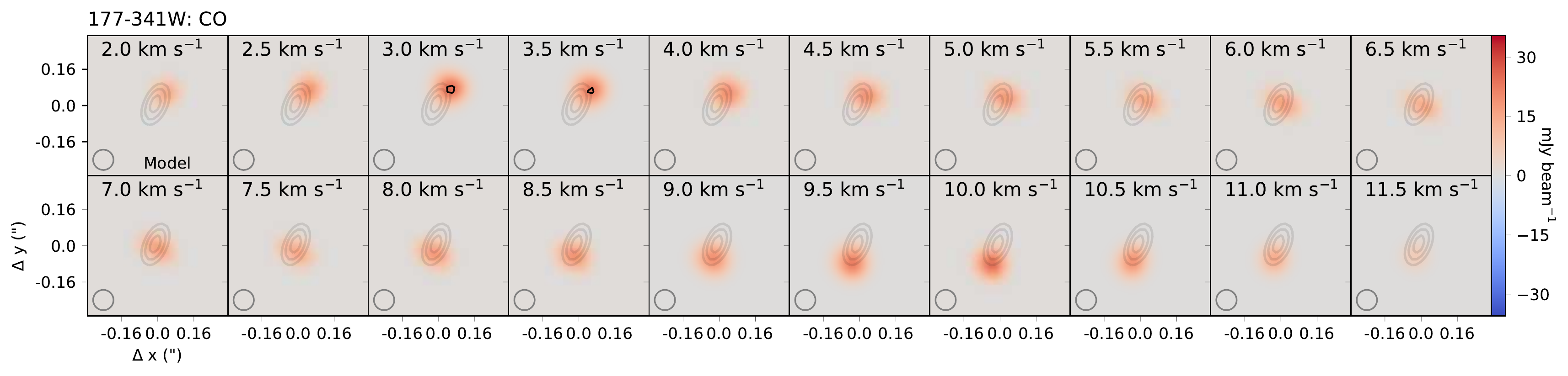}{1.0\textwidth}{}}
	\gridline{\fig{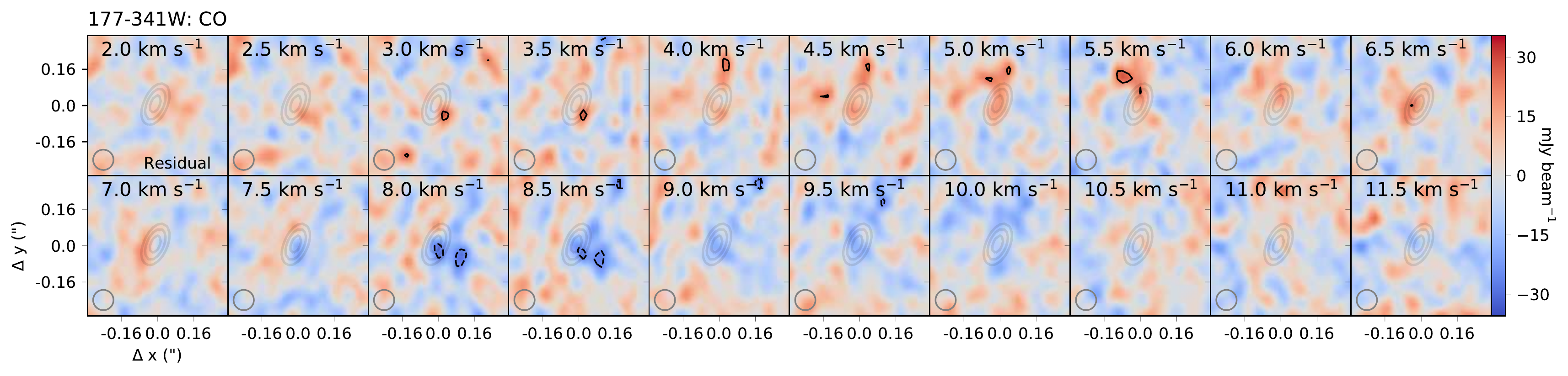}{1.0\textwidth}{}}
	\caption{Modeling results for ONC member 177-341W. The layout of this plot is similar to that of Figure \ref{fig:appendix:fit_65}. \label{fig:appendix:fit_73_CO}}
\end{figure*}

\begin{figure*}[ht!] 
	\epsscale{1.1}
	\centering	
	\gridline{\fig{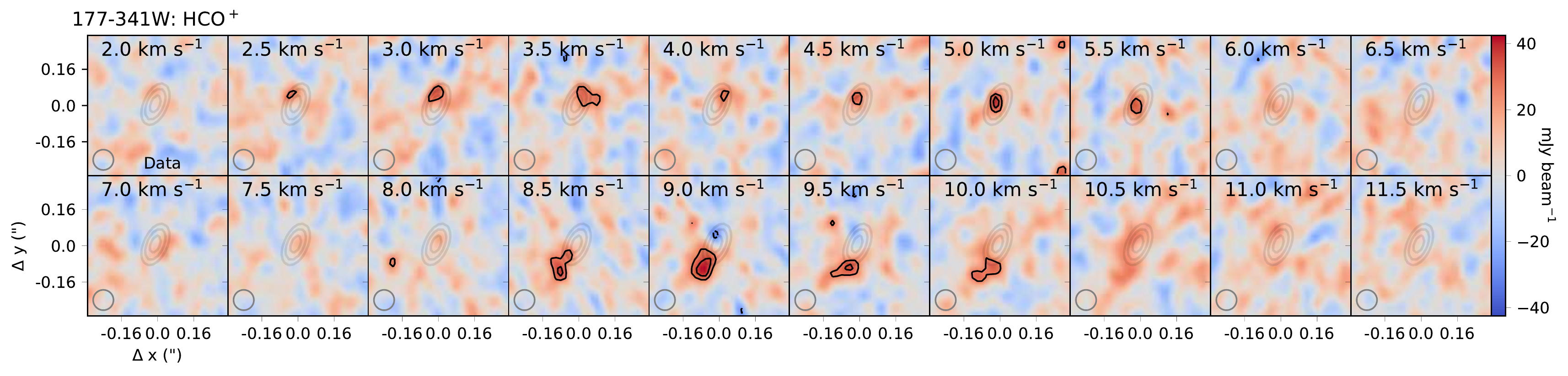}{1.0\textwidth}{}}
	\gridline{\fig{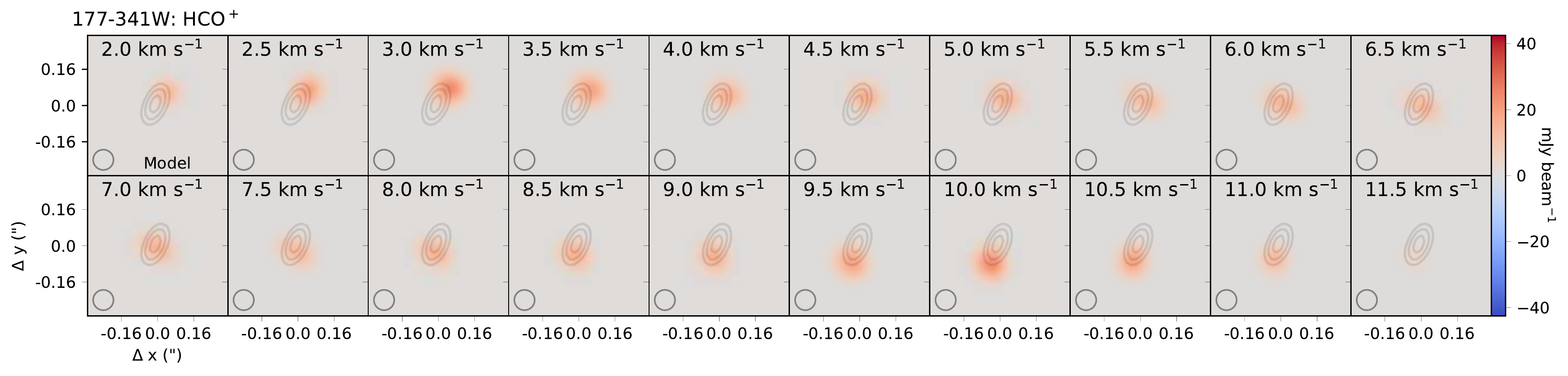}{1.0\textwidth}{}}
	\gridline{\fig{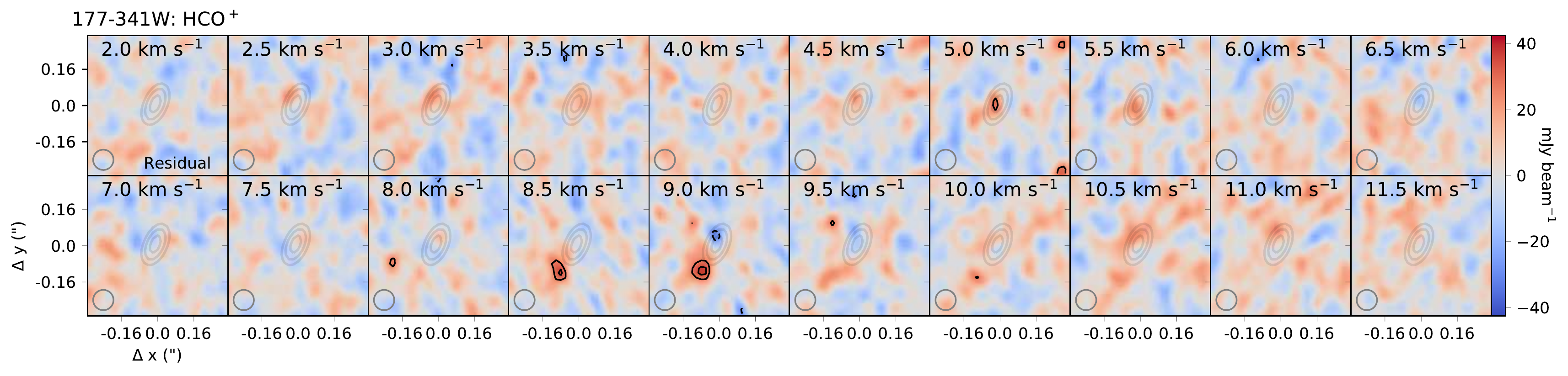}{1.0\textwidth}{}}
	\caption{Modeling results for ONC member 177-341W. The layout of this plot is similar to that of Figure \ref{fig:appendix:fit_80}. \label{fig:appendix:fit_73_HCO}}
\end{figure*}

\begin{figure*}[ht!] 
	\epsscale{1.1}
	\centering	
	\gridline{\fig{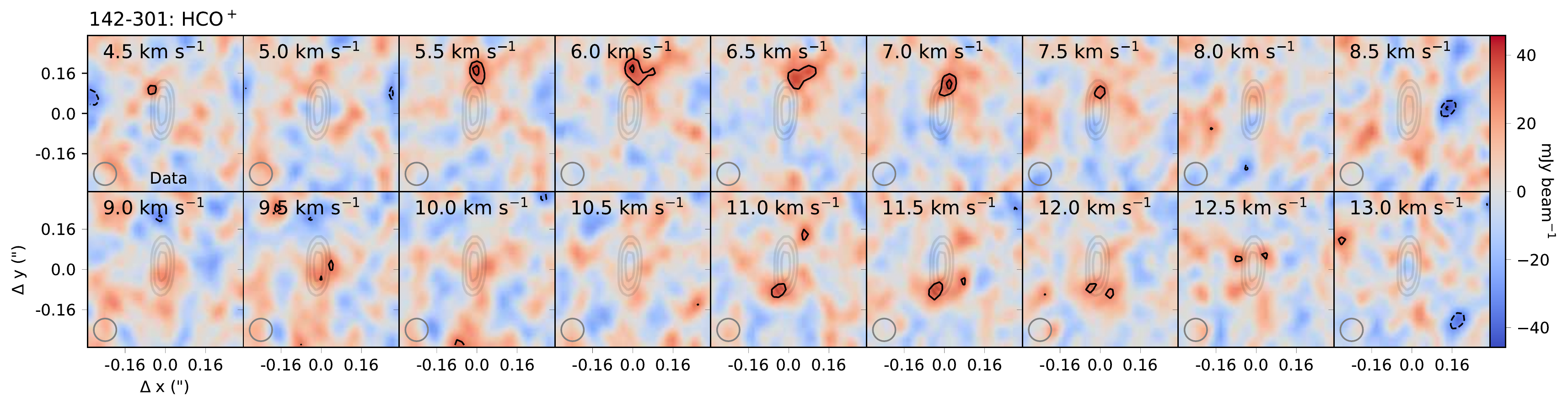}{1.0\textwidth}{}}
	\gridline{\fig{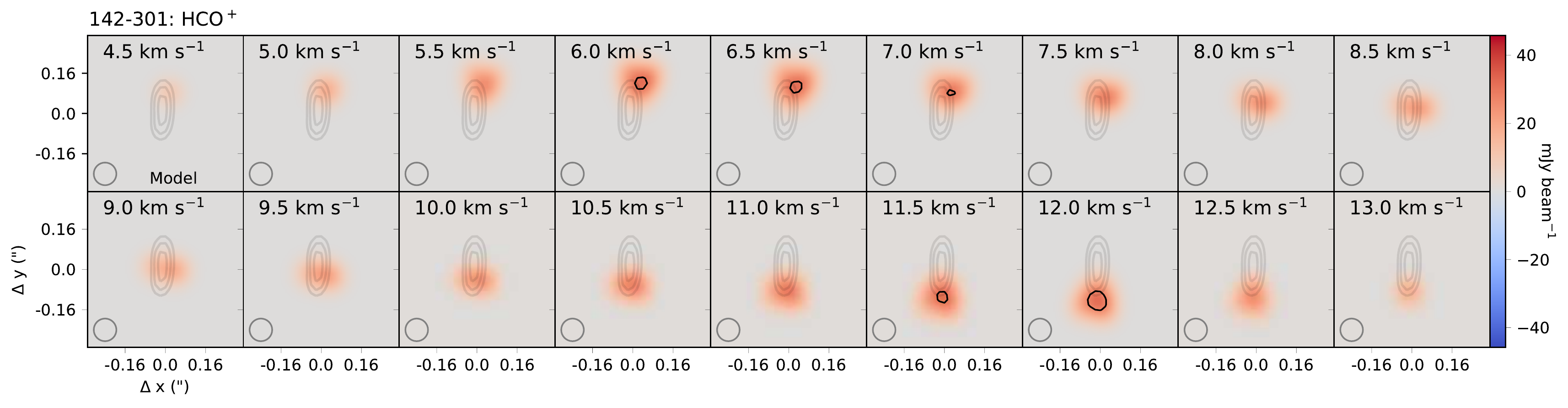}{1.0\textwidth}{}}
	\gridline{\fig{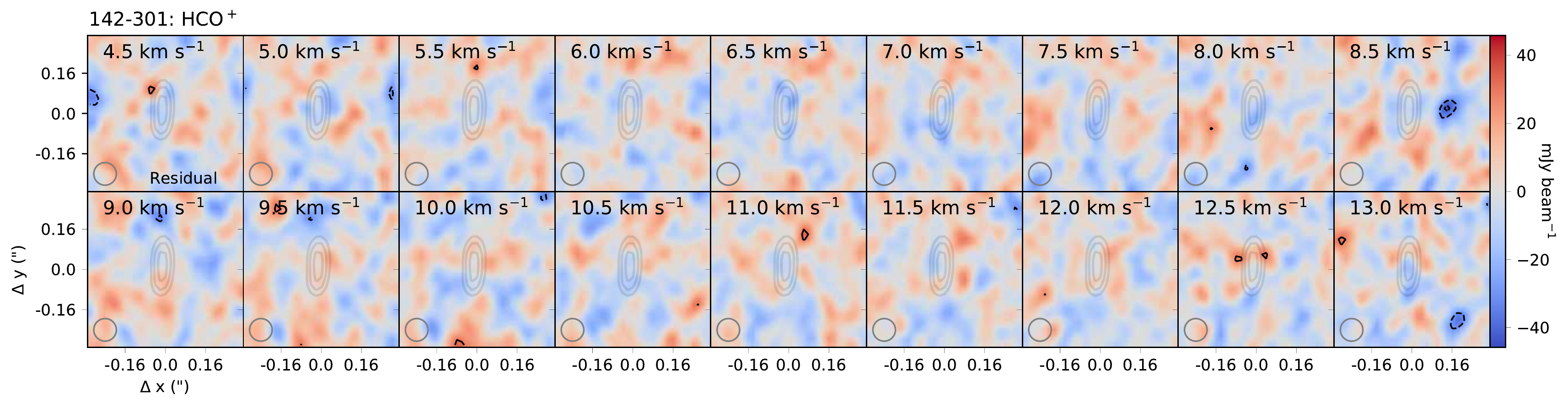}{1.0\textwidth}{}}
	\caption{Modeling results for ONC member 142-301. The layout of this plot is similar to that of Figure \ref{fig:appendix:fit_80}. \label{fig:appendix:fit_4}}
\end{figure*}

\begin{figure*}[ht!] 
	\epsscale{1.1}
	\centering	
	\gridline{\fig{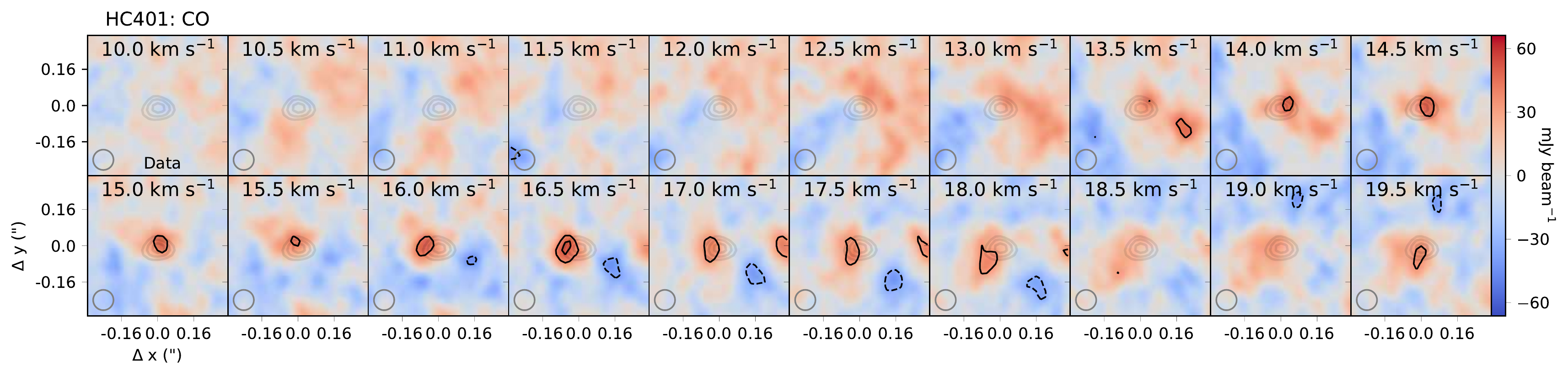}{1.0\textwidth}{}}
	\gridline{\fig{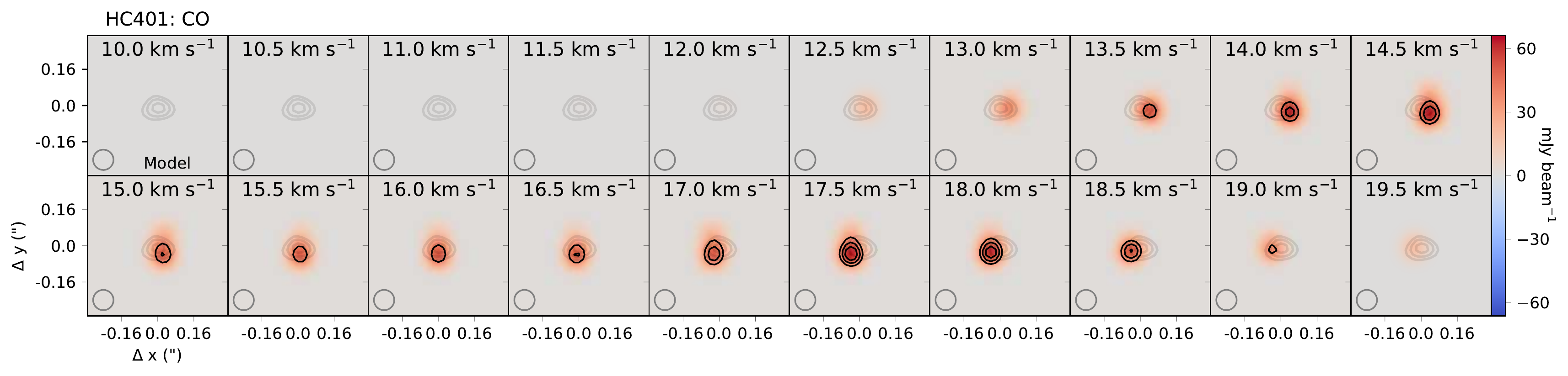}{1.0\textwidth}{}}
	\gridline{\fig{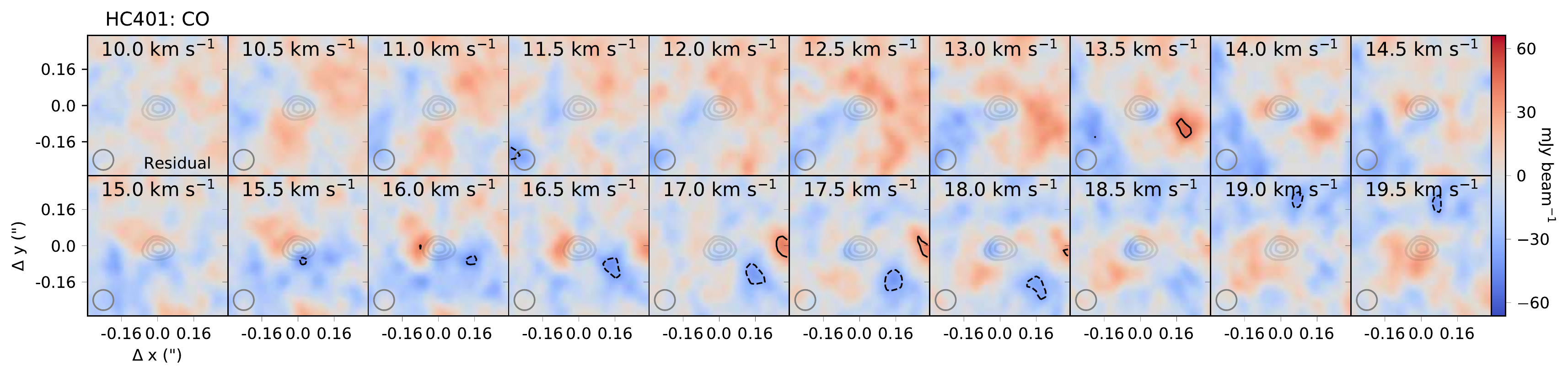}{1.0\textwidth}{}}
	\caption{Modeling results for ONC member HC401. The layout of this plot is similar to that of Figure \ref{fig:appendix:fit_65}. \label{fig:appendix:fit_44}}
\end{figure*}

\begin{figure*}[ht!] 
	\epsscale{1.1}
	\centering	
	\gridline{\fig{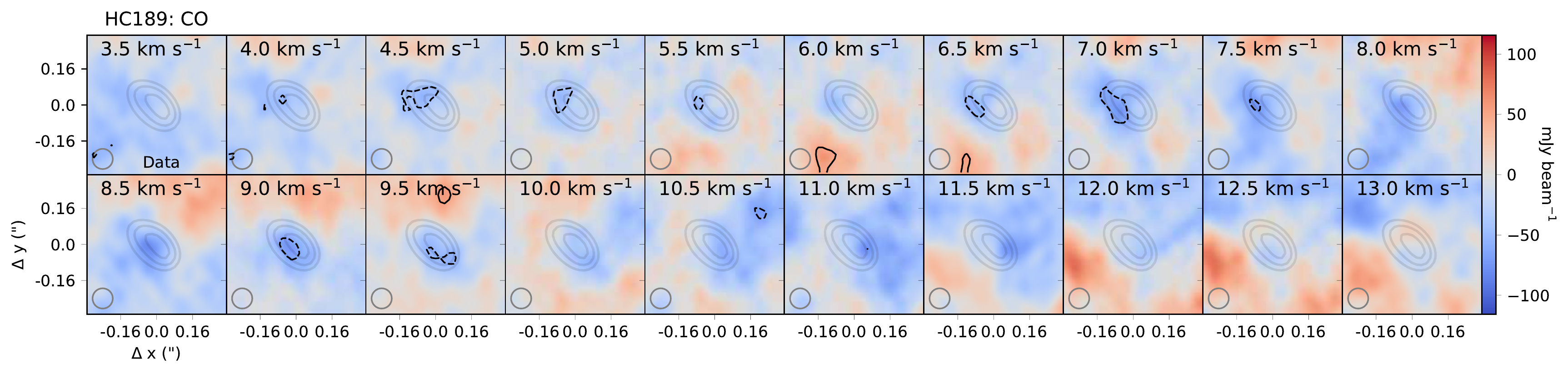}{1.0\textwidth}{}}
	\gridline{\fig{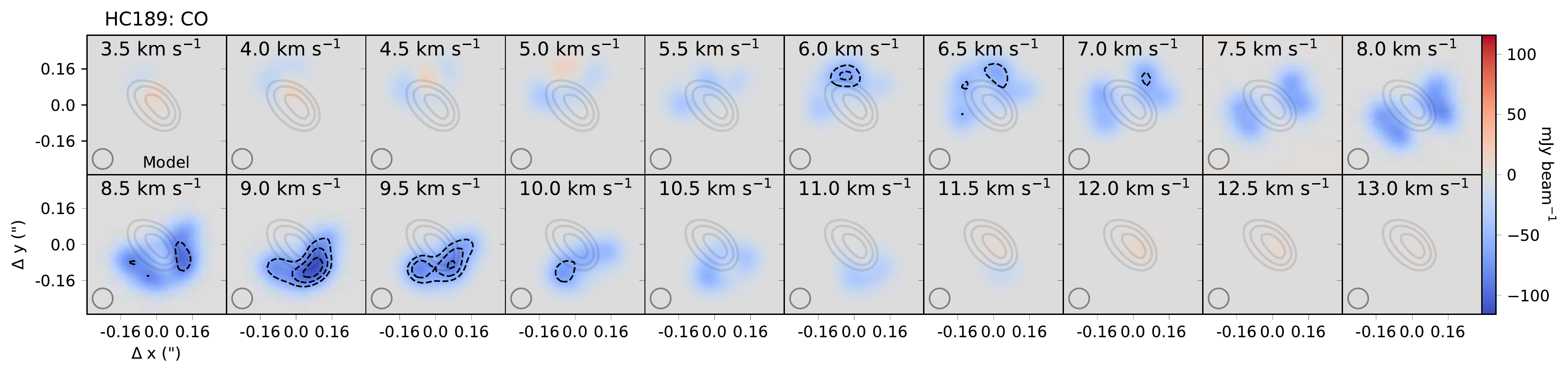}{1.0\textwidth}{}}
	\gridline{\fig{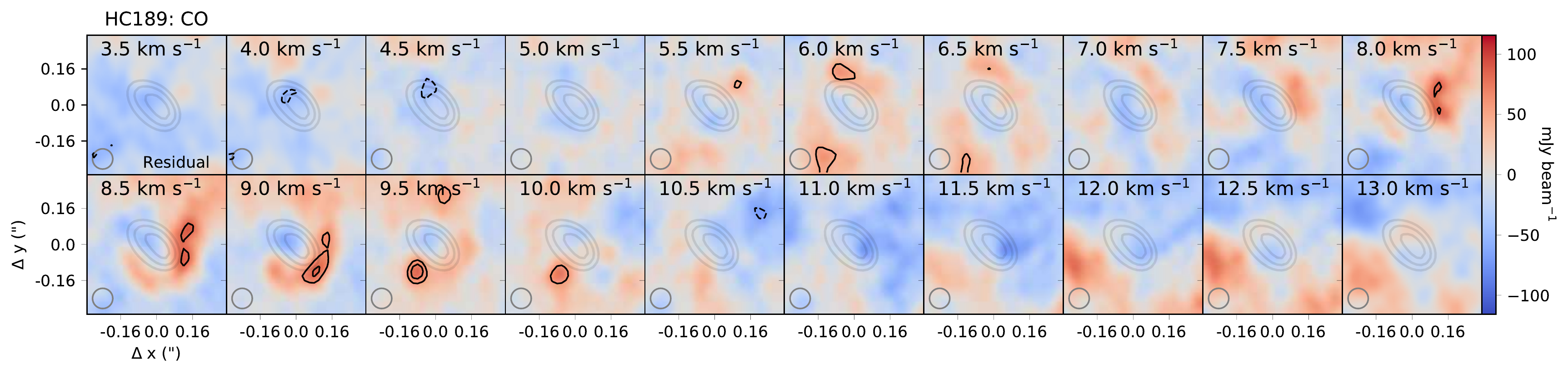}{1.0\textwidth}{}}
	\caption{Modeling results for ONC member HC189. The layout of this plot is similar to that of Figure \ref{fig:appendix:fit_65}. \label{fig:appendix:fit_10}}
\end{figure*}

\begin{figure*}[ht!] 
	\epsscale{1.1}
	\centering	
	\gridline{\fig{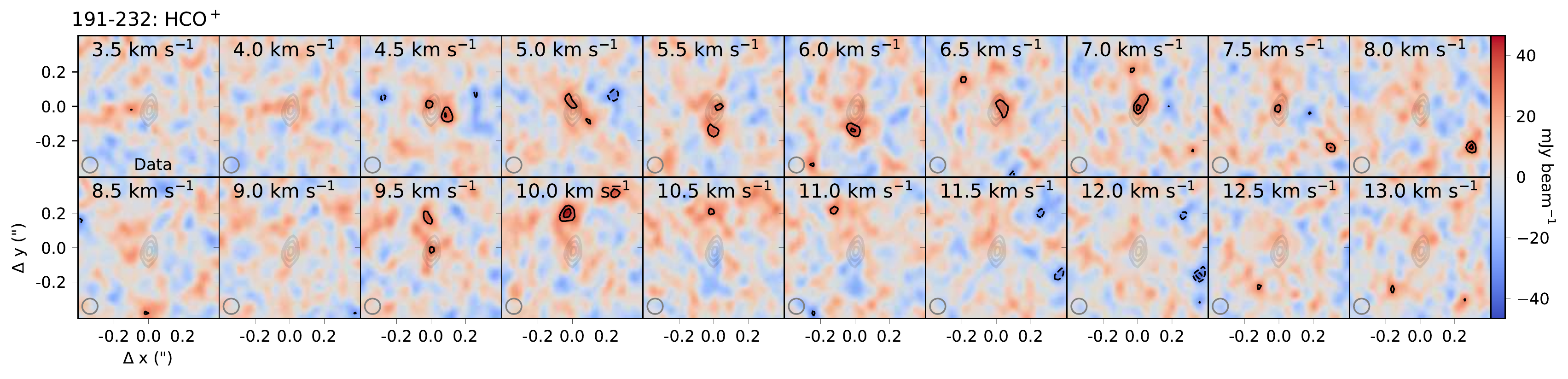}{1.0\textwidth}{}}
	\gridline{\fig{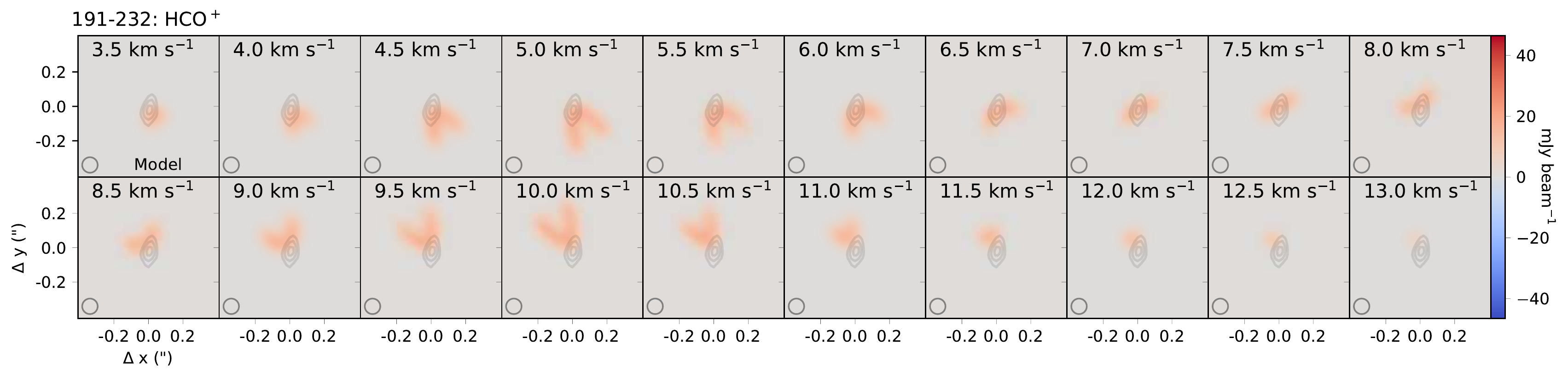}{1.0\textwidth}{}}
	\gridline{\fig{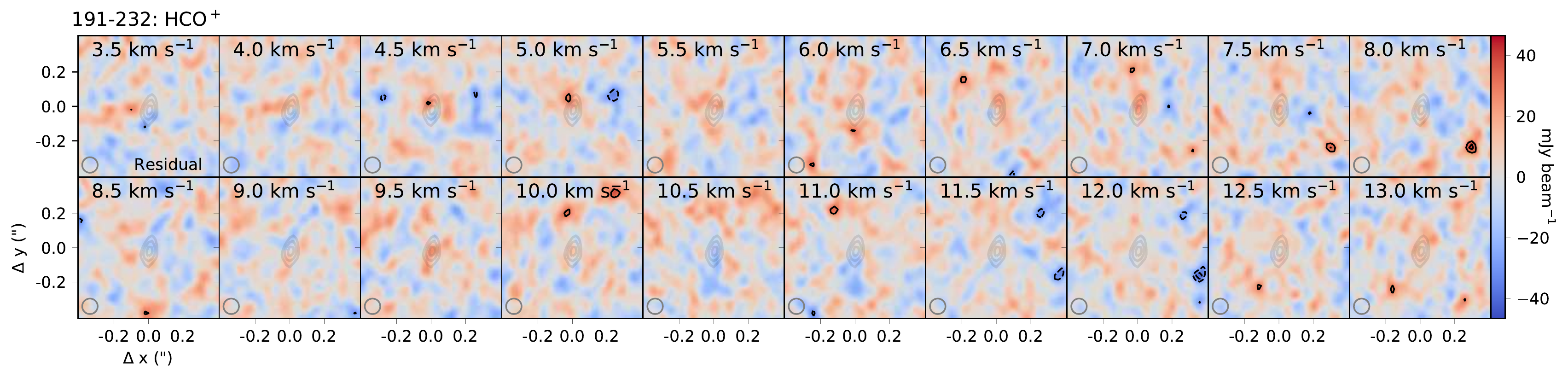}{1.0\textwidth}{}}
	\caption{Modeling results for ONC member 191-232. The layout of this plot is similar to that of Figure \ref{fig:appendix:fit_80}. \label{fig:appendix:fit_96}}
\end{figure*}

\begin{figure*}[ht!] 
	\epsscale{1.1}
	\centering	
	\gridline{\fig{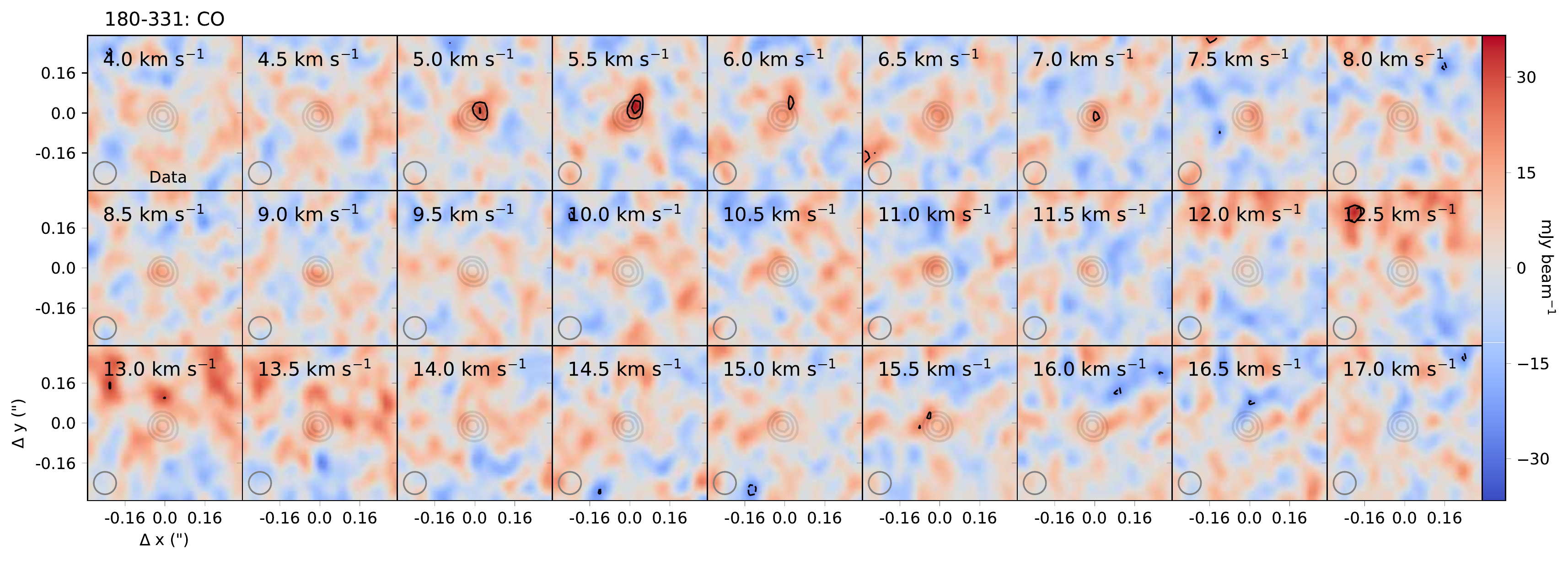}{1.0\textwidth}{}}
	\gridline{\fig{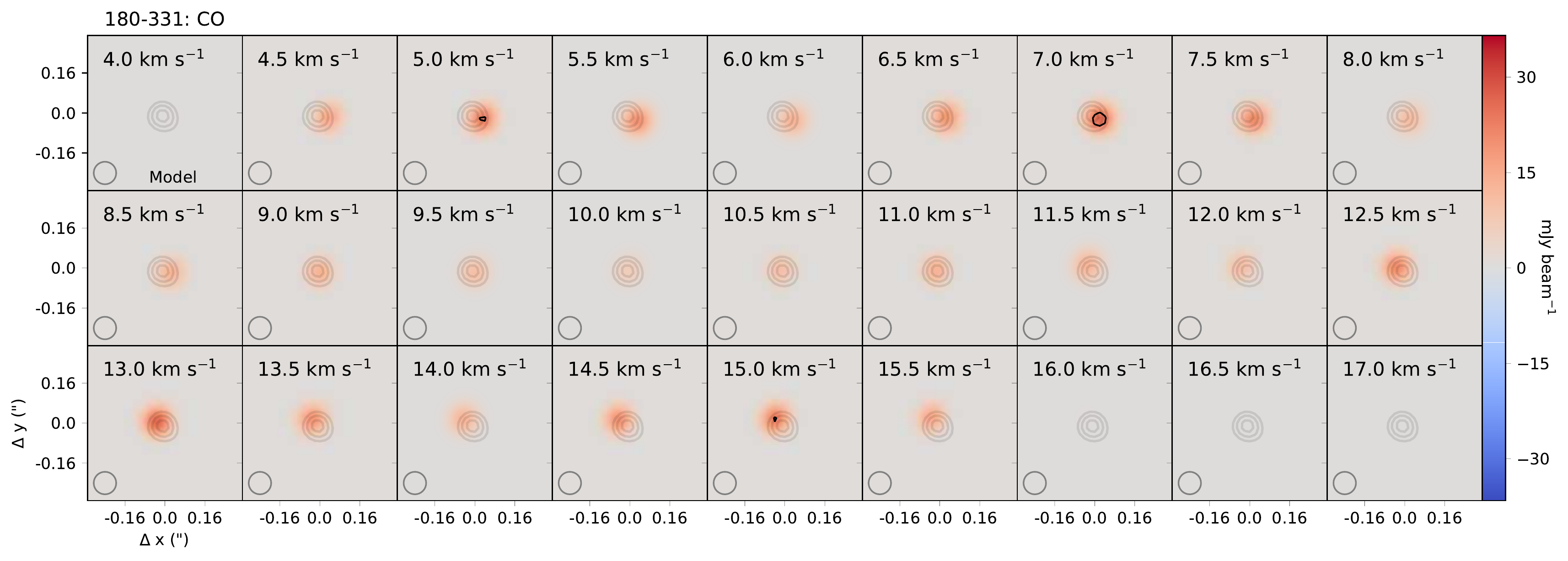}{1.0\textwidth}{}}
	\gridline{\fig{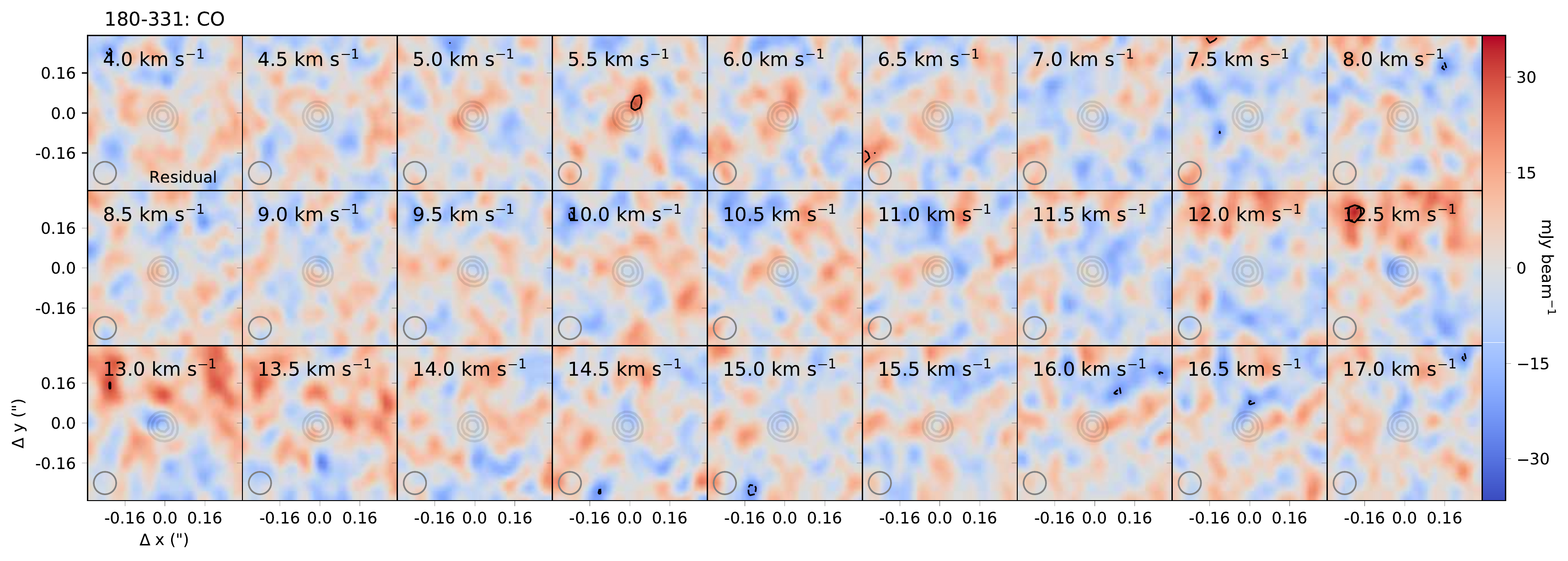}{1.0\textwidth}{}}
	\caption{Modeling results for ONC member 180-331. The layout of this plot is similar to that of Figure \ref{fig:appendix:fit_65}. \label{fig:appendix:fit_78}}
\end{figure*}

\begin{figure*}[ht!] 
	\epsscale{1.1}
	\centering	
	\gridline{\fig{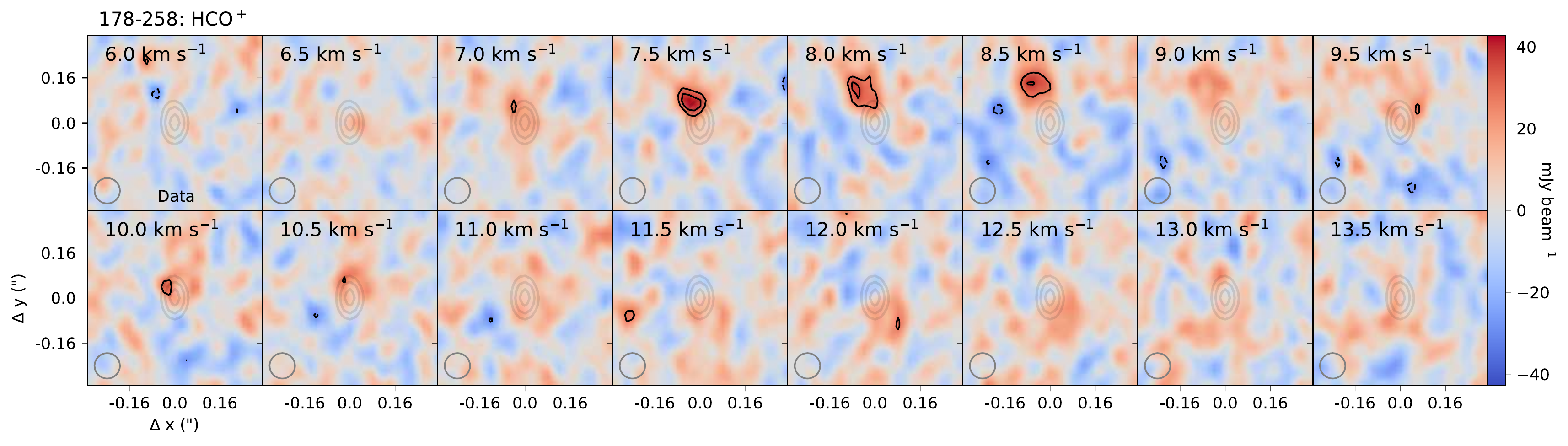}{1.0\textwidth}{}}
	\gridline{\fig{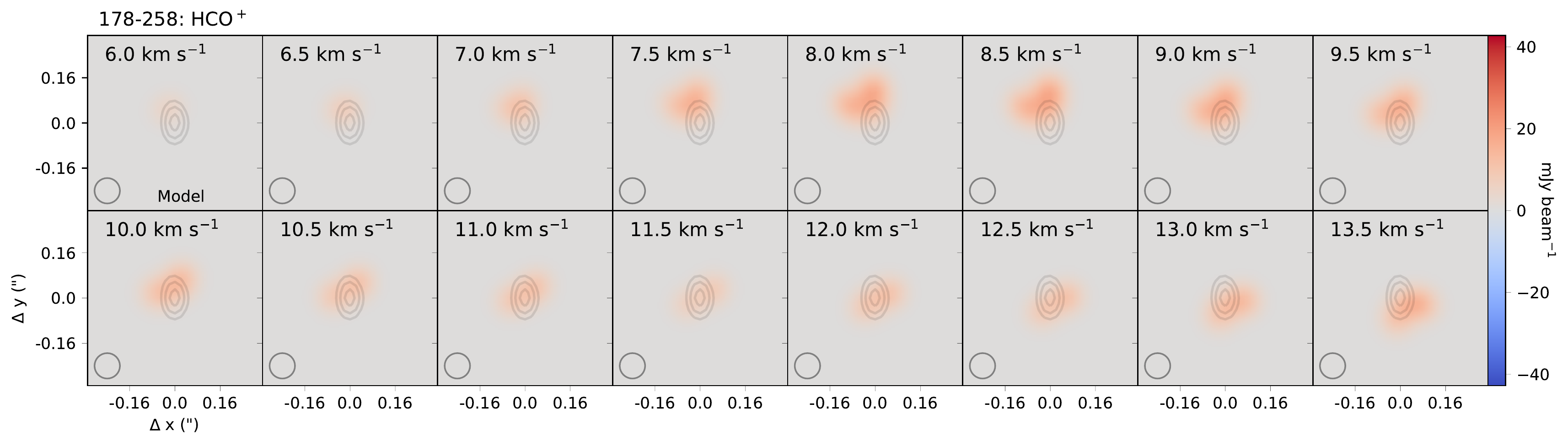}{1.0\textwidth}{}}
	\gridline{\fig{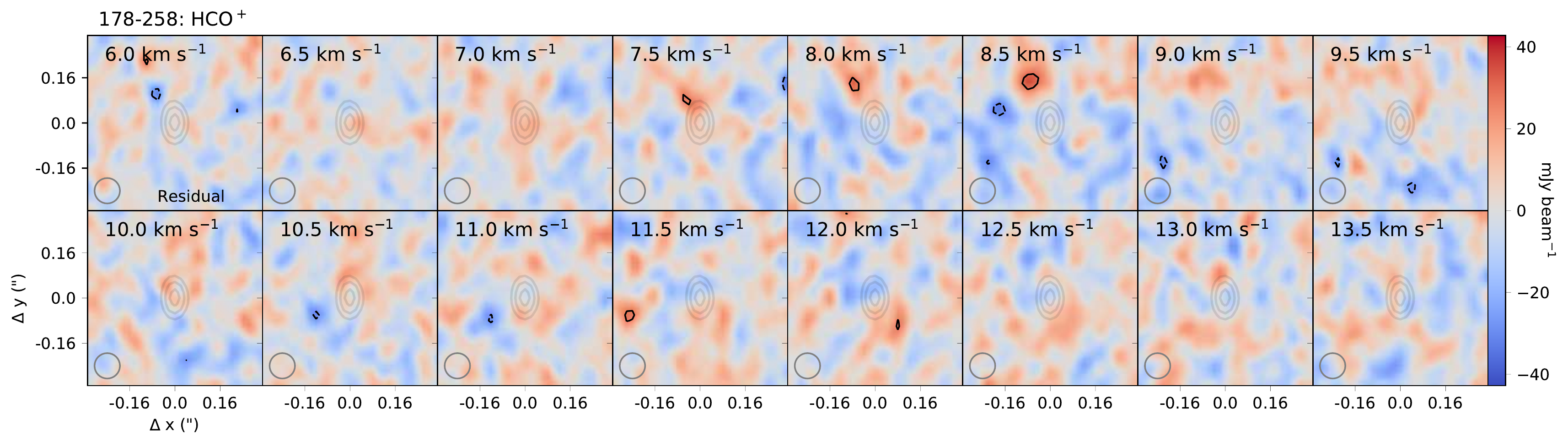}{1.0\textwidth}{}}
	\caption{Modeling results for ONC member 178-258. The layout of this plot is similar to that of Figure \ref{fig:appendix:fit_80}. \label{fig:appendix:fit_76}}
\end{figure*}

\begin{figure*}[ht!] 
	\epsscale{1.1}
	\centering	
	\gridline{\fig{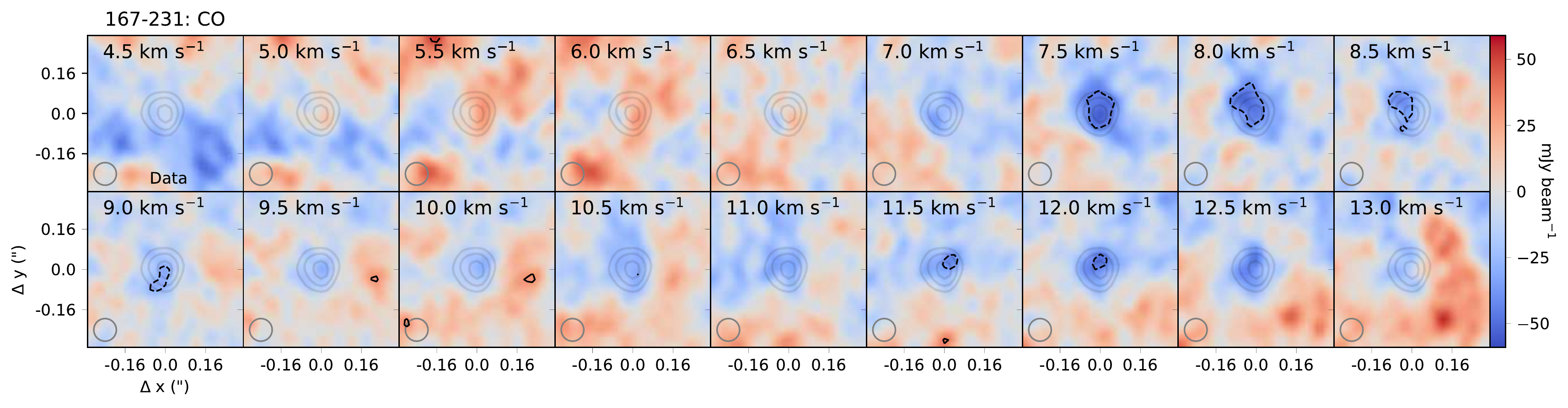}{1.0\textwidth}{}}
	\gridline{\fig{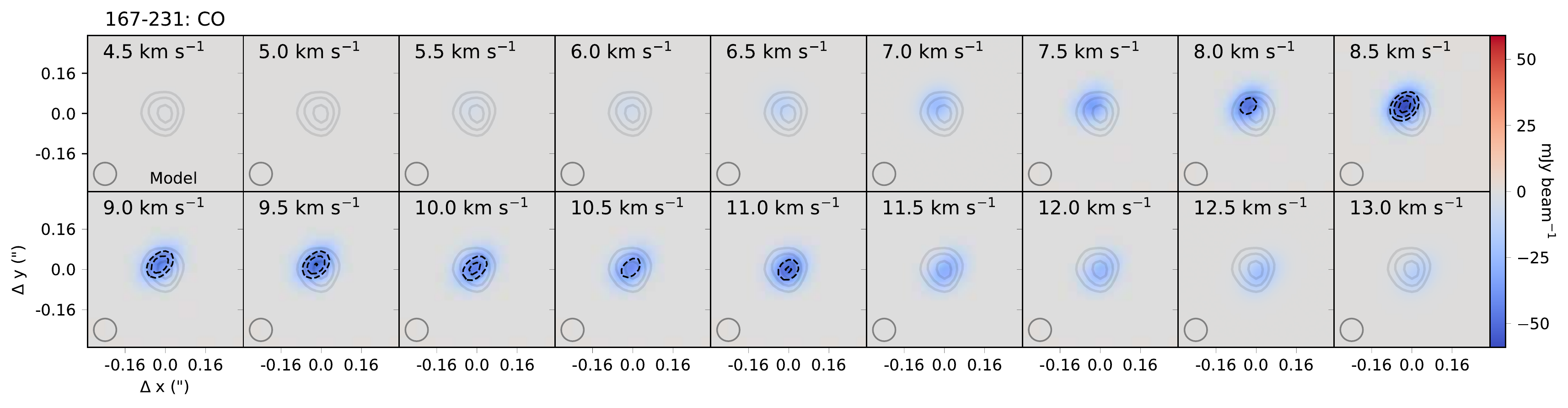}{1.0\textwidth}{}}
	\gridline{\fig{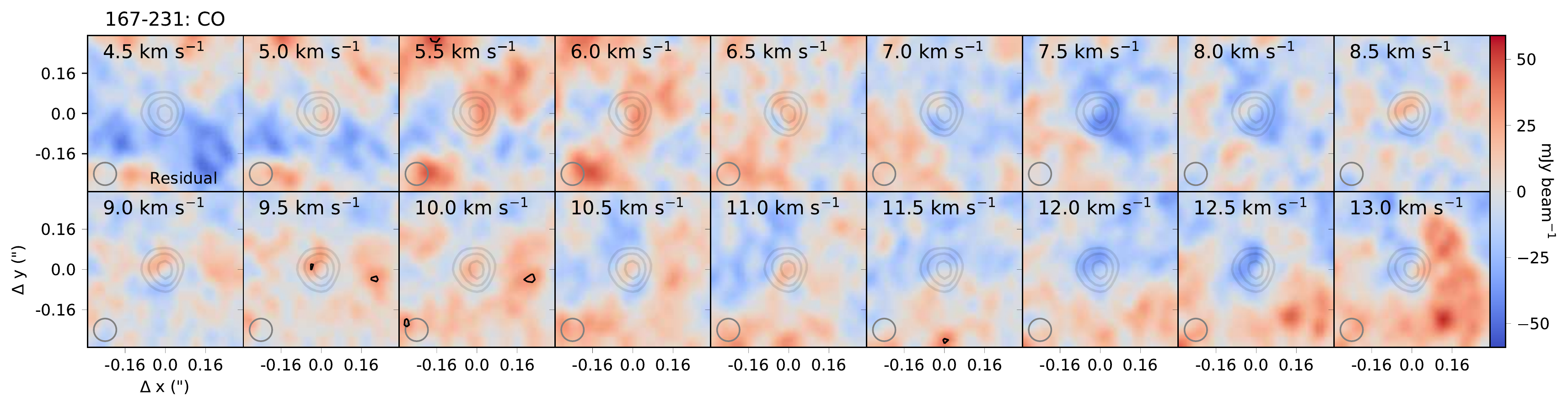}{1.0\textwidth}{}}
	\caption{Modeling results for ONC member 167-231. The layout of this plot is similar to that of Figure \ref{fig:appendix:fit_65}. \label{fig:appendix:fit_53}}
\end{figure*}

\begin{figure*}[ht!] 
	\epsscale{1.1}
	\centering	
	\gridline{\fig{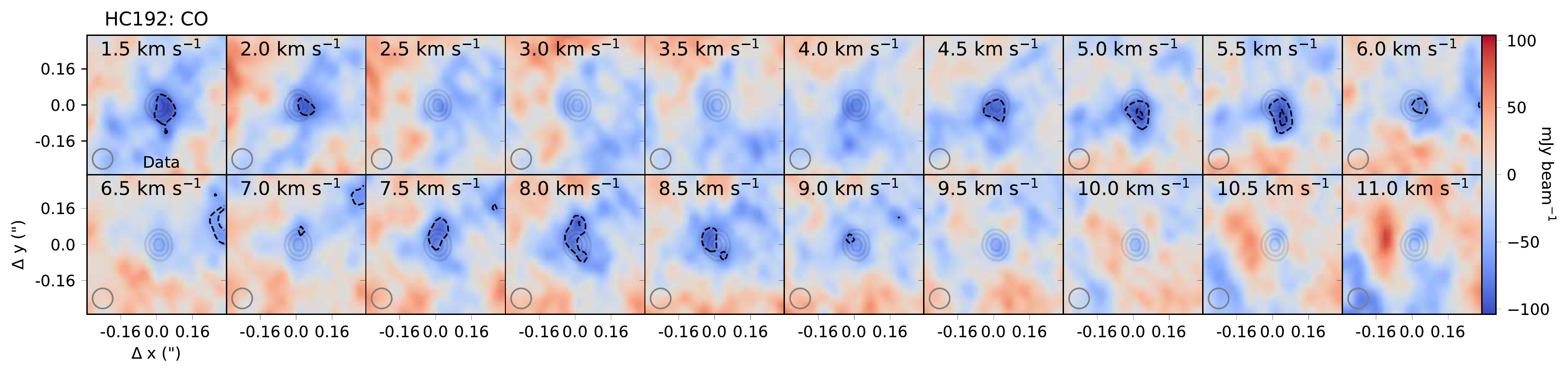}{1.0\textwidth}{}}
	\gridline{\fig{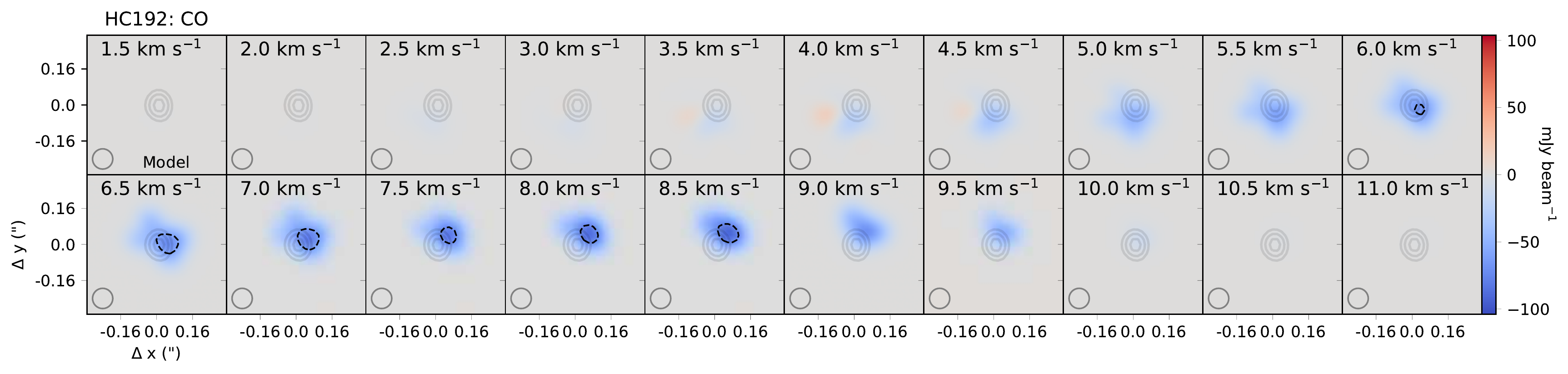}{1.0\textwidth}{}}
	\gridline{\fig{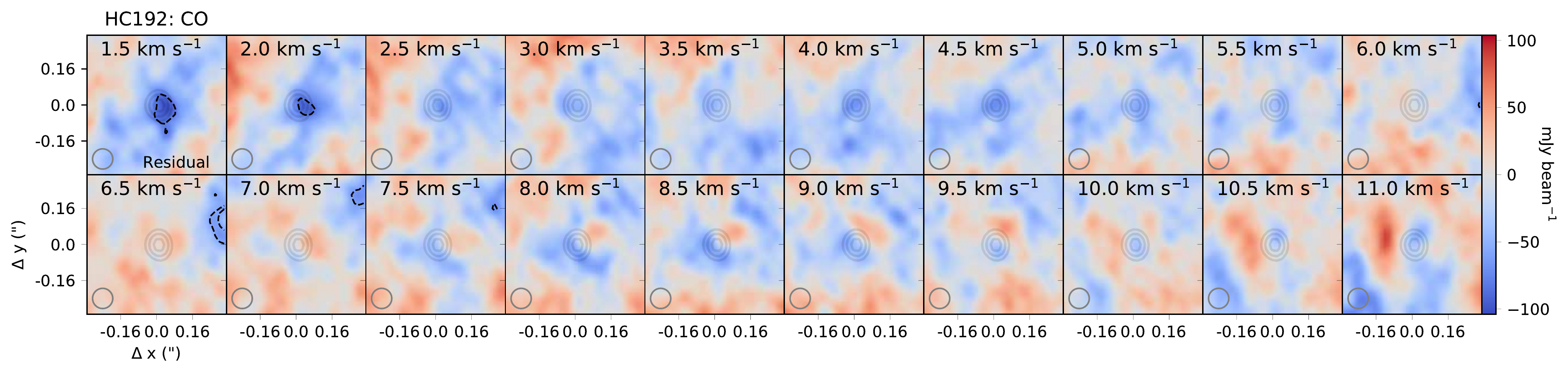}{1.0\textwidth}{}}
	\caption{Modeling results for ONC member HC192. The layout of this plot is similar to that of Figure \ref{fig:appendix:fit_65}. \label{fig:appendix:fit_1}}
\end{figure*}

\end{document}